\pdfoutput=1
\documentclass{elsarticle}
  
\usepackage[margin=1in]{geometry}

\usepackage{bbm}
\usepackage{graphicx}
\usepackage{pstool}
\usepackage{wrapfig}
\usepackage{caption}
\usepackage{subcaption}
\usepackage[section]{placeins} % Defines \FloatBarrier which does not allow

\usepackage{fancyhdr} % Required for custom headers
\usepackage{lastpage} % Required to determine the last page for the footer
\usepackage{extramarks} % Required for headers and footers

\usepackage{xcolor} % Colors
\usepackage{enumerate} % Additional options for \enumerate{}
\usepackage{paralist} % Additional options for lists; inline lists
\usepackage{amsmath, amsthm, amssymb, mathtools}
\usepackage{mathabx, pifont, stmaryrd} % More math symbols, fonts
\usepackage[explicit]{titlesec} % Control over section/subsection styles
\usepackage{etoolbox} % Boolean variables and more
\usepackage{bibentry} % Bibliography entries anywhere in doc
\makeatletter\let\saved@bibitem\@bibitem\makeatother % Do not remove if bibentry
\usepackage[colorlinks, bookmarksopen, bookmarksnumbered,
            citecolor=red,urlcolor=red]{hyperref} % Hyperlinks
\makeatletter\let\@bibitem\saved@bibitem\makeatother % Do not remove if bibentry
\usepackage{rotating}

\usepackage{algorithm}
\usepackage{algorithmic}

\theoremstyle{definition}

\usepackage{bm} % required for: \bm{}
\usepackage{amsfonts} % required for: \mathbb{}, \mathcal{}, ...

\DeclareMathOperator*{\argmin}{arg\,min}

\newcommand{\ds}[1]{\ensuremath{\displaystyle{#1}}}
\newcommand{\func}[3]{\ensuremath{#1 : #2 \rightarrow #3}}

\newcommand{\norm}[1]{\ensuremath{\left\| #1 \right\|}}

\newcommand{\suchthat}{\mathrel{}\middle|\mathrel{}}

\newcommand{\pder}[2]{\ensuremath{\frac{\partial #1}{\partial #2}}}

\newcommand{\Ecal}{\ensuremath{\mathcal{E}}}

\newcommand{\Gcal}{\ensuremath{\mathcal{G}}}
\newcommand{\Hcal}{\ensuremath{\mathcal{H}}}

\newcommand{\Ocal}{\ensuremath{\mathcal{O}}}
\newcommand{\Pcal}{\ensuremath{\mathcal{P}}}

\newcommand{\Vcal}{\ensuremath{\mathcal{V}}}
\newcommand{\Wcal}{\ensuremath{\mathcal{W}}}

\newcommand{\Gbb}{\ensuremath{\mathbb{G}}}

\newcommand{\Rbb}{\ensuremath{\mathbb{R} }}

\newcommand\Rbm{{\ensuremath{\bm{R}}}}

\newcommand\rbm{{\ensuremath{\bm{r}}}}

\newcommand\ubm{{\ensuremath{\bm{u}}}}

\newcommand\xbm{{\ensuremath{\bm{x}}}}
\newcommand\ybm{{\ensuremath{\bm{y}}}}
\newcommand\zbm{{\ensuremath{\bm{z}}}}

\newcommand\nbold{\ensuremath{\mathbf{n}}}

\newcommand\phibold{{\ensuremath{\boldsymbol{\phi}}}}

\newcommand\Psibold{{\ensuremath{\boldsymbol{\Psi}}}}

\newcommand\zerobold{\ensuremath{\mathbf{0}}}

\usepackage{tikz}
\usepackage{pgfplots}
\usepackage{pgfplotstable, filecontents, booktabs}
\pgfplotsset{compat=1.9}

\usetikzlibrary{pgfplots.groupplots}
\usepgfplotslibrary{fillbetween}
\usetikzlibrary{calc,fit,matrix,arrows,automata,positioning,shapes}
\usetikzlibrary{arrows.meta}

\pgfplotsset{select coords between index/.style 2 args={
    x filter/.code={
        \ifnum\coordindex<#1\fi
        \ifnum\coordindex>#2\fi
    }
}}

\tikzset{
 invisible/.style={opacity=0},
 visible on/.style={alt={#1{}{invisible}}},
 alt/.code args={<#1>#2#3}{%
   \alt<#1>{\pgfkeysalso{#2}}{\pgfkeysalso{#3}}
 },
}

\newcommand{\colorbarMatlabParula}[5]{
\begin{tikzpicture}
\begin{axis}[
   hide axis, scale only axis,
   height=0pt, width=0pt,
   colormap={parula}{rgb255=(62,38,168) rgb255=(62,39,172) rgb255=(63,40,175) rgb255=(63,41,178) rgb255=(64,42,180) rgb255=(64,43,183) rgb255=(65,44,186) rgb255=(65,45,189) rgb255=(66,46,191) rgb255=(66,47,194) rgb255=(67,48,197) rgb255=(67,49,200) rgb255=(67,50,202) rgb255=(68,51,205) rgb255=(68,52,208) rgb255=(69,53,210) rgb255=(69,55,213) rgb255=(69,56,215) rgb255=(70,57,217) rgb255=(70,58,220) rgb255=(70,59,222) rgb255=(70,61,224) rgb255=(71,62,225) rgb255=(71,63,227) rgb255=(71,65,229) rgb255=(71,66,230) rgb255=(71,68,232) rgb255=(71,69,233) rgb255=(71,70,235) rgb255=(72,72,236) rgb255=(72,73,237) rgb255=(72,75,238) rgb255=(72,76,240) rgb255=(72,78,241) rgb255=(72,79,242) rgb255=(72,80,243) rgb255=(72,82,244) rgb255=(72,83,245) rgb255=(72,84,246) rgb255=(71,86,247) rgb255=(71,87,247) rgb255=(71,89,248) rgb255=(71,90,249) rgb255=(71,91,250) rgb255=(71,93,250) rgb255=(70,94,251) rgb255=(70,96,251) rgb255=(70,97,252) rgb255=(69,98,252) rgb255=(69,100,253) rgb255=(68,101,253) rgb255=(67,103,253) rgb255=(67,104,254) rgb255=(66,106,254) rgb255=(65,107,254) rgb255=(64,109,254) rgb255=(63,110,255) rgb255=(62,112,255) rgb255=(60,113,255) rgb255=(59,115,255) rgb255=(57,116,255) rgb255=(56,118,254) rgb255=(54,119,254) rgb255=(53,121,253) rgb255=(51,122,253) rgb255=(50,124,252) rgb255=(49,125,252) rgb255=(48,127,251) rgb255=(47,128,250) rgb255=(47,130,250) rgb255=(46,131,249) rgb255=(46,132,248) rgb255=(46,134,248) rgb255=(46,135,247) rgb255=(45,136,246) rgb255=(45,138,245) rgb255=(45,139,244) rgb255=(45,140,243) rgb255=(45,142,242) rgb255=(44,143,241) rgb255=(44,144,240) rgb255=(43,145,239) rgb255=(42,147,238) rgb255=(41,148,237) rgb255=(40,149,236) rgb255=(39,151,235) rgb255=(39,152,234) rgb255=(38,153,233) rgb255=(38,154,232) rgb255=(37,155,232) rgb255=(37,156,231) rgb255=(36,158,230) rgb255=(36,159,229) rgb255=(35,160,229) rgb255=(35,161,228) rgb255=(34,162,228) rgb255=(33,163,227) rgb255=(32,165,227) rgb255=(31,166,226) rgb255=(30,167,225) rgb255=(29,168,225) rgb255=(29,169,224) rgb255=(28,170,223) rgb255=(27,171,222) rgb255=(26,172,221) rgb255=(25,173,220) rgb255=(23,174,218) rgb255=(22,175,217) rgb255=(20,176,216) rgb255=(18,177,214) rgb255=(16,178,213) rgb255=(14,179,212) rgb255=(11,179,210) rgb255=(8,180,209) rgb255=(6,181,207) rgb255=(4,182,206) rgb255=(2,183,204) rgb255=(1,183,202) rgb255=(0,184,201) rgb255=(0,185,199) rgb255=(0,186,198) rgb255=(1,186,196) rgb255=(2,187,194) rgb255=(4,187,193) rgb255=(6,188,191) rgb255=(9,189,189) rgb255=(13,189,188) rgb255=(16,190,186) rgb255=(20,190,184) rgb255=(23,191,182) rgb255=(26,192,181) rgb255=(29,192,179) rgb255=(32,193,177) rgb255=(35,193,175) rgb255=(37,194,174) rgb255=(39,194,172) rgb255=(41,195,170) rgb255=(43,195,168) rgb255=(44,196,166) rgb255=(46,196,165) rgb255=(47,197,163) rgb255=(49,197,161) rgb255=(50,198,159) rgb255=(51,199,157) rgb255=(53,199,155) rgb255=(54,200,153) rgb255=(56,200,150) rgb255=(57,201,148) rgb255=(59,201,146) rgb255=(61,202,144) rgb255=(64,202,141) rgb255=(66,202,139) rgb255=(69,203,137) rgb255=(72,203,134) rgb255=(75,203,132) rgb255=(78,204,129) rgb255=(81,204,127) rgb255=(84,204,124) rgb255=(87,204,122) rgb255=(90,204,119) rgb255=(94,205,116) rgb255=(97,205,114) rgb255=(100,205,111) rgb255=(103,205,108) rgb255=(107,205,105) rgb255=(110,205,102) rgb255=(114,205,100) rgb255=(118,204,97) rgb255=(121,204,94) rgb255=(125,204,91) rgb255=(129,204,89) rgb255=(132,204,86) rgb255=(136,203,83) rgb255=(139,203,81) rgb255=(143,203,78) rgb255=(147,202,75) rgb255=(150,202,72) rgb255=(154,201,70) rgb255=(157,201,67) rgb255=(161,200,64) rgb255=(164,200,62) rgb255=(167,199,59) rgb255=(171,199,57) rgb255=(174,198,55) rgb255=(178,198,53) rgb255=(181,197,51) rgb255=(184,196,49) rgb255=(187,196,47) rgb255=(190,195,45) rgb255=(194,195,44) rgb255=(197,194,42) rgb255=(200,193,41) rgb255=(203,193,40) rgb255=(206,192,39) rgb255=(208,191,39) rgb255=(211,191,39) rgb255=(214,190,39) rgb255=(217,190,40) rgb255=(219,189,40) rgb255=(222,188,41) rgb255=(225,188,42) rgb255=(227,188,43) rgb255=(230,187,45) rgb255=(232,187,46) rgb255=(234,186,48) rgb255=(236,186,50) rgb255=(239,186,53) rgb255=(241,186,55) rgb255=(243,186,57) rgb255=(245,186,59) rgb255=(247,186,61) rgb255=(249,186,62) rgb255=(251,187,62) rgb255=(252,188,62) rgb255=(254,189,61) rgb255=(254,190,60) rgb255=(254,192,59) rgb255=(254,193,58) rgb255=(254,194,57) rgb255=(254,196,56) rgb255=(254,197,55) rgb255=(254,199,53) rgb255=(254,200,52) rgb255=(254,202,51) rgb255=(253,203,50) rgb255=(253,205,49) rgb255=(253,206,49) rgb255=(252,208,48) rgb255=(251,210,47) rgb255=(251,211,46) rgb255=(250,213,46) rgb255=(249,214,45) rgb255=(249,216,44) rgb255=(248,217,43) rgb255=(247,219,42) rgb255=(247,221,42) rgb255=(246,222,41) rgb255=(246,224,40) rgb255=(245,225,40) rgb255=(245,227,39) rgb255=(245,229,38) rgb255=(245,230,38) rgb255=(245,232,37) rgb255=(245,233,36) rgb255=(245,235,35) rgb255=(245,236,34) rgb255=(245,238,33) rgb255=(246,239,32) rgb255=(246,241,31) rgb255=(246,242,30) rgb255=(247,244,28) rgb255=(247,245,27) rgb255=(248,247,26) rgb255=(248,248,24) rgb255=(249,249,22) rgb255=(249,251,21) },
   colorbar horizontal,
   point meta min=#1, point meta max=#5,
   colorbar style={width=10cm, xtick={#1,#2,#3,#4,#5}}
]
\addplot [draw=none] coordinates {(0,0)};
\end{axis}
\end{tikzpicture}
}

\usepackage{setspace}

\newbool{fastcompile}
\setbool{fastcompile}{false}

\begin{document}
\title{A shock-fitting approach to high-order simulation of viscous hypersonic flows based on inviscid implicit shock tracking}

\author[rvt1]{Huijing Dong\fnref{fn1}}
\ead{hdong2@nd.edu}

\author[rvt1]{Alexander M. Perez Reyes\fnref{fn2}}
\ead{aperez23@nd.edu}

\author[rvt1]{Matthew J. Zahr\fnref{fn3}\corref{cor1}}
\ead{mzahr@nd.edu}

\address[rvt1]{Department of Aerospace and Mechanical Engineering, University
               of Notre Dame, Notre Dame, IN 46556, United States}
\cortext[cor1]{Corresponding author}

\fntext[fn1]{Graduate Student, Department of Aerospace and Mechanical
             Engineering, University of Notre Dame}
\fntext[fn2]{Graduate Student, Department of Aerospace and Mechanical
             Engineering, University of Notre Dame}
\fntext[fn3]{Associate Professor, Department of Aerospace and Mechanical
             Engineering, University of Notre Dame}


\begin{abstract}
	This work presents a framework for computing viscous, shock-dominated flows at high Mach and Reynolds
	numbers in which the bow shock is located by inviscid implicit shock tracking and then treated as a boundary
	of the viscous computation. The high-order implicit shock tracking (HOIST) method is applied to the Euler
	equations on a coarse, shock-agnostic mesh to produce a shock-aligned mesh and the shock geometry; the
	elements upstream of the shock are discarded, the post-shock domain is remeshed to resolve the boundary layer,
	and the compressible Navier-Stokes equations are solved on it with a high-order interior penalty discontinuous
	Galerkin method, with the post-shock state prescribed along the shock boundary by the oblique shock relations.
	Because the shock structure is never resolved, the approach requires neither the extreme element compression
	nor the specialized formulations that fully coupled viscous shock tracking demands at these conditions; the
	tracking optimization is confined to the coarse inviscid mesh, and the viscous computation involves no mesh
	unknowns and no additional nonlinear stabilization. The framework is demonstrated on a Mach 10.6 axisymmetric
	blunt cone, a Mach 17.6 cylinder, and a Mach 5 sphere in three dimensions, with surface pressure and heat flux
	in good agreement with experimental measurements and reference computations.
\end{abstract}

\maketitle

\section{Introduction}
\label{sec:intro}

Blunt body flows generate strong detached bow shocks that dominate both
aerodynamic and aerothermal characteristics of hypersonic vehicles, leading to 
severe surface pressure loads and aerodynamic heating~\cite{peery1988blunt,gnoffo2003computational,candler2007unstructured}. 
Consequently, accurate prediction of 
surface pressure and heat transfer is of primary importance. This has motivated computational fluid dynamics researchers to invest substantial effort in the development 
of high-fidelity numerical methods for accurately resolving strong shock waves. However, the presence
of strong shock waves can lead to significant numerical errors in aerothermal simulations when
the computational grid is not aligned with the curvature of the leading shock~\cite{lee1999spurious,gnoffo2003computational,candler2007unstructured,
candler2009current,candler2015next}. Achieving such
alignment typically requires a user-intensive workflow involving repeated iterations between grid
generation and flow solutions~\cite{candler2009current}. These difficulties are exacerbated at higher Mach numbers and increased
vehicle bluntness~\cite{candler2015advances}, where nonlinear instabilities become more pronounced, 
further compromising numerical stability and accuracy.

Shock-dominated flows are most commonly handled using shock-capturing techniques, which
aim to stabilize the numerical solution on a fixed mesh while maintaining reasonable 
shock structures. One class of approaches relies on solution limiting~\cite{van1979towards}, particularly in 
second-order finite volume methods, to suppress nonphysical oscillations near discontinuities. 
High-order methods often employ alternative strategies, such as Weighted Essentially 
Non-Oscillatory (WENO) formulations~\cite{harten1997uniformly,jiang1996efficient,liu1994weighted}, 
which construct stable high-resolution solutions 
by combining multiple reconstruction stencils. WENO schemes are most effective on structured or Cartesian grids and are therefore less well suited to simulations involving complex geometries. Within the high-order 
discretization community, artificial-viscosity-based stabilization has become a prevalent approach. In these methods, additional dissipation is introduced locally in 
regions identified by shock sensors as including strong gradients or discontinuities~\cite{persson2006sub,
fernandez2018physics,barter2010shock,ching2019shock}. 
For hypersonic flows, increasingly large amounts of artificial viscosity are typically needed to maintain stability, which leads to a delicate trade-off between robustness 
and accuracy. Excessive dissipation can significantly modify the shock profile, and 
ultimately degrade the accuracy of surface heat-transfer predictions in the hypersonic regime~\cite{candler2015advances,lee1999spurious}.

In contrast, shock-fitting methods explicitly align the computational mesh with flow
discontinuities, enabling an exact representation of shock positions without the need for 
additional numerical stabilization. When coupled with high-order methods, these techniques 
can achieve high accuracy even on relatively coarse meshes. Traditional shock-fitting techniques 
employ the Rankine-Hugoniot jump conditions to determine the post-shock state and shock propagation, 
from which the shock motion is subsequently evolved~\cite{moretti2002thirty,
salas2009shock,ciallella2020extrapolated,bonfiglioli2016unsteady,geisenhofer2020extended,zou2021embedded}. In particular, in boundary shock
fitting~\cite{moretti1969blunt,moretti1970numerical} the bow shock forms the outer boundary of the computational domain
and the post-shock state is supplied by the Rankine-Hugoniot relations; this treatment was combined with spectral
discretizations for the viscous blunt-body problem by Kopriva~\cite{kopriva1993spectral,kopriva1996spectral}, and
shock fitting has since been extended to unstructured grids, where the shock is a moving internal
front~\cite{paciorri2009shock,bonfiglioli2016unsteady}. A primary limitation of this approach is 
the requirement to generate a mesh aligning with the shock surface; as the complexity of the shock 
topology increases, this requirement becomes more difficult to satisfy, which restricts its applicability to complex flows~\cite{johnsen2010assessment}.

A recently developed class of high-order methods, referred to as implicit shock tracking, 
addresses the primary limitations of classical shock-fitting approaches. This class includes 
the moving discontinuous Galerkin method with interface condition enforcement (MDG-ICE)~\cite{corrigan2019moving,kercher2021moving,
kercher2021least,ching2024moving} 
and the high-order implicit shock tracking (HOIST) method~\cite{zahr2018shktrk,zahr2020implicit,huang2022robust,huang2023high}. These techniques discretize the 
governing conservation laws on a shock-agnostic mesh and simultaneously determine 
both the discrete flow solutions and the mesh coordinates by solving an optimization 
problem. This formulation eliminates the need for explicitly shock-aligned meshes and achieves automatic alignment of the grid with complex shock topologies. As a result, these 
methods are equation- and problem-independent and are capable of recovering the design-order 
accuracy of the underlying high-order discretization for non-smooth flows.

Within implicit shock tracking, viscous shocks have been treated as continuous features of finite thickness
that are resolved by compressing elements into the shock layer. The MDG-ICE method was the first to do so,
for compressible viscous flows~\cite{kercher2021moving} and subsequently for hypersonic viscous flows at high
Mach and Reynolds numbers~\cite{ching2024moving}, where the anisotropic refinement required to resolve the shock
structure is extreme. Our previous work~\cite{dong2025rpadapt} followed the same approach within the HOIST
framework, combining anisotropic $r$-adaptivity with $p$-adaptivity to resolve viscous shock layers at moderate
Mach and Reynolds numbers. Resolving the shock structure in this way is possible, but the required solution
procedure is delicate: elements must be compressed by orders of magnitude without inverting, which demands
careful continuation and solver safeguards~\cite{ching2024moving}, and $p$-adaptivity only partially relieves
the mesh compression. These difficulties grow with Mach and Reynolds number, which is precisely the regime of
interest in this work.

For high Mach and Reynolds number flows considered in this work, the shock thickness becomes negligibly 
small relative to the characteristic length scales, allowing the shock to be treated as a sharp 
discontinuity. Indeed, Moretti~\cite{moretti1969blunt,moretti1970numerical} argued that, for high Reynolds number flows, modeling the shock as a true discontinuity is the natural choice, since the physical shock structure cannot be resolved at any practical mesh resolution. Under the sharp-shock assumption, the location of the bow shock is governed by the inviscid dynamics, and the
viscous effects of interest, namely boundary layers and surface heat transfer, are confined to the post-shock
region. This observation motivates a two-stage strategy. In the first stage, the HOIST method is applied to the
Euler equations to obtain a shock-aligned mesh and the corresponding inviscid solution; Mach number continuation
is used to reach the target freestream conditions robustly~\cite{huang2023high,dong2025rpadapt}. In the second
stage, the elements upstream of the shock are discarded, the post-shock domain is remeshed to resolve the
boundary layer, and the compressible Navier-Stokes equations are solved on this domain with the shock treated as
a boundary on which the post-shock state is prescribed by the oblique shock relations. The viscous problem is
discretized with a high-order interior penalty discontinuous Galerkin (IPDG)
method~\cite{arnold1982interior,arnold2002unified,hartmann2008optimal,hartmann2014higher}, advanced to steady
state with pseudo-transient continuation~\cite{modisette2011automated}, and adapted using a dual-weighted
residual error estimate for the wall heat flux~\cite{fidkowski2011review}. Because the shock is the only
non-smooth feature and it coincides with a boundary of the post-shock domain, no shock-aligned optimization
problem and no nonlinear stabilization are required in the viscous stage.

This two-stage strategy is motivated by the cost and fragility of fully coupled viscous shock tracking at the
conditions of interest. Resolving a viscous shock requires compressing elements to the viscous shock thickness, which
is orders of magnitude smaller than the nose radius at the Reynolds numbers considered here. This has been
achieved for such problems only through a multi-stage workflow built on a specialized least-squares variant of
MDG-ICE~\cite{ching2024moving}. The decoupled
strategy sidesteps the issue entirely: the shock structure is never resolved, the shock position is obtained with the
standard inviscid HOIST solver on a coarse Euler mesh, and the viscous solution is computed with a standard IPDG
discretization on a shock-free, boundary-layer-refined domain with no mesh unknowns. The optimization problem, which
is the expensive and delicate part of shock tracking, is therefore solved only on the coarse inviscid mesh, while the
viscous computation reduces to a conventional nonlinear solve. 

The contributions of this work are: (i) a decoupled framework in which the shock-fitted boundary is obtained
automatically by inviscid implicit shock tracking on a coarse, shock-agnostic Euler mesh, and the viscous problem is
then solved on the post-shock domain with a standard high-order IPDG discretization and no mesh unknowns, so that the
tracking optimization never involves viscous terms, element compression to the shock thickness, or a specialized
tracking formulation~\cite{dong2025rpadapt,ching2024moving}; (ii) demonstration of the framework at realistic
hypersonic conditions, including a Mach 10.6 axisymmetric blunt cone validated against experimental heat transfer
measurements, a Mach 17.6 cylinder compared with coupled viscous tracking results, and a Mach 5 sphere compared with
reference Navier-Stokes computations; and (iii) a shock boundary condition for
the discontinuous Galerkin discretization that prescribes the post-shock state in closed form from the freestream
state and the local shock normal, the discrete counterpart of classical boundary shock
fitting~\cite{moretti1969blunt,kopriva1993spectral} with the shock position frozen at its inviscid location.

The remainder of this paper is organized as follows. Section \ref{sec:govern} introduces a general system of viscous conservation laws and its specialization to the compressible Navier-Stokes equations, its reformulation on a fixed reference domain, and its discretization using an IPDG method;
the section also introduces the treatment of the shock boundary conditions. Section \ref{sec:istshkbc} 
outlines the overall computational framework, including the HOIST formulation originally proposed in \cite{zahr2018shktrk,
zahr2020implicit,huang2022robust}, farfield element removal, viscous flow computation, and the dual weighted 
residual (DWR) error estimate used to drive mesh adaptation. 
Finally, Section \ref{sec:numexp} demonstrates the flexibility of the proposed methodological framework
for two- and three-dimensional shock-dominated viscous problems, including hypersonic flows over a blunt cone, 
a cylinder, and a sphere with accurate heat flux profile predictions. Section 
\ref{sec:conclude} offers conclusions.

\section{Governing equations, high-order discretization, and shock boundary conditions}
\label{sec:govern}
In this section, we introduce a general system of steady viscous conservation laws and its specialization to the compressible Navier-Stokes equations (Section \ref{subsec:ns}), transform the system of conservation laws to a reference domain such that domain
deformations appear explicitly in the governing equations (Section \ref{subsec:transfdom}), discretize the
transformed equations using a high-order DG method (Section \ref{subsec:ipdg}), and finally introduce 
a shock boundary condition (Section \ref{subsec:oblshk}).

\subsection{Navier-Stokes equations}
\label{subsec:ns}
We consider steady compressible, viscous flow through a domain
$\Omega \subset \Rbb^d$  with governing equations given by
\begin{equation}\label{eqn:nsii2d}
	\begin{split}
		\pder{}{x_j}\left(\rho v_j \right) & = 0 \\
		\pder{}{x_j}\left(\rho v_i v_j  + P\delta_{ij}\right) & = \pder{}{x_j} \varpi_{ij}  \\
		\pder{}{x_j}\left(\left(\rho E+P\right)v_j\right) &= \pder{}{x_j} \left(\varpi_{ij} v_i - q_j\right)
	\end{split}
\end{equation}
for $i = 1, \cdots, d$. The density of the fluid $\func{\rho}{\Omega}{\Rbb_{>0}}$,
the fluid velocity $\func{v}{\Omega}{\Rbb^d}$, and the total energy of the fluid
$\func{\rho E}{\Omega}{\Rbb_{>0}}$ are implicitly defined
as the solution of (\ref{eqn:nsii2d}). We assume the fluid is an ideal gas
with thermal and caloric state equations
\begin{equation}\label{eqn:press}
	P = \rho R T, \quad e = \frac{1}{\gamma - 1} \frac{P}{\rho},
\end{equation}
where $P : \Omega \rightarrow \Rbb_{>0}$ and $T : \Omega \rightarrow \Rbb_{>0}$
are the pressure and temperature of the fluid, $e : \Omega \rightarrow \Rbb_{>0}$
is the specific internal energy, $R \in \Rbb_{>0}$ is the specific gas constant,
and $\gamma > 1$ is the ratio of specific heats ($\gamma = 1.4$ in this work). 
The sound speed $c : \Omega \rightarrow \Rbb_{>0}$ and Mach number $M : \Omega \rightarrow \Rbb_{>0}$ 
are defined as 
\begin{equation}
	c^2 = \frac{\gamma P}{\rho}, \quad M = \frac{\sqrt{v_iv_i}}{c}.
\end{equation}
Then for a calorically perfect gas, from the definition of the specific total energy $\func{E}{\Omega}{\Rbb_{>0}}$ 
and total enthalpy $\func{H}{\Omega}{\Rbb_{>0}}$
\begin{equation}
	E = e + \frac{1}{2}v_i v_i, \quad H = \frac{\rho E + P}{\rho},
\end{equation}
and ideal gas caloric state equation (\ref{eqn:press}), we obtain the following direct relationship between pressure and conservative variables for an ideal gas 
\begin{equation}\label{eqn:press2}
	P = \left(\gamma - 1 \right ) \left(\rho E - \rho v_i v_i/2\right).
\end{equation}
Assuming a Newtonian fluid with heat transfer that obeys Fourier's
law, the shear stress,
$\func{\varpi}{\Omega}{\Rbb^{d\times d}}$,
and heat flux,
$\func{q}{\Omega}{\Rbb^d}$, are related to
the primitive flow variables as
\begin{equation}
	\varpi_{ij} = \mu\left(\pder{v_i}{x_j}+\pder{v_j}{x_i}-\frac{2}{3}\pder{v_k}{x_k}\delta_{ij}\right), \qquad
	q_i = -\lambda \pder{T}{x_i},
\end{equation}
where $\func{\mu}{\Omega}{\Rbb_{>0}}$ is the dynamic viscosity, given by Sutherland's law, and
$\func{\lambda}{\Omega}{\Rbb_{>0}}$ is the thermal conductivity, $\lambda = c_P \mu / \text{Pr}$,
where $\text{Pr} = 0.72$ is the Prandtl number and $c_P \in \Rbb_{>0}$ is the specific heat capacity at constant pressure.

All variables are nondimensionalized by the freestream density $\rho_\infty^*$, the freestream sound speed
$c_\infty^*$, the freestream temperature $T_\infty^*$, and a characteristic length $L^*$ (the nose radius for the
problems in this work)~\cite{masatsuka2009like}, where a superscript $*$ denotes a dimensional quantity, i.e.,
\begin{equation}
	\rho = \frac{\rho^*}{\rho_\infty^*}, \quad v = \frac{v^*}{c_\infty^*}, \quad P = \frac{P^*}{\rho_\infty^* (c_\infty^*)^2}, \quad
	T = \frac{T^*}{T_\infty^*}, \quad x = \frac{x^*}{L^*}, \quad \mu = \frac{\mu^*}{\rho_\infty^* c_\infty^* L^*}.
\end{equation}
The nondimensional freestream state is therefore $\rho_\infty = 1$, $P_\infty = 1/\gamma$, $T_\infty = 1$, and
$v_\infty = M_\infty \nbold_\infty$, where $M_\infty$ is the freestream Mach number and $\nbold_\infty$ is the unit vector
in the freestream flow direction, and the nondimensional gas constant and specific heat at constant pressure are
$R = 1/\gamma$ and $c_P = \gamma R/(\gamma-1) = 1/(\gamma-1)$. Let $\text{Re}_\infty = \rho_\infty^* \norm{v_\infty^*} L^* / \mu_\infty^*$
denote the freestream Reynolds number, where $\mu_\infty^*$ is the dimensional freestream viscosity. Then the
nondimensional freestream viscosity is $\mu_\infty = \mu_\infty^* / (\rho_\infty^* c_\infty^* L^*) = M_\infty/\text{Re}_\infty$
and Sutherland's law takes the nondimensional form
\begin{equation}\label{eqn:mu-nondim}
	\mu = \mu_\infty \frac{1 + C}{T + C}\, T^{3/2}, \qquad C = \frac{C^*}{T_\infty^*},
\end{equation}
where $C^* = 110$~K is the Sutherland constant. In this work, $T_\infty^* = 293.15$~K is used, giving $C = 110/293.15$.
The parameter $\mu_\infty$ is the only quantity through which the freestream conditions enter the viscous terms, and
the inviscid (Euler) equations correspond to $\mu_\infty = 0$.

The Navier-Stokes equations are a specific instance of a general system of viscous conservation laws. Consider a general system of $m$ conservation laws over $\Omega \subset \Rbb^d$
\begin{equation} \label{eqn:vclaw-phys0}
	\nabla \cdot F(U, \nabla U; \mu_\infty) = S(U, \nabla U) \quad \text{in}~~ \Omega,
\end{equation}
where $\func{U}{\Omega}{\Rbb^m}$ is implicitly defined as the solution of (\ref{eqn:vclaw-phys0}),
$\func{F}{\Rbb^m \times \Rbb^{m \times d}}{\Rbb^{m\times d}}$
is the flux function, and $\func{S}{\Rbb^m\times\Rbb^{m\times d}}{\Rbb^{m}}$ is
the source term, $\mu_\infty \in \Rbb_{\geq 0}$ is the freestream viscosity that parametrizes the viscous flux (the viscosity field $\mu$ is determined from the state through Sutherland's law \eqref{eqn:mu-nondim}), and $\ds{\nabla \coloneqq (\partial_{x_1},\dots,\partial_{x_d})}$
is the gradient operator in the physical domain. We assume the flux function 
can be split into an inviscid and viscous term if $\mu_\infty \neq 0$ as
\begin{equation}\label{eqn:vclaw-phys1}
	\nabla \cdot F_{\text{inv}}(U) - \nabla \cdot F_{\text{visc}}(U,\nabla U; \mu_\infty) = S(U,\nabla U), \end{equation} where $\func{F_\text{inv}}{\Rbb^m}{\Rbb^{m\times d}}$ is the inviscid flux function,
 $\func{F_\text{visc}}{\Rbb^m\times\Rbb^{m\times d}}{\Rbb^{m\times d}}$ is
the viscous flux function. 
Furthermore, we assume the viscous flux function is linear
in its state gradient
\begin{equation}
	F_{\text{visc}}(U, \nabla U; \mu_\infty) = D(U; \mu_\infty) : \nabla U,
\end{equation}
where
$\func{D}{\Rbb^m}{\Rbb^{m \times d \times m \times d}}$ is the viscous tensor, and 
the tensor $D(U; 0) = 0$ if the viscosity is zero.

The steady compressible Navier-Stokes equations in (\ref{eqn:nsii2d}) can be written in the form (\ref{eqn:vclaw-phys1}) ($m = d + 2$) as
\begin{equation}
	\begin{aligned}
		U = \begin{bmatrix}
			\rho \\
			\rho v \\
			\rho E
		\end{bmatrix},
		\qquad
		F_{\text{inv}}(U) = \begin{bmatrix}
			\rho v^T\\
			\rho v v^T + P I \\
			\rho H v^T
		\end{bmatrix},
		\qquad
		F_{\text{visc}}(U, \nabla U; \mu_\infty) = \begin{bmatrix}
			0\\
			\varpi \\
			v^T \varpi - q ^T
		\end{bmatrix},
		\qquad
		S(U, \nabla U) = 0.
	\end{aligned}
\end{equation}
For axisymmetric flows without swirl, which arise for the blunt cone problem in Section~\ref{subsec:cone2d}, the equations
are solved in the meridional plane ($d = 2$) with $x_2$ the axis of symmetry and $x_1$ the radial coordinate. In this
setting, the Navier-Stokes equations retain the form \eqref{eqn:vclaw-phys1} with the fluxes above, provided the
dilatation in the shear stress is replaced by its axisymmetric counterpart,
$\nabla\cdot v = \partial v_1/\partial x_1 + \partial v_2/\partial x_2 + v_1/x_1$, and the source term is
\begin{equation}\label{eqn:axisrc}
	S(U, \nabla U) = -\frac{1}{x_1}\left( F(U, \nabla U)\, e_1 - \left(P - \varpi_{\theta\theta}\right) e_2 \right), \qquad
	\varpi_{\theta\theta} = \mu\left(2\frac{v_1}{x_1} - \frac{2}{3}\nabla\cdot v\right),
\end{equation}
where $e_1 \in \Rbb^d$ and $e_2 \in \Rbb^m$ are canonical unit vectors (so $F e_1$ is the radial flux and $e_2$ selects the
radial momentum component of the state) and $\varpi_{\theta\theta}$ is the circumferential normal stress. The source term
depends explicitly on $x_1$, which is suppressed in the notation of \eqref{eqn:vclaw-phys0}; on the reference domain
this dependence enters through the mapping $\Gcal$.

\subsection{Transformed conservation laws on a fixed reference domain}
\label{subsec:transfdom}

Before discretizing (\ref{eqn:vclaw-phys1}), it is convenient to transform the domain of
the conservation law $\Omega$ to a fixed reference domain $\Omega_0\subset\Rbb^d$.
 Let $\Gbb$ be the collection of diffeomorphisms from
the reference domain to the physical domain; i.e., for any
$\Gcal\in\Gbb$, $\func{\Gcal}{\Omega_0}{\Omega}$.
Then the conservation law on the physical domain is transformed to
a conservation law on the reference domain as
\begin{equation} \label{eqn:vclaw-ref}
	\bar\nabla\cdot \bar{F}_\text{inv}(\bar{U};G) - \bar\nabla\cdot \bar{F}_\text{visc}(\bar{U}, \bar\nabla \bar{U};G, \mu_\infty) = \bar{S}(\bar{U}, \bar\nabla \bar{U};G) \quad \text{in}~~ \Omega_0,
\end{equation}
where $\func{\bar{U}}{\Omega_0}{\Rbb^m}$, $\bar{U} : X \mapsto \bar{U}(X)$ is the solution
of the transformed conservation law,
$\func{\bar{F}_\text{inv}}{\Rbb^m\times\Rbb^{d \times d}}{\Rbb^{m\times d}}$
is the transformed inviscid flux function, $\func{\bar{F}_\text{visc}}{\Rbb^m \times \Rbb^{m \times d} \times \Rbb^{d \times d}}{\Rbb^{m\times d}}$ is the transformed viscous flux
function,
$\func{\bar{S}}{\Rbb^m \times \Rbb^{m \times d} \times \Rbb^{d\times d}}{\Rbb^{m}}$ 
is the transformed source term, and
$\ds{\bar\nabla \coloneqq (\partial_{X_1},\dots,\partial_{X_d})}$
is the gradient operator in the reference domain. The deformation gradient
$\func{G}{\Omega_0}{\Rbb^{d \times d}}$ and Jacobian are defined as
\begin{equation}
	G : X \mapsto \bar\nabla \Gcal(X), \quad g : X \mapsto \det G(X).
\end{equation}
The reference domain solution is related to the solution over the physical domain as
\begin{equation}
	\bar{U}(X) = U(\Gcal(X)), \quad
	\bar{\nabla} \bar{U}(X) = \nabla U(\Gcal(X)) G(X),
\end{equation}
for any $X \in \Omega_0$. The unit normal in the reference and physical domain are related by
\begin{equation}\label{eqn:nrmltransf}
	n = \frac{gG^{-T}N}{\left\lVert gG^{-T}N \right\rVert},
\end{equation}
where $N$ and $n$ are the unit normal vectors in the reference and physical domains~\cite{zahr2020implicit}, respectively. 
If we take the reference and physical domains to be the same, then the transformation
is an identity mapping, i.e., $G = I_d$.
The fluxes and source terms transform as 
\begin{equation} \label{eqn:transf}
	\begin{aligned}
		\bar{F}_\text{inv} : (\bar{U}; G) &\mapsto
		g F_\text{inv}(\bar{U}) G^{-T} \\
		\bar{F}_\text{visc} : (\bar{U}, \bar{\nabla} \bar{U}; G, \mu_\infty) &\mapsto
		g F_\text{visc}(\bar{U}, \bar{\nabla} \bar{U} G^{-1}; \mu_\infty) G^{-T} \\
		\bar{S} : (\bar{U}, \bar{\nabla} \bar{U}; G) &\mapsto
		g S(\bar{U}, \bar{\nabla} \bar{U}G^{-1}).
	\end{aligned}
\end{equation}
The transformed viscous flux still maintains linearity with respect to its second argument~\cite{dong2025rpadapt}
\begin{equation} \label{eqn:vflx-ref}
	\bar{F}_\text{visc}(\bar{U}, \bar{\nabla} \bar{U}; G,\mu_\infty) = \bar{D}(\bar{U}; G,\mu_\infty) : \bar{\nabla} \bar{U}.
\end{equation}

\subsection{Interior penalty discontinuous Galerkin discretization of the transformed conservation law}
\label{subsec:ipdg}

We use an IPDG method~\cite{arnold1982interior,arnold2002unified,hartmann2008optimal,hartmann2014higher} to discretize the transformed
conservation law (\ref{eqn:vclaw-ref}), which reduces to the standard nodal discontinuous Galerkin method when
the viscosity vanishes, i.e., $\mu_\infty = 0$. Let $\Ecal_{h}$ represent a discretization of the reference domain $\Omega_0$
into non-overlapping computational elements. We define the space of discontinuous piecewise polynomials as \begin{equation} \label{eqn:trial} \Vcal^p_h = \left\{v \in [L^2(\Omega_0)]^m \suchthat \left.v\right|_K \in [\Pcal_p(K)]^m,~\forall K\in\Ecal_{h}\right\}, \end{equation} where $\Vcal^p_h$ is the trial space for the flow solution ($\bar{U}$) and $\Pcal_p(K)$ is the space of polynomial functions of degree at most $p \geq 0$ on the element $K$. Furthermore, we define the space of globally continuous piecewise 
polynomials of degree $q$ associated with the mesh $\Ecal_{h}$ as
\begin{equation} \label{eqn:gfcnsp}
	\Wcal_{h} = \left\{v \in H^1(\Omega_0) \suchthat
	\left.v\right|_K \in
	\Pcal_q(K),~\forall K\in\Ecal_{h}\right\},
\end{equation}
and discretize the domain mapping with the corresponding vector-valued space $[\Wcal_{h}]^d$.
Taking the DG test space $\Vcal^{p'}_h$, where $p' \geq p$, the elemental residual form is given by $\func{r_K^{p',p}}{\Vcal_h^{p'} \times \Vcal_h^p \times [\Wcal_h]^d}{\Rbb}$
\begin{equation}\label{eqn:eleres}
	\begin{aligned}
		r_K^{p',p} :
		(\bar\psi_h, \bar{U}_h; \Gcal_h) \mapsto
		& \int_{\partial K \setminus \partial \Omega_0} \bar\psi_h^+ \cdot \left(\bar\Hcal(\bar{U}_h^+,\bar{U}_h^-,N_h;\bar\nabla\Gcal_h) - \hat{\sigma}(\bar{U}_h^+,\bar{U}_h^-, \bar\nabla \bar{U}_h^+, \bar\nabla \bar{U}_h^-; \bar\nabla\Gcal_h )N_h\right) \, dS \\
		& - \frac{1}{2}\int_{\partial K \setminus \partial \Omega_0}  \left(\bar\nabla\bar\psi_h^+ : \bar{D}(\bar{U}_h^+;\bar\nabla\Gcal_h,\mu_\infty)\right)  :  \left[(\bar{U}_h^+ - \bar{U}_h^-)\otimes N_h\right] \, dS \\
		& + \int_{\partial K \cap \partial \Omega_0} \bar\psi_h^+ \cdot \left(\bar{F}^\partial_\text{inv}(\bar{U}_h^+, \bar{U}^{\partial}, N_h; \bar\nabla\Gcal_h) - \bar{\sigma}^\partial(\bar{U}^{\partial},\bar{U}_h^+, \bar\nabla \bar{U}_h^+, N_h; \bar\nabla\Gcal_h ) \right) \, dS \\
		& - \int_{\partial K \cap \partial \Omega_0}  \left(\bar\nabla\bar\psi_h^+ : \bar{D}(\bar{U}^{\partial};\bar\nabla\Gcal_h,\mu_\infty)\right) : \left[(\bar{U}_h^+ - \bar{U}^{\partial})\otimes N_h\right]
		\, dS \\
		& - \int_{K} \bar\nabla \bar\psi_h : \left ( \bar{F}_\text{inv}(\bar{U}_h; \bar\nabla\Gcal_h) - \bar{D}(\bar{U}_h;\bar\nabla\Gcal_h,\mu_\infty) : \bar\nabla\bar{U}_h \right ) \, dV \\
		& - \int_K \bar\psi_h \cdot \bar{S}(\bar{U}_h, \bar\nabla\bar{U}_h; \bar\nabla\Gcal_h) \, dV,
	\end{aligned}
\end{equation}
where, for $A, B \in \Rbb^{m\times d}$ and $C \in \Rbb^{m\times d\times m\times d}$, the contractions are defined as $A : B = A_{ij}B_{ij}$, $(C : B)_{ij} = C_{ijkl}B_{kl}$, and $(A : C)_{kl} = A_{ij}C_{ijkl}$ (summation over repeated indices), $N_h$ is the unit outward normal to element $K \in \Ecal_{h}$, $\bar{U}_h^+(\bar{U}_h^-)$
denotes the interior (exterior) trace of $\bar{U}_h \in \Vcal_{h}^p$
to the element, $\bar{U}^{\partial} = \bar{U}^{\partial}(\bar{U}_h^+, N_h)$ denotes the boundary state on $\partial \Omega_0$, which depends on the boundary condition, the interior trace, and the unit normal, and is discussed in Section \ref{subsec:oblshk}, $\func{\bar{F}^\partial_\text{inv}}{\Rbb^m \times \Rbb^m \times \Rbb^d \times \Rbb^{d\times d}}{\Rbb^m}$ is the inviscid boundary flux, which for all boundary types except the shock boundary is the transformed inviscid flux evaluated at the boundary state and contracted with the normal, $\bar{F}^\partial_\text{inv}(\bar{U}_h^+, \bar{U}^{\partial}, N_h; \bar\nabla\Gcal_h) = \bar{F}_\text{inv}(\bar{U}^{\partial}; \bar\nabla\Gcal_h) N_h$ (the shock boundary is treated in Section~\ref{subsec:oblshk}), and $\bar\Hcal$ is the transformed inviscid numerical flux function~\cite{zahr2020implicit}, which is related to the physical numerical flux function $\Hcal$ as
\begin{equation}\label{eqn:nflx}
	\bar\Hcal(\bar{U}_h^+,\bar{U}_h^-,N_h;\bar\nabla\Gcal_h) = \left\lVert gG^{-T}N_h \right\rVert \Hcal(U_h^+,U_h^-,n_h).
\end{equation}
The transformed viscous numerical flux function,
$\hat\sigma : \Rbb^m \times \Rbb^m \times \Rbb^{m\times d} \times \Rbb^{m\times d} \times \Rbb^{d\times d} \rightarrow \Rbb^{m\times d}$,
is defined using the approach in \cite{hartmann2008optimal}
\begin{equation}
	\hat\sigma : (\bar{U}_h^+,\bar{U}_h^-,\bar\nabla\bar{U}_h^+,\bar\nabla\bar{U}_h^-; \bar\nabla\Gcal_h) \mapsto \left\{\!\!\left\{\bar{D}(\bar{U}_h; \bar\nabla\Gcal_h,\mu_\infty) : \bar\nabla\bar{U}_h\right\}\!\!\right\} - \delta_\mathrm{IP}(\bar{U}_h^+, \bar{U}_h^-; \bar\nabla\Gcal_h),
\end{equation}
where $\{\!\!\{v\}\!\!\} = (v^+ + v^-)/2$ is the average operator and
$\delta_\mathrm{IP} : \Rbb^m \times \Rbb^m \times \Rbb^{d\times d} \rightarrow \Rbb^{m\times d}$ is
the penalty term
\begin{equation}
		\left.\delta_\mathrm{IP}\right|_{\partial K \cap \partial K'} : (\bar{U}_h^+, \bar{U}_h^-; \bar\nabla\Gcal_h) \mapsto C_\mathrm{IP}\frac{p(K)^2}{h} \{\!\!\{\bar{D}(\bar{U}_h; \bar\nabla\Gcal_h,\mu_\infty)\}\!\!\}: \left[(\bar{U}_h^+ - \bar{U}_h^-)\otimes N_h\right],
\end{equation}
with $h = \min\{|K|, |K'|\} / |\partial K \cap \partial K'|$ being the linear mesh size
associated with the face $\overline K \cap \overline K'$
and $C_\mathrm{IP} \in \Rbb_{> 0}$ a user-defined stabilization parameter. The transformed viscous boundary flux 
$\bar{\sigma}^\partial : \Rbb^m \times \Rbb^m \times \Rbb^{m\times d} \times \Rbb^d \times \Rbb^{d\times d} \rightarrow \Rbb^{m}$ is defined as 
\begin{equation}
	\bar{\sigma}^\partial :(\bar{U}^{\partial},\bar{U}_h^+, \bar\nabla \bar{U}_h^+, N_h; \bar\nabla\Gcal_h) \mapsto \bar{F}_{\text{visc}}^\partial(\bar{U}^{\partial}, \bar\nabla \bar{U}_h^+, N_h; \bar\nabla\Gcal_h) - \delta_\mathrm{IP}^\partial(\bar{U}^{\partial}, \bar{U}_h^+; \bar\nabla\Gcal_h) N_h
\end{equation}
where $\bar{F}_{\text{visc}}^\partial : \Rbb^m \times \Rbb^{m\times d} \times \Rbb^d \times \Rbb^{d\times d} \rightarrow \Rbb^{m}$ is the transformed boundary viscous flux (the normal component of the viscous flux on the boundary), which is specified for each boundary type in Section \ref{sec:numexp}, and 
$\delta_\mathrm{IP}^\partial : \Rbb^m \times \Rbb^m \times \Rbb^{d\times d} \rightarrow \Rbb^{m\times d}$ is
the boundary penalty term 
\begin{equation}
	\left.\delta_\mathrm{IP}^\partial\right|_{\partial K \cap \partial \Omega_0} : (\bar{U}^{\partial}, \bar{U}_h^+; \bar\nabla\Gcal_h) \mapsto C_\mathrm{IP}\frac{p(K)^2}{h} \bar{D}(\bar{U}^{\partial}; \bar\nabla\Gcal_h,\mu_\infty): \left[(\bar{U}_h^+ - \bar{U}^{\partial})\otimes N_h\right],
\end{equation}
where $h$ is the linear mesh size
associated with the face $\partial K \cap \partial \Omega_0$, $h = |K| / |\partial K \cap \partial \Omega_0|$. 

We also split the elemental residual form into two parts,
\begin{equation}
	r_K^{p',p}(\bar\psi_h, \bar{U}_h; \Gcal_h) = r_{K,\text{inv}}^{p',p} + r_{K,\text{vis}}^{p',p},
\end{equation}
where $r_{K,\text{inv}}^{p',p}$ is the inviscid residual, and $r_{K,\text{vis}}^{p',p}$ is
the viscous residual. From these definitions, the primal residual of the IPDG method is given by
$\func{r_h^{p',p}}{\Vcal_h^{p'} \times \Vcal_h^p \times [\Wcal_h]^d}{\Rbb}$ 
\begin{equation}
	r_h^{p',p} :
	(\bar\psi_h, \bar{U}_h; \Gcal_h) \mapsto
	\sum_{K\in\Ecal_h}  r_{K}^{p',p}(\bar\psi_h,\bar{U}_h,\Gcal_h).
\end{equation}
Finally, after introducing a basis for the test, trial, and domain mapping spaces, the governing equations with $p' = p$ reduce
to the algebraic residual form 
\begin{equation} \label{eqn:dg-alg}
	\rbm : \Rbb^{N_\ubm}\times\Rbb^{N_\xbm} \rightarrow \Rbb^{N_\ubm}, \qquad
	\rbm : (\ubm,\xbm;\mu_\infty) \mapsto \rbm_{\text{inv}}(\ubm,\xbm) + \rbm_{\text{vis}}(\ubm,\xbm;\mu_\infty),
\end{equation}
where $N_\ubm = \dim\Vcal_{h}^p$ and $N_\xbm = \dim([\Wcal_{h}]^d)$, $\ubm\in\Rbb^{N_\ubm}$ are the coefficients of the flow solution ($\bar{U}_h$),
and $\xbm\in\Rbb^{N_\xbm}$ are the coefficients of the domain mapping ($\Gcal_h$). In addition, when the test space has degree $p' \geq p$, we denote the algebraic enriched residual as
\begin{equation} \label{eqn:dg-plus1-alg}
	\Rbm : \Rbb^{N_\ubm}\times\Rbb^{N_\xbm} \rightarrow \Rbb^{N_{\ubm}^{'}}, \qquad
	\Rbm : (\ubm,\xbm;\mu_\infty) \mapsto \Rbm_{\text{inv}}(\ubm,\xbm) + \Rbm_{\text{vis}}(\ubm,\xbm;\mu_\infty),
\end{equation}
where $N_{\ubm}^{'} = \dim\Vcal_{h}^{p'}$, which will be used to formulate the implicit shock tracking method.

\subsection{Shock boundary conditions}
\label{subsec:oblshk}
The IPDG elemental residual \eqref{eqn:eleres} involves boundary integrals over $\partial K \cap \partial \Omega_0$, 
whose evaluation depends on the prescribed boundary states entering the numerical fluxes. 
In this work, the shock is modeled as a computational boundary. The corresponding boundary 
states are therefore determined by the shock boundary conditions derived from the oblique 
shock relations. We now describe the construction of these shock boundary states and their 
incorporation into the boundary numerical fluxes, which provide the exterior states required in 
IPDG residual evaluation.

\begin{figure}[!htbp]
	\centering
	\begin{tikzpicture}[scale=2.25]

\draw[dashed, gray] (0.0,0) -- (2.0,0);
\draw[dashed, gray] (1.0, 1.0) -- (2.5, 1.0);

\draw[red, very thick]
  (0,0) -- ({3*cos(45)},{3*sin(45)})
  node[pos=1.0, text=black, above right] {Oblique shock};

\draw[blue, very thick, -{Stealth[length=3mm,width=2mm]}]
  (-0.8, 1.0) -- (1.0, 1.0);

\draw[blue, thick, -{Stealth[length=3mm,width=2mm]}]
  (-0.8, 1.0) -- (0.1, 1.9)
  node[midway, sloped, above, fill=white,
  inner sep=1pt, font=\footnotesize] {$w_1$, $M_{t, 1}$};

\draw[blue, thick, -{Stealth[length=3mm,width=2mm]}]
  (0.1, 1.9) -- (1.0, 1.0)
  node[midway, sloped, above, fill=white,
  inner sep=1pt, font=\footnotesize] {$u_1$, $M_{n, 1}$};

\draw[blue, thick, -{Stealth[length=3mm,width=2mm]}]
  (1.0, 1.0) -- (1.0 + 0.3674234614174768, 1.0 - 0.3674234614174768)
  node[midway, sloped, below,
  inner sep=1pt, font=\footnotesize, fill opacity=0.85,
  text opacity=1] {$u_2$, $M_{n, 2}$};

\draw[blue, thick, -{Stealth[length=3mm,width=2mm]}]
  (1.0 + 0.3674234614174768, 1.0 - 0.3674234614174768) -- (1.0 + 0.3674234614174768 + 0.6363961030678928, 1.0 - 0.3674234614174768 + 0.6363961030678928)
  node[midway, sloped, below, fill=white,
  inner sep=1pt, font=\footnotesize] {$w_2$, $M_{t, 2}$};

\draw[blue, very thick, -{Stealth[length=3mm,width=2mm]}]
  (1.0, 1.0) -- (1.0 + 0.3674234614174768 + 0.6363961030678928, 1.0 - 0.3674234614174768 + 0.6363961030678928)
  node[midway, sloped, above, fill=white,
  inner sep=1pt, font=\footnotesize] {$V_{2}$, $M_2$};

\node at (1.55, 1.75) {$P_1$};
\node at (1.85, 1.65) {$P_2$};

\node[
fill=white,
text=blue,
inner sep=1pt
] at (0.0, 0.9) {$V_1,M_1$};

\end{tikzpicture}
	\caption{
		Oblique shock schematic~\cite{anderson2011ebook}.
	}
	\label{fig:oblshks}
\end{figure}
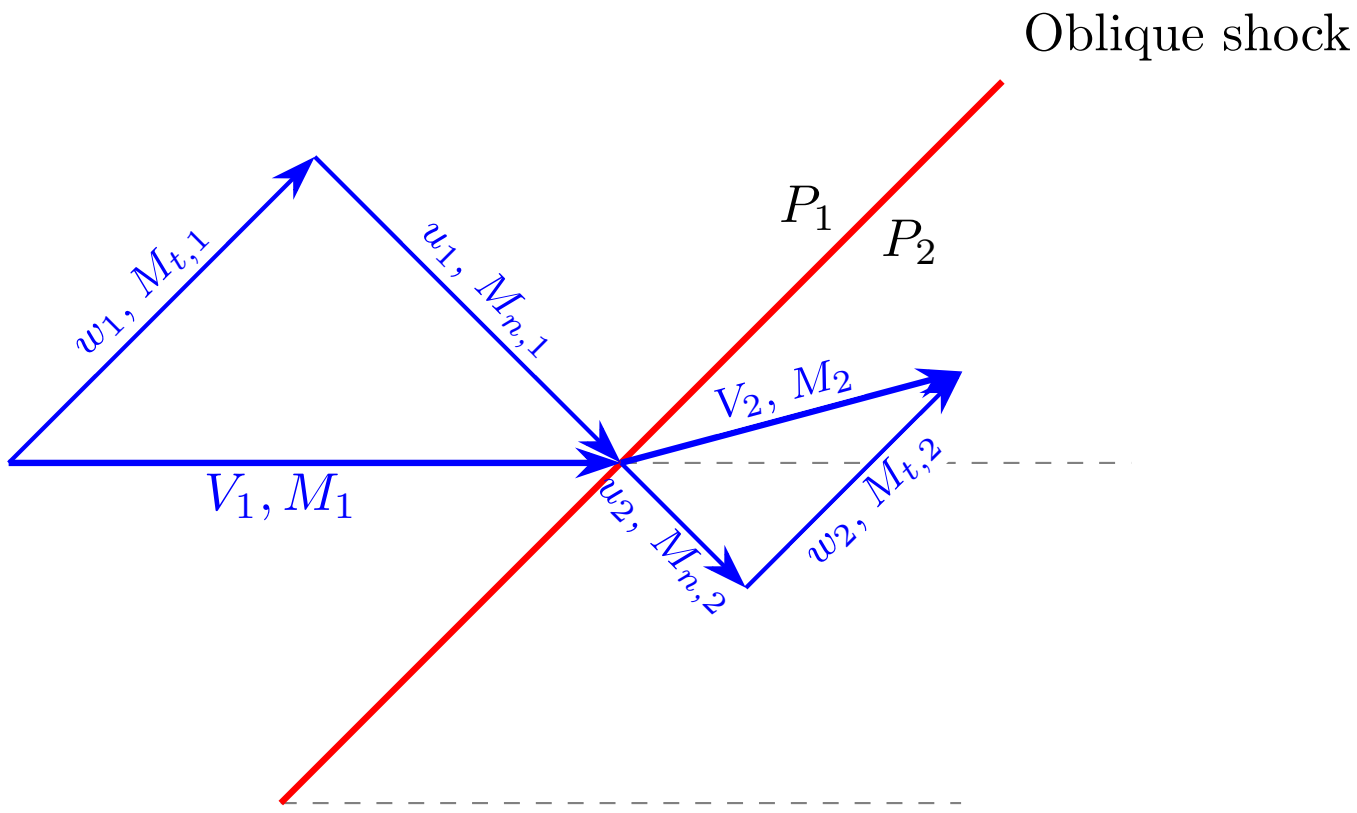

Consider the oblique shock configuration shown in Figure~\ref{fig:oblshks}. The 
upstream flow with velocity $V_1$ is decomposed into normal and tangential components 
relative to the shock surface, denoted by $u_1$ and $w_1$, respectively. Similarly, 
the downstream velocity $V_2$ consists of normal and tangential components $u_2$ and $w_2$. 

For an oblique shock in compressible flow, the tangential component of the velocity remains 
continuous across the shock, while the normal component satisfies the one-dimensional normal 
shock relations. Under these assumptions, the flow quantities across the oblique shock are related 
by the following expressions (see \cite{anderson2011ebook}):
\begin{equation}\label{eqn:oblshk}
	\begin{aligned}
		M_{n,1} &= \frac{\lVert u_1 \rVert}{c_1}, &\qquad
		\frac{\rho_2}{\rho_1} &= \frac{(\gamma+1) M_{n,1}^2}{(\gamma-1) M_{n,1}^2 + 2}, \\[0.4em]
		\frac{P_2}{P_1} &= 1 + \frac{2\gamma}{\gamma+1}\left(M_{n,1}^2 - 1\right), &\qquad
		V_2 &= \frac{\rho_1}{\rho_2} u_1 + w_1,
	\end{aligned}
\end{equation}
where $M_{n,1}$ is the upstream Mach number normal to the shock and $c_1$ is the upstream speed of sound. The last relation combines continuity of the tangential velocity, $w_2 = w_1$, with conservation of mass across the shock, $\rho_2 u_2 = \rho_1 u_1$. The relations require $M_{n,1} > 1$, which holds on the shock-aligned boundary since the fitted shock is, by construction, an entropy-satisfying discontinuity of the inviscid solution.
When the computational mesh is accurately aligned with the shock, the upstream flow components 
can be directly identified with the freestream values. The boundary condition is formulated in the physical domain in terms of the unit outward normal $n_h$, which is obtained from the mesh geometry and related to the reference normal $N_h$ by \eqref{eqn:nrmltransf}; the boundary state entering \eqref{eqn:eleres} is $\bar{U}^\partial(\bar{U}_h^+, N_h) = U^\partial(\bar{U}_h^+, n_h)$. On the fixed post-shock mesh used for the viscous computation, the reference and physical domains coincide and $n_h = N_h$. The same convention applies to the other boundary states specified in Section~\ref{sec:numexp}. 

The outward unit normal $n_h$ on the shock boundary points into the upstream (freestream) region, so the shock normal oriented from the upstream to the downstream side is $n_h' = -n_h$. The upstream velocity has a positive component along $n_h'$, $V_1 \cdot n_h' > 0$, but is in general not aligned with it. 
The upstream velocity is then decomposed into normal and tangential components with respect to $n_h'$: 
\begin{equation}
	u_1 = \left(V_1 \cdot n_h'\right) n_h', \quad w_1 = V_1 - u_1,
\end{equation}
so that $M_{n,1} = (V_1 \cdot n_h')/c_1$, and the downstream flow quantities ($\rho_2, V_2, P_2$) are evaluated 
explicitly from the oblique shock relations in \eqref{eqn:oblshk}. 
The resulting downstream primitive variables are then converted to conservative form and imposed 
as the boundary data 
\begin{equation}
	U^\partial = \Psibold(\rho_2, V_2, P_2),
\end{equation}
where $\func{\Psibold}{\Rbb^m}{\Rbb^{m}}$ is the mapping that converts primitive variables $(\rho, v, P)$ into 
conservative variables $(\rho, \rho v, \rho E)$, thereby defining the boundary state appearing in \eqref{eqn:eleres}. Unlike the other boundary types, where the inviscid boundary flux is the flux of the boundary state, the post-shock state is imposed weakly through the Roe numerical flux between the interior trace and the boundary state, \begin{equation}\label{eqn:shkflx} \bar{F}^\partial_\text{inv}(\bar{U}_h^+, \bar{U}^{\partial}, N_h; \bar\nabla\Gcal_h) = \bar\Hcal(\bar{U}_h^+, \bar{U}^{\partial}, N_h; \bar\nabla\Gcal_h), \end{equation} with $\bar\Hcal$ the transformed numerical flux of \eqref{eqn:nflx}. 

This procedure avoids the need for solving the nonlinear Rankine-Hugoniot conditions at the discrete level, as the 
shock jump relations are enforced analytically through closed-form expressions. Consequently, the shock boundary 
conditions can be applied in a straightforward and robust manner without introducing additional nonlinear iterations. 
The resulting downstream states are used as exterior states in the evaluation of the numerical fluxes on the shock boundary.

We close this section by stating the modeling assumptions underlying the shock boundary condition. First, the shock position and shape are fixed by the inviscid HOIST solution and do not respond to the viscous post-shock solution. This neglects the displacement effect of the boundary layer on the shock and the viscous structure of the shock
itself. The former shifts the shock by an amount of the order of the boundary-layer displacement thickness, which
scales as $\text{Re}_\infty^{-1/2}$ relative to the nose radius and is well below the resolution of the quantities
compared in Section~\ref{sec:numexp}; the latter is of the order of the mean free path. The assumption is checked
directly for the sphere problem in Section~\ref{subsec:sph3d}, where the standoff distance of the tracked Euler shock
agrees with the viscous computation of Blottner~\cite{blottner1989accurate} along the entire body
(Figure~\ref{fig:ns3_cmp}). Second, the closed-form relations \eqref{eqn:oblshk} require the upstream state to be known, which restricts the approach to the leading bow shock in a uniform freestream; embedded shocks, shock-shock interactions, and shocks induced by boundary-layer separation would require either shock tracking within the post-shock domain or an upstream state obtained from a separate computation. Third, the boundary state $U^\partial$ depends only on the freestream state and the local shock normal, not on the interior solution, so the full post-shock state is prescribed even over the portion of the bow shock near the stagnation streamline where the post-shock flow is subsonic and fewer boundary conditions are admissible. The Roe flux in \eqref{eqn:shkflx} resolves this: it is an upwind flux between the interior trace and the prescribed state, so only the incoming characteristic information is enforced and the outgoing characteristics are taken from the interior solution, as in a weakly imposed Dirichlet condition.

\section{Shock-aligned framework for viscous hypersonic flows}
\label{sec:istshkbc}
In this section, we describe the numerical framework used to align the computational mesh 
with the discontinuities and to subsequently recover viscous solutions. We begin by 
summarizing the HOIST method for the inviscid problem (Section \ref{subsec:ist}), which provides the core shock tracking 
mechanism through a mesh-solution optimization procedure. We then describe the removal of farfield 
elements required to accommodate shock-aligned boundaries (Section \ref{subsec:elerem}). Finally, we extend the framework to 
viscous computations using a remeshing strategy combined with pseudo-transient continuation (PTC) (Section \ref{subsec:ptc}), 
and introduce a dual-weighted residual error estimate to drive further mesh adaptation (Section \ref{subsec:dwr}).

\subsection{High-order implicit shock tracking (HOIST) method}
\label{subsec:ist}
We first summarize the HOIST method in the inviscid setting, which serves as the 
foundation for the subsequent viscous and adaptive extensions. Recall that the IPDG 
discretization derived in Section \ref{subsec:ipdg} leads to the algebraic residual 
form \eqref{eqn:dg-alg}. We consider the inviscid formulation by setting $\mu_\infty = 0$, 
so that the residual reduces to $\rbm(\ubm,\xbm;\mu_\infty)=\rbm_{\text{inv}}(\ubm,\xbm)$. The objective 
of the HOIST method is to align element faces with flow discontinuities, thereby enabling a stable 
representation of shocks. To achieve this, the method is formulated as a PDE-constrained 
optimization problem in which both the discrete solution $\ubm$ and nodal mesh coordinates $\xbm$ are 
treated as optimization variables,
\begin{equation} \label{eqn:pde-opt}
	\left(\ubm^{\star}(\Xi), \ybm^{\star}(\Xi)\right) \coloneqq
	\argmin_{\ubm\in\Rbb^{N_\ubm},\ybm\in\Rbb^{N_\ybm}} f\left(\ubm, \ybm;\Xi \right) \quad \text{subject to:} \quad \rbm_{\text{inv}}\left(\ubm,\phibold(\ybm);\Xi \right) = \zerobold,
\end{equation}
where $\func{f}{\Rbb^{N_\ubm}\times\Rbb^{N_\ybm}}{\Rbb}$ is the objective function, $\ybm \in \Rbb^{N_\ybm}$ are the unconstrained mesh degrees of freedom, $\func{\phibold}{\Rbb^{N_\ybm}}{\Rbb^{N_\xbm}}$ is the mapping from the unconstrained degrees of freedom to the nodal coordinates of the mesh that enforces geometric boundary constraints, e.g., nodes on a boundary slide along it~\cite{huang2022robust}, and $\Xi \in \Rbb$ is a continuation parameter, which is used to improve the robustness of the shock tracking solver~\cite{huang2023high,dong2025rpadapt} by defining a sequence of optimization problems. The nodal coordinates of the aligned mesh are $\xbm^\star = \phibold(\ybm^\star(\Xi))$. In this work, $\Xi$ is the freestream Mach number $M_\infty$, which enters the residual through the freestream state prescribed at the inflow boundary. 
The objective function consists of two terms,
\begin{equation}\label{eqn:obj0}
	f : \left(\ubm, \ybm;\Xi \right) \mapsto f_\text{err}\left(\ubm, \ybm;\Xi \right) + \kappa^2 f_\text{msh}(\ybm),
\end{equation}
where $\func{f_\text{err}}{\Rbb^{N_\ubm}\times\Rbb^{N_\ybm}}{\Rbb_{>0}}$ is the mesh alignment term and 
$f_\text{msh} : \Rbb^{N_\ybm} \rightarrow \Rbb_{>0}$ is a mesh quality term, defined as
\begin{equation}\label{eqn:obj1}
	f_\text{err} : \left(\ubm, \ybm;\Xi \right) \mapsto \frac{1}{2}\norm{\Rbm_{\text{inv}}\left(\ubm,\phibold(\ybm);\Xi \right)}_2^2, \quad
	f_{\text{msh}} : \ybm \mapsto \frac{1}{2}\norm{\Rbm_{\text{msh}}(\phibold(\ybm))}_2^2,
\end{equation}
where $\Rbm_\text{msh} : \Rbb^{N_\xbm} \rightarrow \Rbb_{>0}^{|\Ecal_{h}|}$ is the
elemental mesh distortion defined in \cite{zahr2020implicit}, and $\kappa\in\Rbb_{>0}$
is the mesh quality penalty parameter.

\begin{figure}[!htbp]
	\centering
	\input{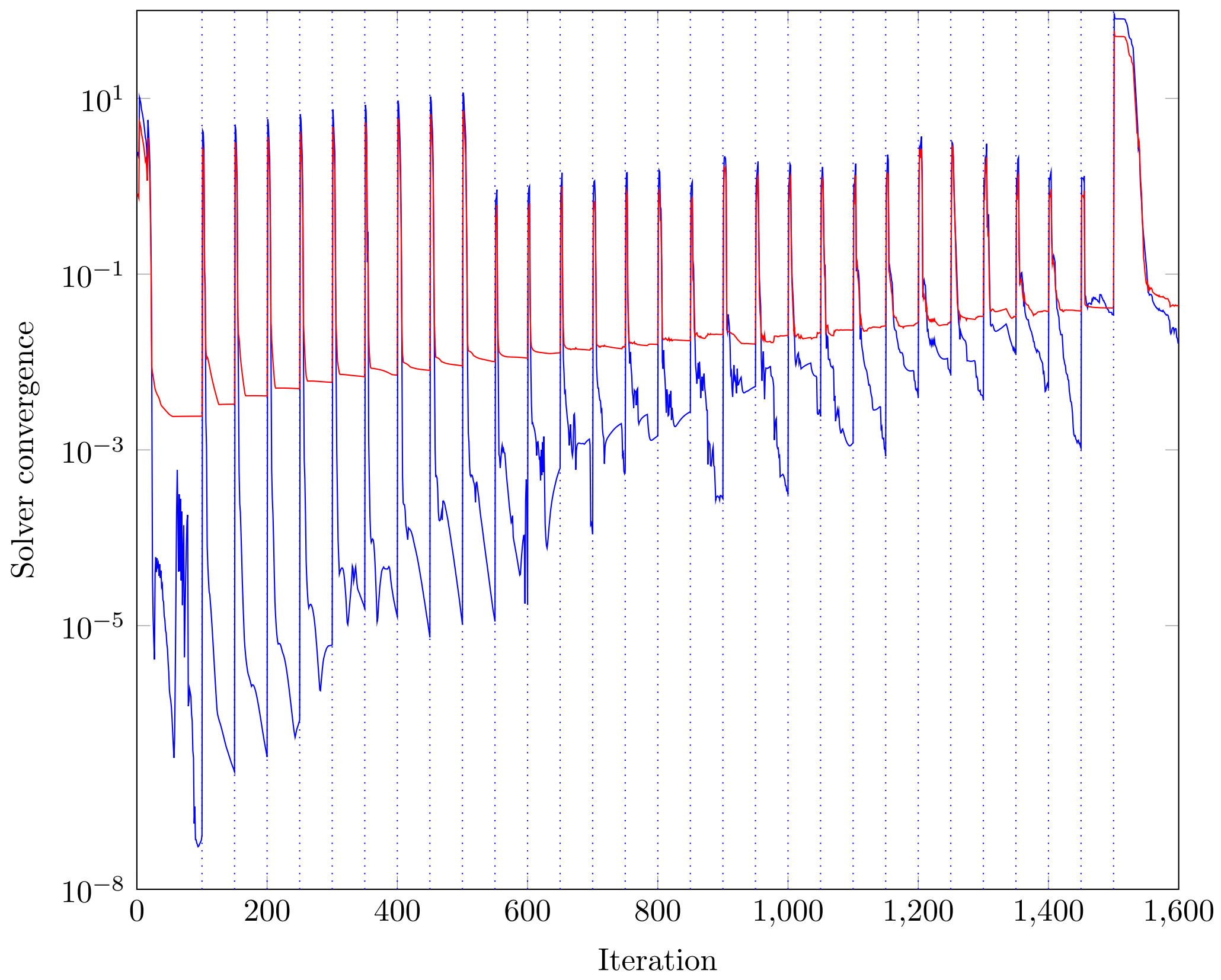}
	\caption{
		HOIST convergence of the DG residual $\norm{\rbm_{\text{inv}}(\ubm_k,\phibold(\ybm_k); \Xi_i)}$ (\ref{line:fcon}) 
	    and enriched DG residual $\norm{\Rbm_{\text{inv}}(\ubm_k,\phibold(\ybm_k); \Xi_i)}$ (\ref{line:ferr}) 
        across all Mach number continuation stages $\{\Xi_i\}_{i=1}^{30}$ for a cylinder problem from $M_\infty = 5$ to $M_\infty= 17.605$. 
        Vertical lines indicate the end of a continuation stage (\ref{line:MachCon_nrm}).
	}
	\label{fig:ns2_opthist}
\end{figure}

\begin{figure}[!htbp]
	\centering
	\begin{tikzpicture}[scale=1.0]
\begin{axis}[
  name=shk,
  scale only axis, width=0.92\textwidth, height=0.30\textwidth,
  xmin=-2.8, xmax=2.8, ymin=-0.1, ymax=1.55,
  xtick={-2.5,-1,0,1,2.5}, ytick={0,0.5,1,1.5},
  xlabel={$x_1$}, ylabel={$x_2$},
]
\def\cylshklabels{1}
\input{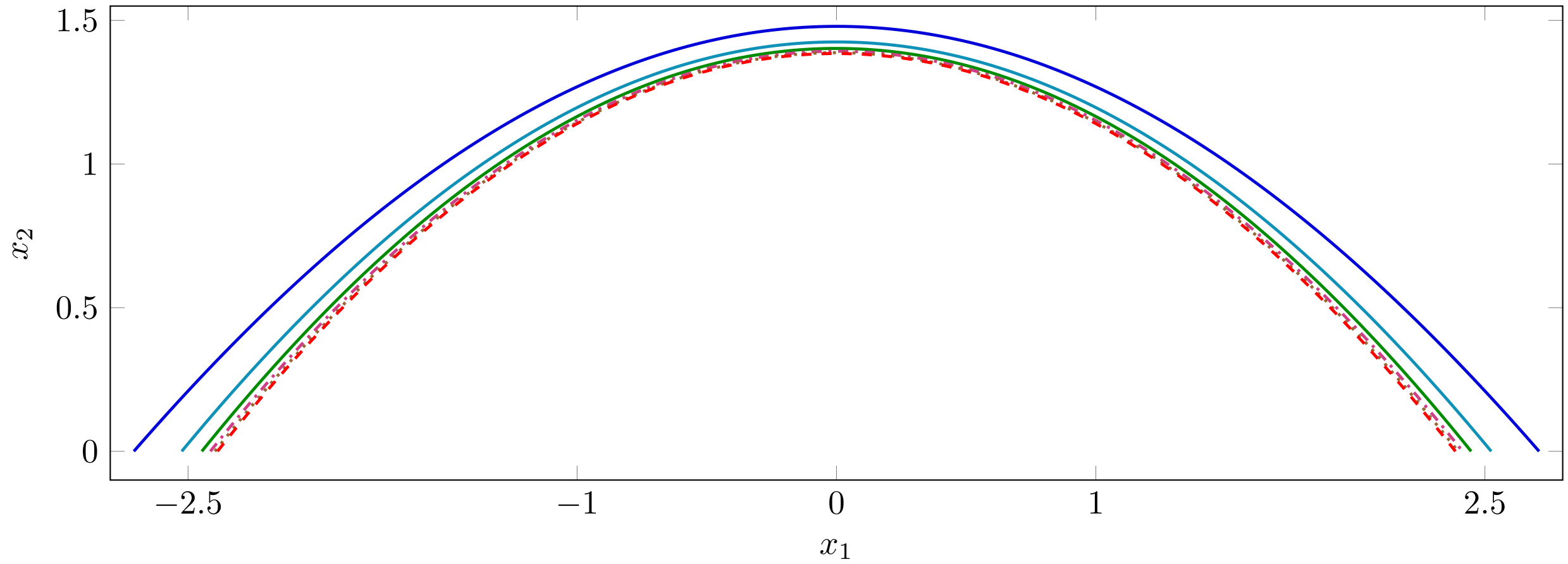}
\end{axis}

\end{tikzpicture}
	\caption{
		Shock positions tracked by HOIST for the cylinder problem at selected Mach number continuation stages
		$\Xi_i$, $i \in \{1,6,12,18,24,30\}$, i.e., $M_\infty = 5$ (\ref{pl:cylM5}), $7.17$ (\ref{pl:cylM7}),
		$9.78$ (\ref{pl:cylM9}), $12.39$ (\ref{pl:cylM12}), $15.0$ (\ref{pl:cylM15}), and $17.605$ (\ref{pl:cylM17}).
		The final-stage mesh and Mach number field are shown in Figure~\ref{fig:ns2_soln1}.
	}
	\label{fig:cylMachcon}
\end{figure}

A sequential quadratic programming (SQP) method with a modified Levenberg-Marquardt Hessian approximation
is used to solve the optimization problem (\ref{eqn:pde-opt}). The SQP solver 
simultaneously converges the flow solution $\ubm$ and mesh degrees of freedom
$\ybm$ to their optimal values. The SQP solver defines a sequence of
iterations $\{\zbm_0(\Xi), \zbm_1(\Xi), \dots\}$, where each iterate
contains both the flow and mesh degrees of freedom, i.e.,
$\zbm_k(\Xi) = (\ubm_k(\Xi), \ybm_k(\Xi))$ that converges to
the solution of (\ref{eqn:pde-opt}): $\zbm_k(\Xi) \rightarrow (\ubm^\star(\Xi), \ybm^\star(\Xi))$,
and the iteration is updated by
\begin{equation} \label{eqn:stpupd}
	\zbm_{k+1}(\Xi) = \zbm_{k}(\Xi) + \alpha_k(\Xi) \Delta \zbm_k(\Xi),
\end{equation}
where $\Delta \zbm_k(\Xi) \in \Rbb^{N_\ubm + N_\ybm}$ is the search direction,
and $\alpha_k(\Xi) \in \Rbb_{\geq 0}$ is the step length at $k$-th iteration.
The search direction $\Delta \zbm_k(\Xi)$ is determined by solving a quadratic program, and
the step length is obtained from a backtracking line search~\cite{nocedal2006numerical}.
A complete description of the HOIST method can be found in \cite{zahr2020implicit,huang2022robust}, 
while its extension with continuation parameters is detailed in \cite{huang2023high,dong2025rpadapt}. 
The behavior of the SQP solver for the cylinder problem of Section~\ref{subsec:cyl2d} across all Mach number continuation stages is shown in Figure~\ref{fig:ns2_opthist}, 
and the tracked shock positions at selected Mach number continuation stages $\Xi_i$, $i \in \{1, 6, 12, 18, 24, 30\}$, are overlaid in Figure~\ref{fig:cylMachcon}, which shows the shock standoff distance and shape converging as $M_\infty$ increases.

\subsection{Farfield element removal and post-shock remeshing}
\label{subsec:elerem}
Since the HOIST method obtains a shock-aligned mesh, farfield elements can be removed and 
the mesh can subsequently be refined through remeshing. Element classification is performed 
based on the computed density field, exploiting the fact that the freestream density is 
nondimensionalized to $\rho_\infty = 1.0$. For each element, we extract the local degrees 
of freedom corresponding to density and evaluate the deviation from the freestream value. 
If the maximum of the absolute difference $|\rho - \rho_\infty|$ over the element remains below a prescribed tolerance ($10^{-2}$ used in this work), the 
element is identified as upstream; otherwise, the element is classified as downstream. Since flow 
variables undergo sharp changes across a discontinuity, this criterion provides a reliable 
mechanism for separating upstream and downstream regions. 

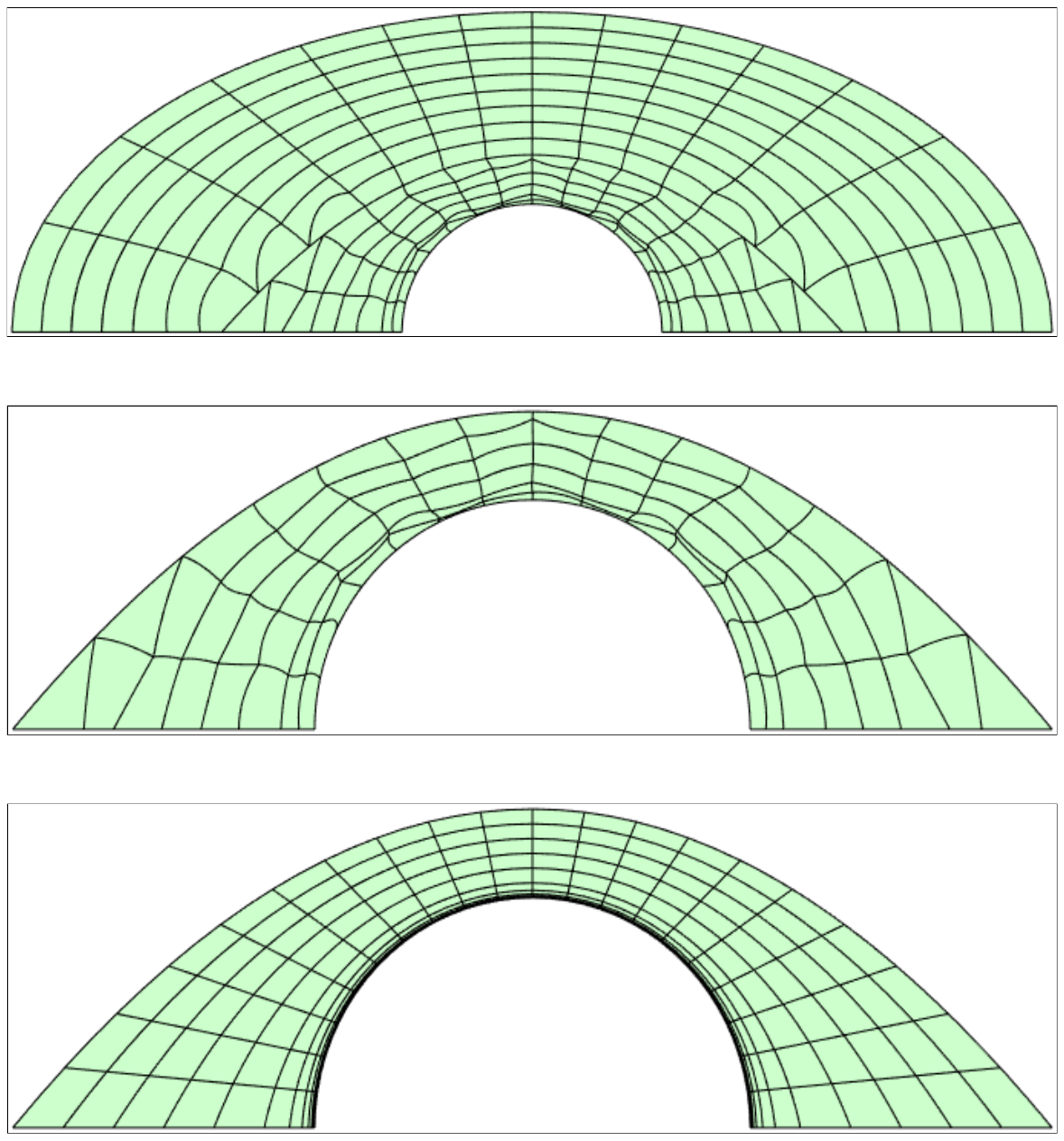
\begin{figure}[!htbp]
	\centering
	\begin{tikzpicture}[scale=1.15]
\begin{groupplot}[
  group style={
      group size=1 by 3,
      horizontal sep=0.45cm,
      vertical sep=0.45cm
  },
  width=0.5\textwidth,
  axis equal image,
  xlabel={$x_1$},
  ylabel={$x_2$},
  xmin=-4.0, xmax=4.0,
  ymin=0.0, ymax=2.5
]

\nextgroupplot[xlabel={}, ylabel={}, xtick=\empty, ytick=\empty]
\addplot graphics [xmin=-4.0, xmax=4.0, ymin=0.0, ymax=2.5] {{_img/cyl_M17p6_ist_whmsh}.png};

\nextgroupplot[xlabel={}, ylabel={}, xtick=\empty, ytick=\empty]
\addplot graphics [xmin=-4.0, xmax=4.0, ymin=0.0, ymax=2.5] {{_img/cyl_M17p6_ist_shkdwnmsh}.png};

\nextgroupplot[xlabel={}, ylabel={}, xtick=\empty, ytick=\empty]
\addplot graphics [xmin=-4.0, xmax=4.0, ymin=0.0, ymax=2.5] {{_img/cyl_NSM17p6_remshs1}.png};

\end{groupplot}

\end{tikzpicture}
	\caption{
		HOIST mesh after Mach continuation, post-shock mesh after farfield element removal, and remeshed post-shock domain for the $M_\infty = 17.605$, $\text{Re}_\infty = 376930$ cylinder problem (\textit{top-to-bottom}).
	}
	\label{fig:cylmshcomp}
\end{figure}

Once the post-shock domain has been identified, the mesh can be locally modified to 
better support viscous flow computations. In particular, elements associated with the upstream 
and farfield regions are removed, and the remaining downstream domain is remeshed to generate 
a grid suitable for resolving boundary layer effects (Figure~\ref{fig:cylmshcomp}). The new mesh is constructed with strong 
anisotropy near wall boundaries, featuring refined spacing in the wall-normal direction to accurately 
capture steep velocity and temperature gradients. The boundary-layer meshing strategy significantly improves 
the resolution of viscous phenomena while maintaining computational efficiency, and it enables 
a smooth transition from the initial inviscid computation used to locate shocks to subsequent 
viscous computations.

\subsection{Downstream viscous computation}
\label{subsec:ptc}
\begin{figure}[!htbp]
	\centering
	\begin{tikzpicture}[scale=1.0]
\begin{groupplot}[
  group style={
      group size=2 by 3,
      horizontal sep=0.45cm,
      vertical sep=0.45cm
  },
  width=0.5\textwidth,
  axis equal image,
  xlabel={$x_1$},
  ylabel={$x_2$},
  xmin=-2.4, xmax=2.4,
  ymin=0, ymax=1.4
]

\nextgroupplot[xlabel={}, ylabel={}, xtick=\empty, ytick=\empty]
\addplot graphics [xmin=-2.4, xmax=2.4, ymin=0, ymax=1.4] {{_img/cyl_M17p6_istEuler_Mach_wmsh}.png};

\nextgroupplot[xlabel={}, ylabel={}, xtick=\empty, ytick=\empty]
\addplot graphics [xmin=-2.4, xmax=2.4, ymin=0, ymax=1.4] {{_img/cyl_M17p6_istEuler_Mach_womsh}.png};

\nextgroupplot[xlabel={}, ylabel={}, xtick=\empty, ytick=\empty]
\addplot graphics [xmin=-2.4, xmax=2.4, ymin=0, ymax=1.4] {{_img/cyl_M17p6_p0Mach_wmsh}.png};

\nextgroupplot[xlabel={}, ylabel={}, xtick=\empty, ytick=\empty]
\addplot graphics [xmin=-2.4, xmax=2.4, ymin=0, ymax=1.4] {{_img/cyl_M17p6_p0Mach_womsh}.png};

\nextgroupplot[xlabel={}, ylabel={}, xtick=\empty, ytick=\empty]
\addplot graphics [xmin=-2.4, xmax=2.4, ymin=0, ymax=1.4] {{_img/cyl_M17p6shkdwns1msh1_Mach_wmsh}.png};

\nextgroupplot[xlabel={}, ylabel={}, xtick=\empty, ytick=\empty]
\addplot graphics [xmin=-2.4, xmax=2.4, ymin=0, ymax=1.4] {{_img/cyl_M17p6shkdwns1msh1_Mach_womsh}.png};

\end{groupplot}
\node[anchor=north] at ($(group c1r3.south)!0.5!(group c2r3.south)$) {\colorbarMatlabParula{0.0}{0.65}{1.30}{1.95}{2.60}};

\end{tikzpicture}
	\caption{
		Post-shock mesh (\textit{left}) and Mach number (\textit{right}) for the $M_\infty = 17.605$, $\text{Re}_\infty = 376930$ cylinder problem: the HOIST Euler solution, the viscous solution with $p = 0$, and the fully converged viscous solution using PTC (\textit{top-to-bottom}).
	}
	\label{fig:cylp0ptc}
\end{figure}
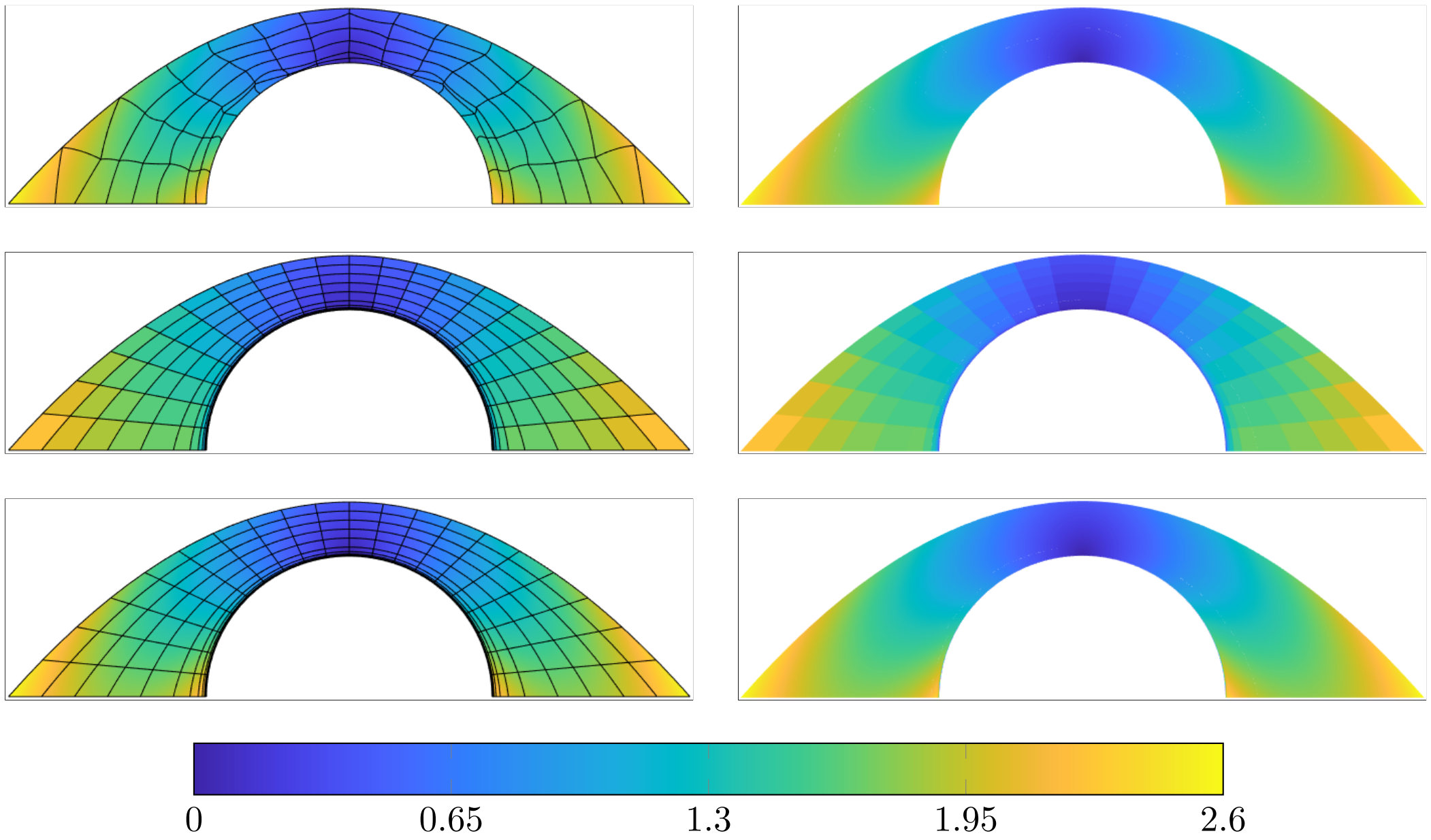

For the downstream viscous computation, the flow field is smooth and continuous, as the dominant 
shock discontinuities have already been isolated during the inviscid stage. Consequently, standard IPDG 
discretization can be employed for the viscous fluxes without additional shock stabilization to solve 
\begin{equation}
	\rbm(\ubm, \phibold(\ybm^\star); \mu_\infty) = \zerobold.
\end{equation}
The resulting nonlinear system is solved using a pseudo-transient continuation (PTC) approach~\cite{modisette2011automated}, 
which provides robust convergence to the steady-state solution. 
The computation is initialized from the $p=0$ viscous solution rather than from the HOIST solution $\ubm^\star$ 
in order to ensure the starting point contains a boundary layer (Figure~\ref{fig:cylp0ptc}). 
The combination of IPDG discretization and PTC nonlinear solution 
strategy offers an efficient and stable framework for resolving viscous effects in the downstream region 
on the boundary-layer-refined mesh.

\subsection{Dual-weighted residual error estimate and mesh adaptation}
\label{subsec:dwr}
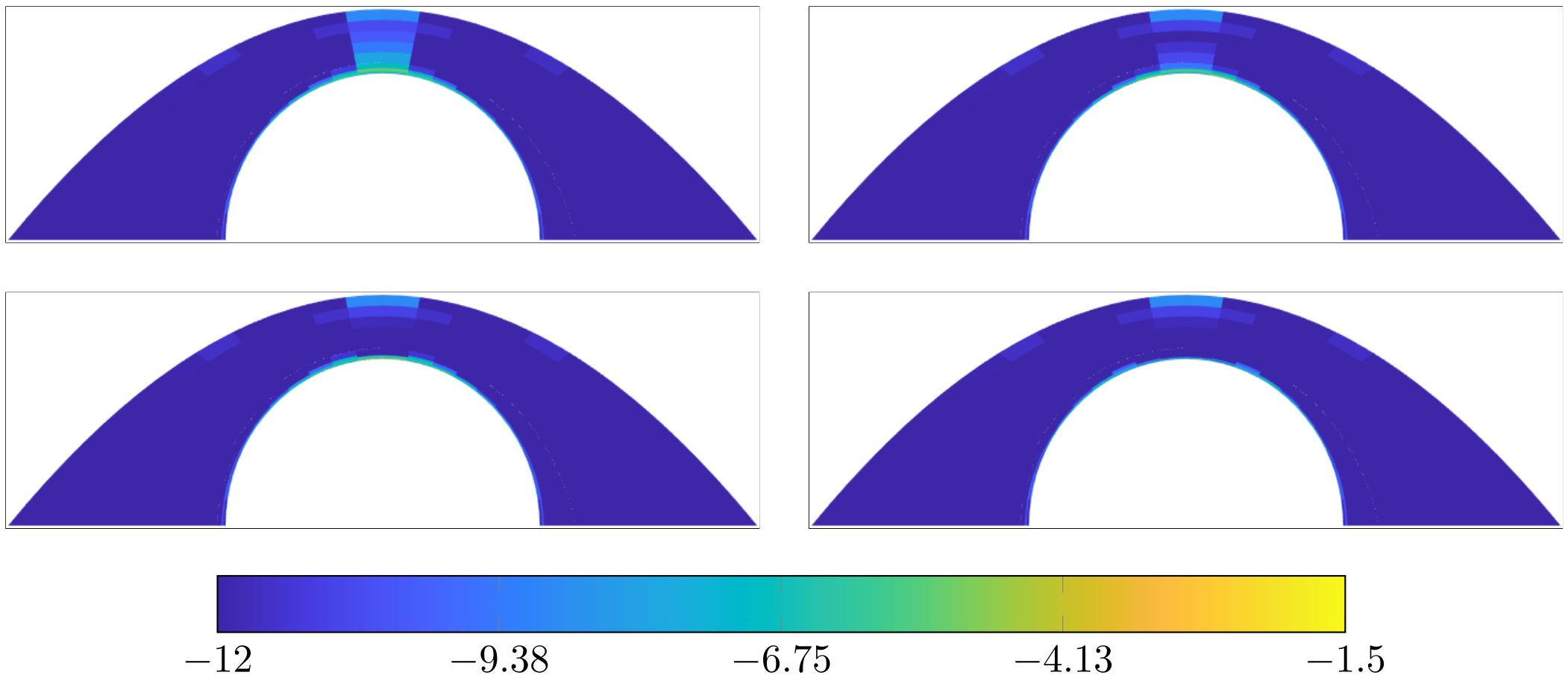
\begin{figure}[!htbp]
	\centering
	\begin{tikzpicture}[scale=1.0]
\begin{groupplot}[
  group style={
      group size=2 by 2,
      horizontal sep=0.45cm,
      vertical sep=0.45cm
  },
  width=0.5\textwidth,
  axis equal image,
  xlabel={$x_1$},
  ylabel={$x_2$},
  xmin=-4.0, xmax=4.0,
  ymin=0.0, ymax=2.5
]

\nextgroupplot[xlabel={}, ylabel={}, xtick=\empty, ytick=\empty]
\addplot graphics [xmin=-4.0, xmax=4.0, ymin=0.0, ymax=2.5] {{_img/cyl_NSM17p6_dwrSs1}.png};

\nextgroupplot[xlabel={}, ylabel={}, xtick=\empty, ytick=\empty]
\addplot graphics [xmin=-4.0, xmax=4.0, ymin=0.0, ymax=2.5] {{_img/cyl_NSM17p6_dwrSs4}.png};

\nextgroupplot[xlabel={}, ylabel={}, xtick=\empty, ytick=\empty]
\addplot graphics [xmin=-4.0, xmax=4.0, ymin=0.0, ymax=2.5] {{_img/cyl_NSM17p6_dwrSs6}.png};

\nextgroupplot[xlabel={}, ylabel={}, xtick=\empty, ytick=\empty]
\addplot graphics [xmin=-4.0, xmax=4.0, ymin=0.0, ymax=2.5] {{_img/cyl_NSM17p6_dwrSs8}.png};

\end{groupplot}
\node[anchor=north] at ($(group c1r2.south)!0.5!(group c2r2.south)$) {\colorbarMatlabParula{-12}{-9.38}{-6.75}{-4.13}{-1.5}};
\end{tikzpicture}
	\caption{
		Elemental dual-weighted residual error estimate ($\log_{10}$ of the absolute value) prior to mesh adaptation (\textit{top left}) and after three (\textit{top right}), five (\textit{bottom left}), and seven (\textit{bottom right}) mesh refinement cycles for the $M_\infty = 17.605$, $\text{Re}_\infty = 376930$ cylinder problem.
	}
	\label{fig:cyldwr}
\end{figure}

To further enhance the resolution of viscous effects and accurately capture surface heat 
transfer, mesh adaptation is driven by a dual-weighted residual (DWR) error indicator~\cite{becker1996feed,Becker_2001_DWR,fidkowski2011review}. 
Heat flux over the blunt body wall is selected as the quantity of interest in this work, allowing the adaptive procedure to 
target regions that most strongly influence thermal effects. The resulting error estimates naturally 
concentrate refinement in areas with large temperature gradients and near-wall viscous layers where 
heat transfer is most sensitive to mesh resolution (Figure~\ref{fig:cyldwr}). This goal-oriented adaptation strategy ensures 
efficient use of computational resources while improving the accuracy of heat predictions in the downstream 
viscous region.

\section{Numerical experiments}
\label{sec:numexp}
In this section, we present a set of numerical examples to demonstrate the effectiveness of the 
proposed framework for shock-dominated viscous hypersonic problems. The selected test cases 
involve strong shocks, thin boundary layers, and sharp thermal gradients, providing a rigorous 
assessment of the method's ability to robustly track discontinuities and recover downstream viscous 
effects, including accurate heating predictions. 

\subsection{Mach 10.6 flow over a blunt cone}
\label{subsec:cone2d}

First, we consider steady, axisymmetric hypersonic flow over a blunt $15^\circ$ cone at zero angle of attack
with a final freestream Mach number $M_{\infty} = 10.6$. The problem is solved in the meridional plane ($d = 2$)
using the axisymmetric formulation \eqref{eqn:axisrc}, with $x_2$ the axis of symmetry. This case is selected to
assess the solver for axisymmetric flow conditions, and the experimental heat flux measurements of
Cleary~\cite{cleary1969effects} are used for validation.
The computational geometry and boundary conditions
are illustrated in Figure~\ref{fig:ns1_geom0}, and the original dimensional geometry is 
described in \cite{cleary1969effects}. Lengths are nondimensionalized by the nose radius $L^*$
(Section~\ref{subsec:ns}), and the axial coordinate $x_2$ is measured from the nose tip. The entire primitive state is prescribed at the supersonic inlet, and the cone nose and conical surface are modeled as slip walls.

The domain is discretized using a shock-agnostic mesh consisting of 551 curved ($q = 2$) quadrilateral
elements with quadratic solution approximation ($p = 2$). The HOIST solver is initialized from a shock capturing
DG solution using PDE-based artificial viscosity from \cite{barter2010shock} on this coarse discretization
(Figure~\ref{fig:ns1_soln_ic}). Mach number continuation is performed over 10 stages with equal 
increments beginning with $M_{\infty} = 8$ and terminating at $M_{\infty} = 10.6$. The first continuation stage
is allotted $200$ SQP iterations, while each of the intermediate stages is advanced using $100$ SQP iterations.

The high-order elements align with the detached bow shock, which yields a highly accurate inviscid 
solution when combined with high-order flow field approximation (Figure~\ref{fig:ns1_soln}) at both the initial and final Mach 
numbers, $M_\infty = 8.0$ and $M_\infty = 10.6$. The farfield
elements are subsequently removed to isolate the post-shock domain, as shown in Figure~\ref{fig:ns1_geom1}. 
To facilitate structured mesh generation, this post-shock domain is mapped to a block topology using transfinite 
interpolation~\cite{gordon1973transfinite,eriksson1982generation}; the corresponding computational geometry 
and boundary conditions are also illustrated in Figure~\ref{fig:ns1_geom1}. 
Finally, the IPDG method is applied to solve the Navier-Stokes equations on this post-shock domain.

For the viscous computation, the Reynolds number based on the nose radius
is $\text{Re}_\infty = 165000$. The shock boundary condition of Section~\ref{subsec:oblshk} is applied at the inlet with zero boundary viscous flux, $\bar{F}_{\text{visc}}^\partial = 0$. The cone surface is a no-slip, isothermal wall with $T_w / T_{\infty} = 6.22$. Its boundary state has the interior density, zero velocity, and the wall temperature, and its boundary viscous flux is the physical viscous flux evaluated with this state and the interior gradient, contracted with the normal~\cite{hartmann2008optimal}; the reported heat flux is computed from this flux. A symmetry condition is applied on the axis. The outlet is a supersonic outflow with zero boundary viscous flux; in the subsonic portion of the boundary layer, the pressure is prescribed to the value obtained from the corresponding inviscid solution. The post-shock domain is discretized using 950 curved quartic ($q = 4$)
elements with a polynomial approximation of order $p = 4$, providing sufficient resolution to capture both the viscous 
boundary layer and the near-wall thermal gradients (Figure~\ref{fig:ns1_msh}). Figure~\ref{fig:ns1_dwstr_soln} illustrates 
the $p = 0$ solution serving as the initial solution for the PTC solver, and the 
converged Mach number field in the post-shock domain.

Figure~\ref{fig:ns1_cp} compares the surface pressure coefficient, defined as
\begin{equation}\label{eqn:Cp}
	C_p = \frac{P - P_\infty}{\frac{1}{2}\rho_\infty V_\infty^2},
\end{equation}
from the inviscid HOIST solution and the viscous IPDG solution.
At the stagnation point, the two solutions give $C_{p,\max} = 1.833$ and $1.836$, respectively, within $0.01\%$ and
$0.2\%$ of the theoretical value for a normal shock followed by isentropic compression ($1.833$).
Over the spherical nose, both solutions follow modified Newtonian theory~\cite{anderson2006hypersonic},
$C_p = C_{p,\max}\cos^2\phi = C_{p,\max}(1 - x_2)^2$, where $\phi$ is the angle along the nose measured from the
stagnation point. The agreement is close near the stagnation point and deteriorates toward the sphere-cone junction
($x_2 = 1-\sin 15^\circ \approx 0.74$), where Newtonian theory is known to be less accurate~\cite{anderson2006hypersonic}.
Downstream of the junction, the pressure overexpands to a minimum near $x_2 \approx 5$ and then gradually recovers,
the characteristic behavior of blunted cones~\cite{cleary1965experimental}. Along the flank, the viscous pressure lies
$2$--$3\%$ above the inviscid one, consistent with a weak viscous-interaction effect.

Finally, Figure~\ref{fig:ns1_hflx} compares the predicted surface heat flux with the measurements
of Cleary~\cite{cleary1969effects} along the cone flank. Both are normalized by the stagnation-point heating
rate of the wind-tunnel model, whose measured nose radius differs slightly from the nominal value used in the
computation due to oblateness introduced during electroformation~\cite{cleary1969effects}. The prediction agrees with all nine measurements to
within $3.9\%$, inside the $\pm 6\%$ experimental repeatability, with no systematic bias.

\ifbool{fastcompile}{}{
	
	\begin{figure}[!htbp]
		\centering
		\input{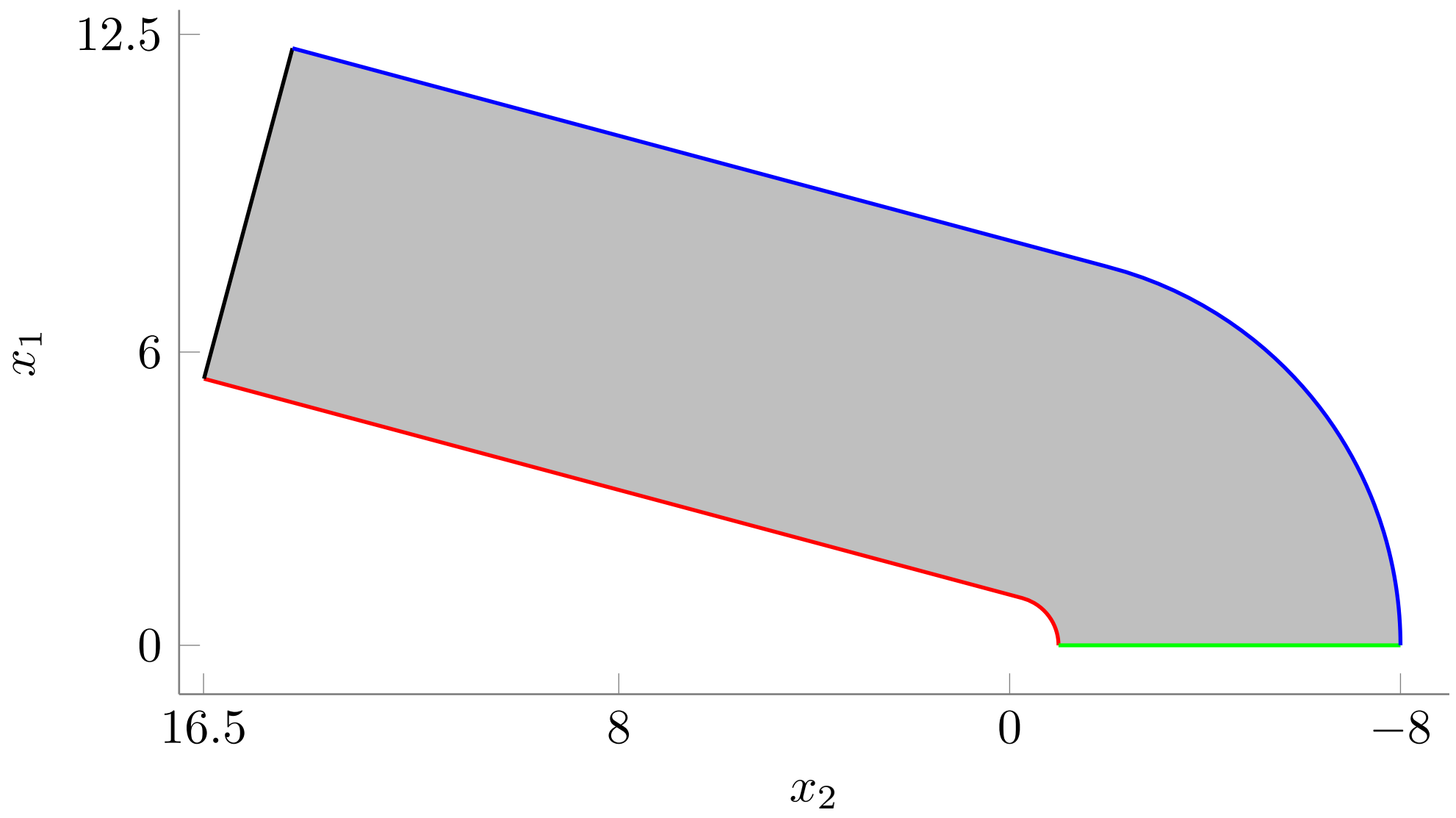}
		\caption{
			Schematic of blunt cone domain and boundary conditions. Boundary 
			conditions: Mach $8$ supersonic inlet (\ref{line:cone0:inlet}), supersonic outlet (\ref{line:cone0:outlet}), slip walls (\ref{line:cone0:wall}), and symmetry axis (\ref{line:cone0:axis}).
		}
		\label{fig:ns1_geom0}
	\end{figure}

    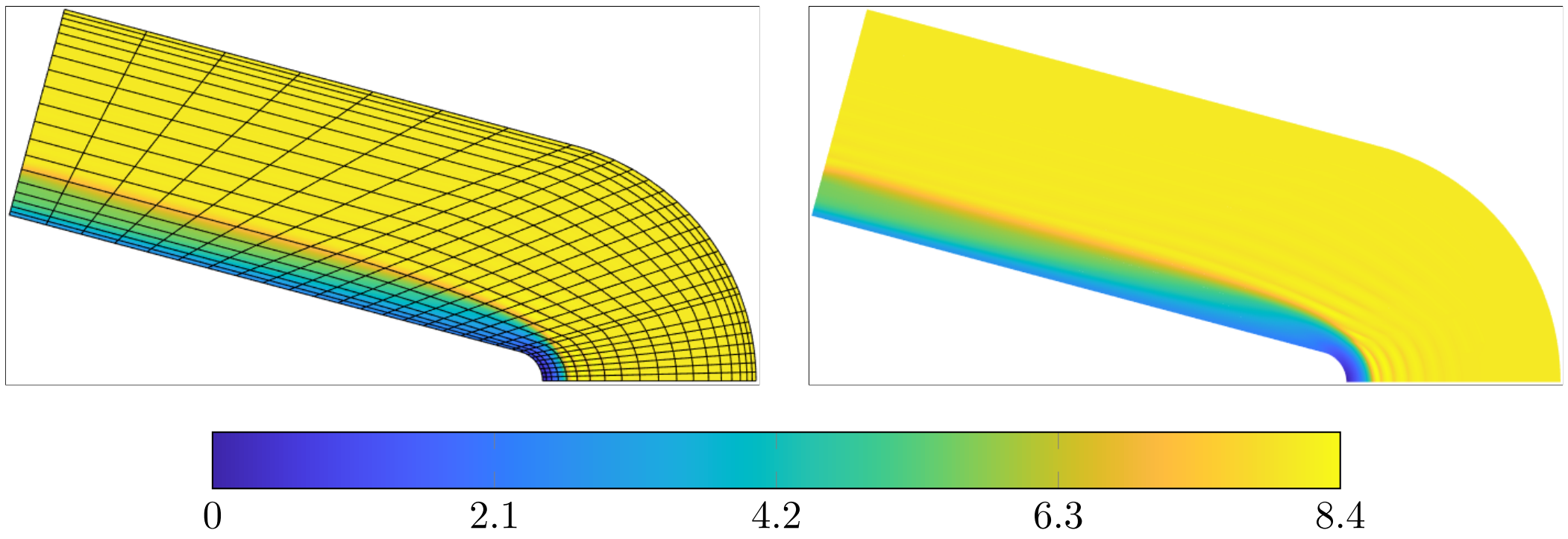
\begin{figure}[!htbp]
    	\centering
    	\begin{tikzpicture}[scale=1.0]
\begin{groupplot}[
  group style={
      group size=2 by 1,
      horizontal sep=0.45cm,
      vertical sep=1.5cm
  },
  width=0.5\textwidth,
  axis equal image,
  xtick = {-16.5, -8.0, 8.0},
  xticklabels={-16.5, -8.0, 8.0},
  ytick = \empty,
  xmin=-16.5, xmax=8.0,
  ymin=0.0, ymax=12.3
]

\nextgroupplot[xtick=\empty, xlabel={}]
\addplot graphics [xmin=-16.5,xmax=8.0, ymin=0.0, ymax=12.3] {{_img/cone_Euler2dM8_Mach_wmsh}.png};

\nextgroupplot[xtick=\empty, xlabel={}]
\addplot graphics [xmin=-16.5, xmax=8.0, ymin=0.0, ymax=12.3] {{_img/cone_Euler2dM8_Mach_womsh}.png};

\end{groupplot}

\node[anchor=north] at ($(group c1r1.south)!0.5!(group c2r1.south)$) {\colorbarMatlabParula{0.0}{2.1}{4.2}{6.3}{8.4}};

\end{tikzpicture}
    	\caption{
    		Mesh and Mach number field of the shock-capturing solution at $M_\infty = 8$ used to initialize the HOIST method for the blunt cone problem.
    	}
    	\label{fig:ns1_soln_ic}
    \end{figure}

	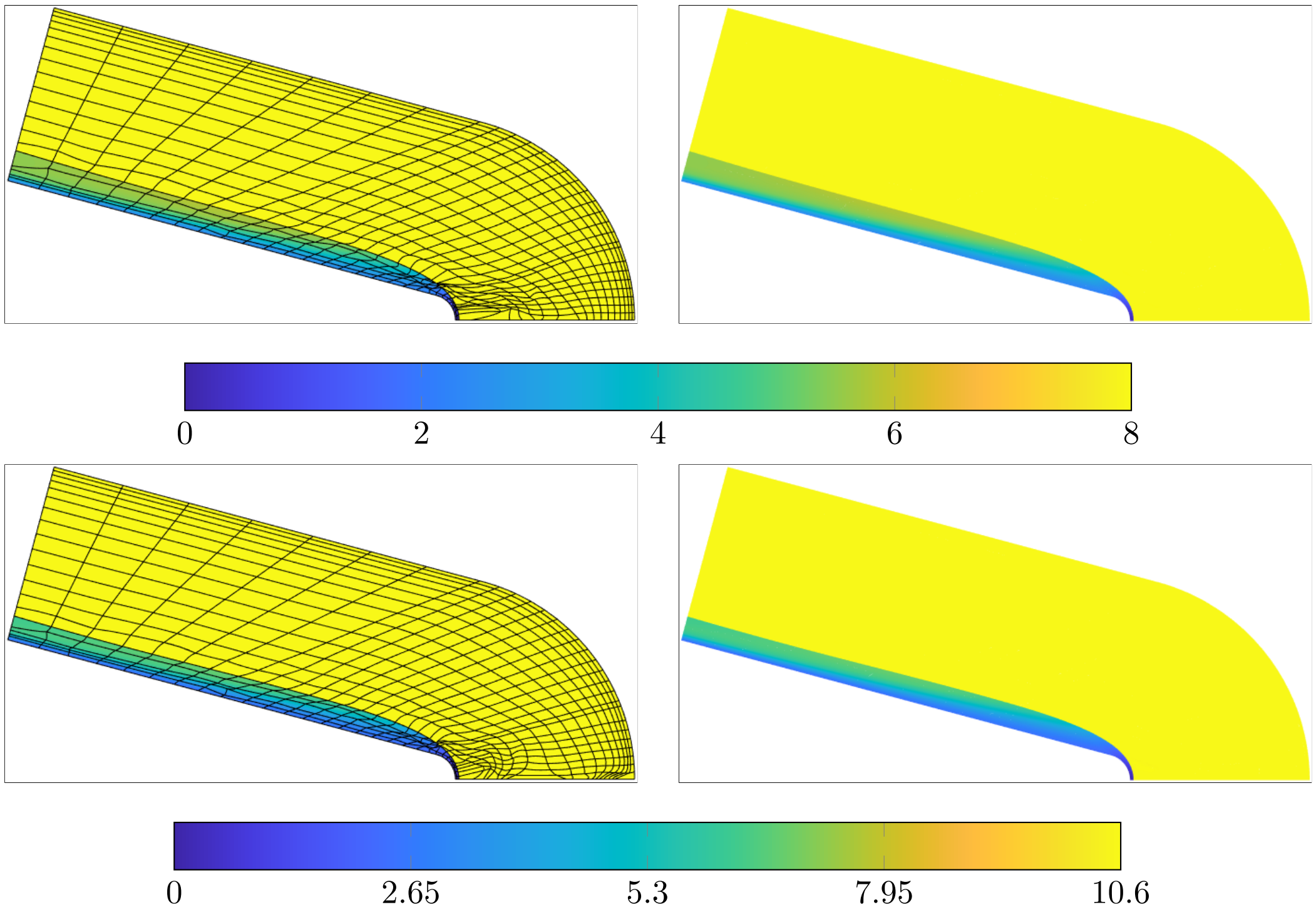
\begin{figure}[!htbp]
		\centering
		\begin{tikzpicture}[scale=1.0]
\begin{groupplot}[
  group style={
      group size=2 by 2,
      horizontal sep=0.45cm,
      vertical sep=1.5cm
  },
  width=0.5\textwidth,
  axis equal image,
  xtick = {-16.5, -8.0, 8.0},
  xticklabels={-16.5, -8.0, 8.0},
  ytick = \empty,
  xmin=-16.5, xmax=8.0,
  ymin=0.0, ymax=12.3
]

\nextgroupplot[xtick=\empty, xlabel={}]
\addplot graphics [xmin=-16.5, xmax=8.0, ymin=0.0, ymax=12.3] {{_img/cone_Euler2dM8IST_Mach_wmsh}.png};

\nextgroupplot[xtick=\empty, xlabel={}]
\addplot graphics [xmin=-16.5, xmax=8.0, ymin=0.0, ymax=12.3] {{_img/cone_Euler2dM8IST_Mach_womsh}.png};

\nextgroupplot[xtick=\empty, xlabel={}]
\addplot graphics [xmin=-16.5, xmax=8.0, ymin=0.0, ymax=12.3] {{_img/cone_Euler2dM10p6IST_Mach_wmsh}.png};

\nextgroupplot[xtick=\empty, xlabel={}]
\addplot graphics [xmin=-16.5, xmax=8.0, ymin=0.0, ymax=12.3] {{_img/cone_Euler2dM10p6IST_Mach_womsh}.png};

\end{groupplot}

\node[anchor=north] at ($(group c1r1.south)!0.5!(group c2r1.south)$) {\colorbarMatlabParula{0.0}{2}{4}{6}{8}};
\node[anchor=north] at ($(group c1r2.south)!0.5!(group c2r2.south)$) {\colorbarMatlabParula{0.0}{2.65}{5.3}{7.95}{10.6}};

\end{tikzpicture}
		\caption{
			HOIST mesh (\textit{left}) and solution (Mach number, \textit{right}) at the initial ($M_\infty = 8$, \textit{top}) and final ($M_\infty = 10.6$, \textit{bottom}) continuation stages for the blunt cone problem.
		}
		\label{fig:ns1_soln}
	\end{figure}

	\begin{figure}[!htbp]
		\centering
		\input{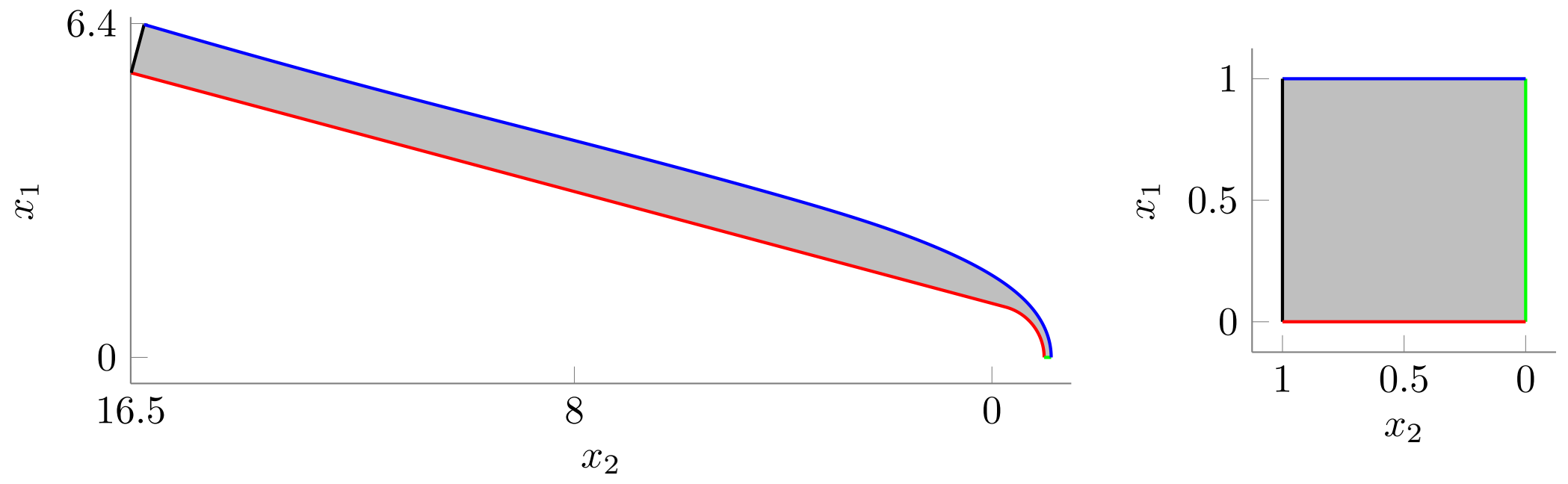}
		\caption{
			Schematic of blunt cone post-shock domain (\textit{left}), and the corresponding transfinite block reference domain (\textit{right}). 
			Boundary conditions: shock boundary condition in Section \ref{subsec:oblshk} (\ref{line:cone1:shkbc}), supersonic outlet (\ref{line:cone1:outlet}), isothermal no-slip wall (\ref{line:cone1:wall}), and symmetry axis (\ref{line:cone1:axis}).
		}
		\label{fig:ns1_geom1}
	\end{figure}

	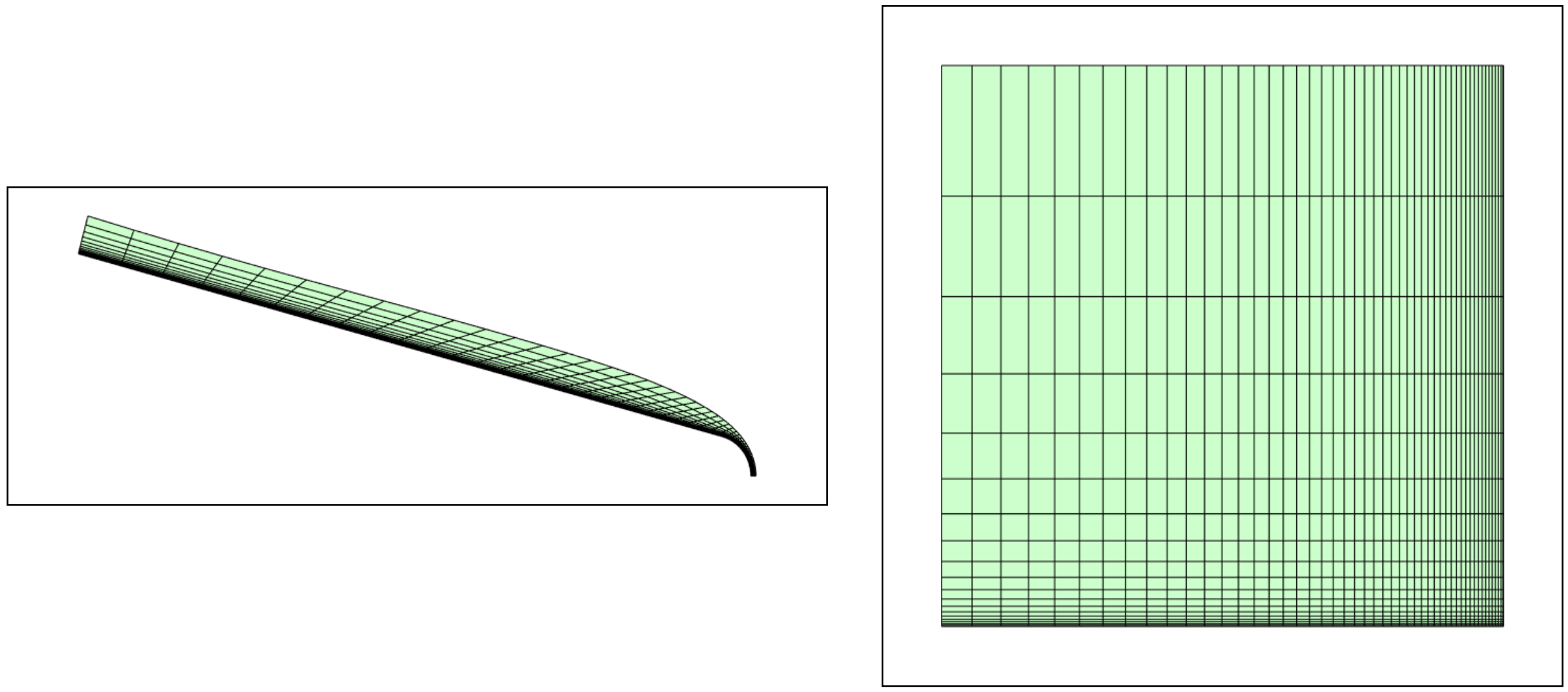
\begin{figure}[!htbp]
		\centering
		\begin{tikzpicture}[scale=1.25]
\begin{groupplot}[
  group style={
      group size=2 by 1,
      horizontal sep=0.45cm,
      vertical sep=1.5cm
  },
  width=0.5\textwidth,
  axis equal image,
  xlabel={$x_1$},
  ylabel={$x_2$},
]

\nextgroupplot[xlabel={}, ylabel={}, xtick=\empty, ytick=\empty]
\addplot graphics [xmin=-16.5, xmax=0.0, ymin=0, ymax=6.4] {{_img/cone_NSM10p6_phymshgls1}.png};

\nextgroupplot[xlabel={}, ylabel={}, xtick=\empty, ytick=\empty]
\addplot graphics [xmin=0.0, xmax=1.0, ymin=0.0, ymax=1.0] {{_img/cone_NSM10p6_refmshgls1}.png};

\end{groupplot}

\end{tikzpicture}
		\caption{
			$M_\infty = 10.6$ blunt cone post-shock mesh in physical domain (\textit{left}) and in a transfinite unit block reference domain (\textit{right}).
		}
		\label{fig:ns1_msh}
	\end{figure}
	
	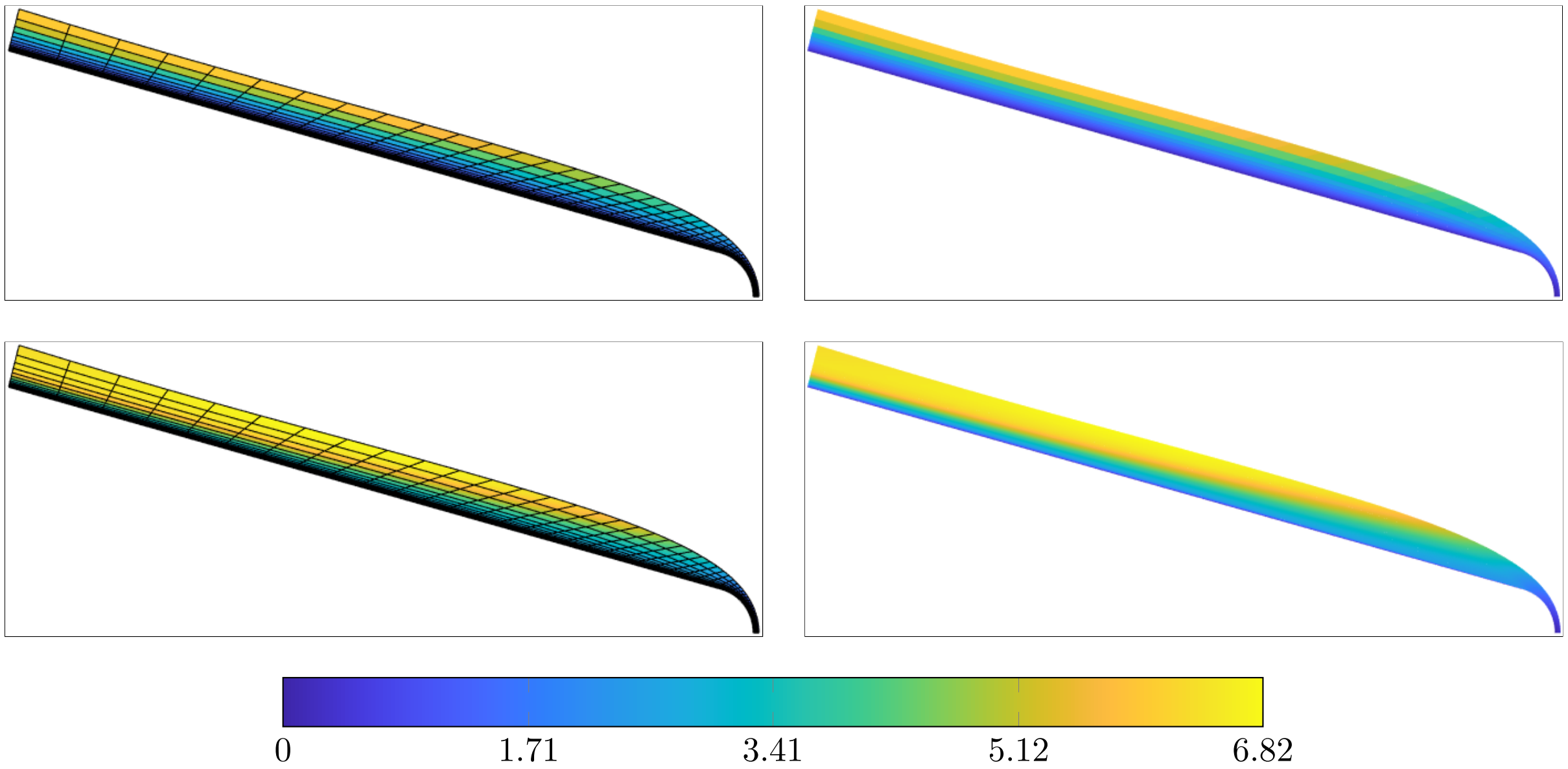
\begin{figure}[!htbp]
		\centering
		\begin{tikzpicture}[scale=1.25]
\begin{groupplot}[
  group style={
      group size=2 by 2,
      horizontal sep=0.35cm,
      vertical sep=0.35cm
  },
  width=0.47\textwidth,
  axis equal image,
  xtick = {-16.5, -8.0, 0.0},
  xticklabels={-16.5, -8.0, 0.0},
  ytick = \empty,
  xmin=-16.5, xmax=0.0,
  ymin=0, ymax=6.4
]

\nextgroupplot[xtick=\empty, xlabel={}]
\addplot graphics [xmin=-16.5, xmax=0.0, ymin=0, ymax=6.4] {{_img/cone_NSshkdwn_Mach_p0soln_wmsh}.png};

\nextgroupplot[xtick=\empty, xlabel={}]
\addplot graphics [xmin=-16.5, xmax=0.0, ymin=0, ymax=6.4] {{_img/cone_NSshkdwn_Mach_p0soln_womsh}.png};

\nextgroupplot[xtick=\empty, xlabel={}]
\addplot graphics [xmin=-16.5, xmax=0.0, ymin=0, ymax=6.4] {{_img/cone_NSshkdwns1_Mach_wmsh}.png};

\nextgroupplot[xtick=\empty, xlabel={}]
\addplot graphics [xmin=-16.5, xmax=0.0, ymin=0, ymax=6.4] {{_img/cone_NSshkdwns1_Mach_womsh}.png};

\end{groupplot}
\node[anchor=north] at ($(group c1r2.south)!0.5!(group c2r2.south)$) {\colorbarMatlabParula{0.0}{1.71}{3.41}{5.12}{6.82}};

\end{tikzpicture}
		\caption{
		Mach number field in the post-shock domain for the $M_\infty = 10.6$, $\text{Re}_\infty = 165000$ blunt cone problem: the $p = 0$ viscous solution used to initialize the PTC solver (\textit{top}) and the converged $p = 4$ PTC solution (\textit{bottom}).
		}
		\label{fig:ns1_dwstr_soln}
	\end{figure}
		
	\begin{figure}[!htbp]
		\centering
		\input{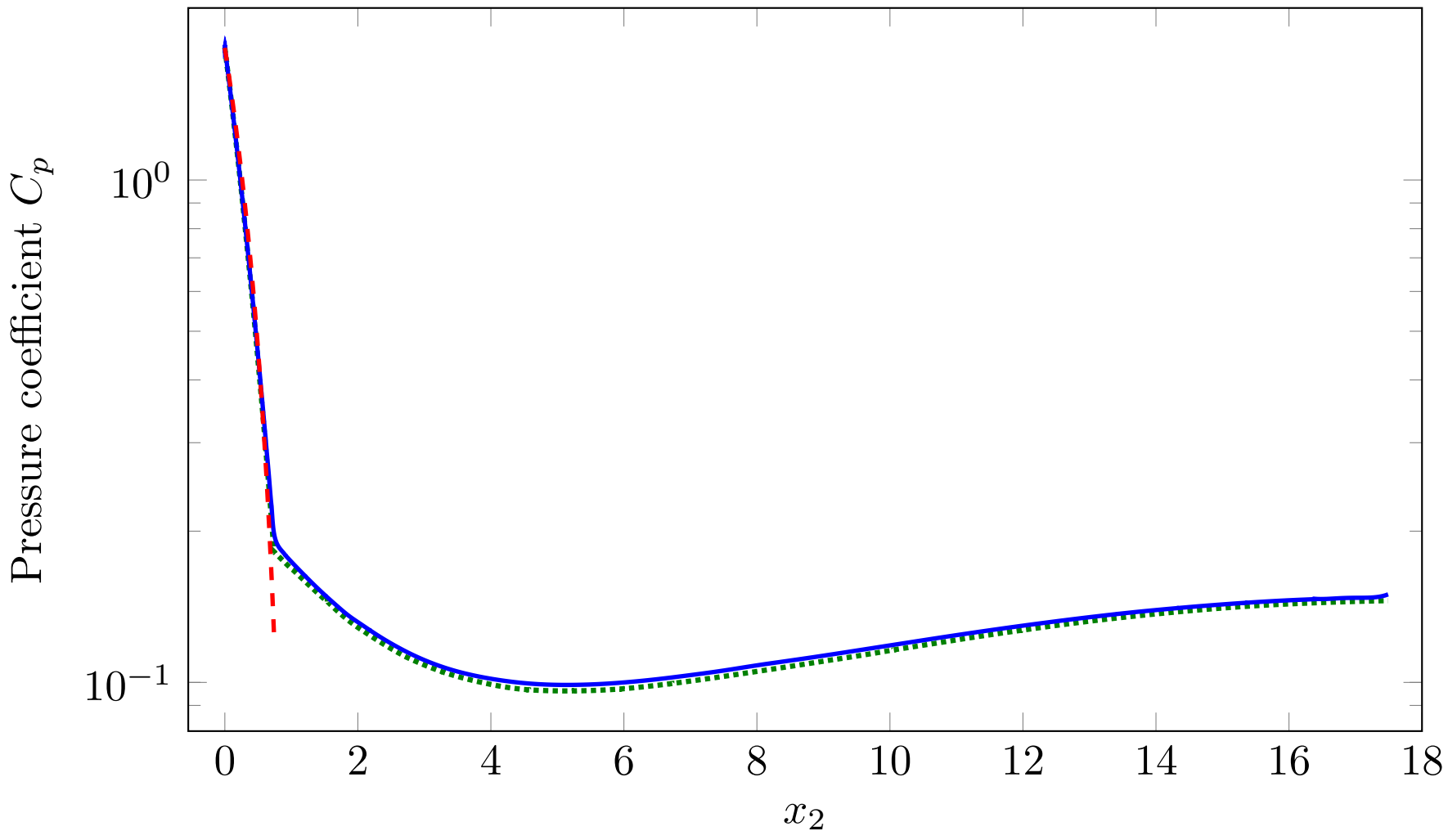}
			\caption{
				Surface pressure coefficient from inviscid HOIST solution (\ref{line:cone:cp:euler}), IPDG
				solution (\ref{line:cone:cp:ipdg}), and modified Newtonian
				theory~\cite{anderson2006hypersonic} over the spherical nose, $C_p = C_{p,\max}(1 - x_2)^2$ with $C_{p,\max} = 1.833$ (\ref{line:cone:cp:newt}),
				for the blunt cone problem.
				The axial coordinate $x_2$, nondimensionalized by the nose radius $L^*$, is measured from the nose tip.
				}
		\label{fig:ns1_cp}
	\end{figure}

	\begin{figure}[!htbp]
		\centering
		\input{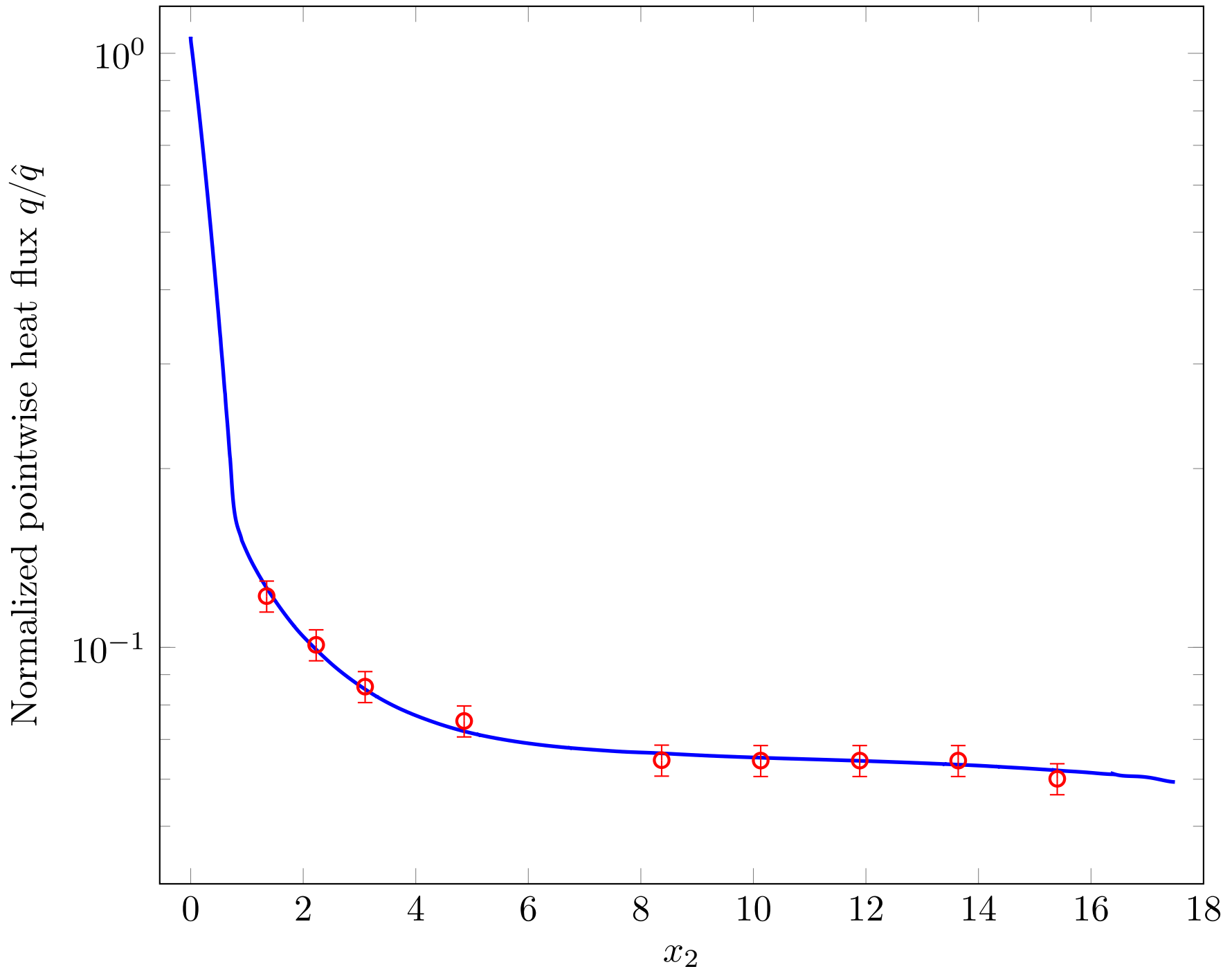}
			\caption{
				Surface heat flux normalized by the stagnation-point value, $q/\hat{q}$, from the IPDG solution
				(\ref{dots:cone:ipdg}) and experiment~\cite{cleary1969effects} (\ref{dots:cone:exp}) for the blunt cone
				problem. Error bars indicate the $\pm 6\%$ experimental repeatability.
				}
		\label{fig:ns1_hflx}
	\end{figure}

}

\subsection{Mach \texorpdfstring{$17.605$}{17.605} flow over a cylinder}
\label{subsec:cyl2d}
Next, we consider steady hypersonic flow over a circular cylinder.
The final freestream flow conditions are $M_{\infty} = 17.605$, $\text{Re}_\infty = 376930$. This problem has been widely adopted as a benchmark for evaluating the predictive capability of
numerical techniques in hypersonic flow simulations~\cite{gnoffo2004computational,nompelis2004development}. The cylinder wall is defined as the unit circle and the farfield boundary as the ellipse
$(x_1 / 4)^2 + (x_2 / 2.5)^2 = 1$. A Mach continuation strategy starting at $M_{\infty} = 5.0$ is also employed. 
The geometry and boundary conditions are illustrated in Figure~\ref{fig:ns2_geom0}.

We discretize the domain using a structured mesh of 210 curved ($q = 2$) quadrilateral elements
with quadratic approximation. The HOIST solver is initialized from a shock capturing DG solution (Figure \ref{fig:ns2_soln0}).
The Mach continuation is performed using $30$ stages. Each intermediate stage is allotted $50$ SQP iterations,
while the initial and final stages are solved using $100$ SQP iterations. The continuation framework produces well-resolved
shock profiles (Figure~\ref{fig:ns2_soln1}), and the SQP solver convergence history is shown in Figure~\ref{fig:ns2_opthist}. 
Even though the DG residual does not reach a tight tolerance at the final continuation stage, it is reduced 
by approximately four orders of magnitude (from $10^2$ to $10^{-2}$), and a nonlinear solver can be employed to further drive the residual to $\Ocal(10^{-9})$ on the fixed mesh without additional nonlinear stabilization~\cite{huang2023high,dong2025rpadapt}. This is unnecessary here, because only the shock-aligned mesh is carried forward: the inviscid flow solution computed with HOIST is discarded in favor of the $p = 0$ viscous solution used to initialize the viscous computation (Section~\ref{subsec:ptc}).

After Mach number continuation, the post-shock domain is extracted by removing upstream elements; the resulting domain and the corresponding block domain for transfinite mapping are shown in Figure~\ref{fig:ns2_geom1}. For this domain, the shock-boundary inlet
Mach number is $M_{\infty} = 17.605$ and the Reynolds number based on the radius of the cylinder is
$\text{Re}_\infty = 376930$. The shock boundary condition of Section~\ref{subsec:oblshk} is applied at the inlet and a supersonic outflow condition at the outlet, both with zero boundary viscous flux, $\bar{F}_{\text{visc}}^\partial = 0$; the cylinder surface is a no-slip, isothermal wall with $T_w/T_{\infty} = 2.5$, where the boundary state has the interior density, zero velocity, and the wall temperature, and the boundary viscous flux is the physical viscous flux evaluated with this state and the interior gradient, contracted with the normal~\cite{hartmann2008optimal}.

The post-shock domain is initially discretized using $340$ curved elements with a polynomial approximation of degree 
$p = 3$. An anisotropic $h$-adaptive refinement procedure is then applied to improve resolution in regions that most strongly 
influence the quantity of interest. The refinement strategy is driven by
a dual weighted residual (DWR) error estimator based on the integrated wall heat flux, which serves as 
the target functional. Elements are selectively refined according to their estimated contribution to 
the error, resulting in enhanced resolution near the boundary layer region. After five refinement cycles, the 
total number of elements increases to $496$. Figure~\ref{fig:ns2_soln2} shows the downstream flow solution before and after mesh refinement, while Figure~\ref{fig:ns2_msh} illustrates the corresponding block-domain meshes at the initial and final 
refinement levels.
To assess solution accuracy, the pressure coefficient (\ref{eqn:Cp}) and the Stanton number
are used as key validation metrics for hypersonic flow predictions, where the Stanton number is defined as
\begin{equation}\label{eqn:ChCp}
	C_h = \frac{q_n}{c_P\rho_{\infty}V_\infty(T_{t,\infty} - T_w)},
\end{equation}
where $q_n$ is the wall-normal heat flux and $T_{t,\infty}$ is the freestream total temperature.

Figures~\ref{fig:ns2_precoeff} and \ref{fig:ns2_hflx} demonstrate that the final computed pressure coefficient 
and surface heat flux profiles converge with refinement and show good agreement with the MDG-ICE results of Ching et al.~\cite{ching2024moving}. Those results were obtained with fully coupled viscous shock tracking and aggressive $r$-refinement that completely resolves the viscous shock structure as well as the boundary layer, so they constitute a high-fidelity reference for both the surface pressure and the surface heat flux.

\ifbool{fastcompile}{}{
	
    \begin{figure}[!htbp]
        \centering
        \begin{tikzpicture}
\begin{axis}[
width=0.65\textwidth,
axis equal image,
axis line style={gray},
axis x line*=bottom,
axis y line*=left,
xtick={-4, -1, 0, 1,4},
ytick={0, 2.5},
xlabel={$x_1$},
ymax=3.0,
xmax=4.5,
ylabel={$x_2$},
xmin=-4.5,
ymin=-0.5]

\addplot[domain=0:180, samples=200, fill=gray!50, draw=none] 
({4*cos(x)}, {2.5*sin(x)});
\addplot[domain=0:180, samples=200, fill=white, draw=none] 
({cos(x)}, {sin(x)});

\addplot[domain=0:180, samples=200, thick, blue] 
({4*cos(x)}, {2.5*sin(x)});\label{line:cyl0:inflow}

\addplot[domain=0:180, samples=200, thick, red] 
({cos(x)}, {sin(x)});\label{line:cyl0:wall}

\addplot[thick, black] coordinates {(-4,0) (-1,0)};\label{line:cyl0:outflow}
\addplot[thick, black] coordinates {(1,0) (4,0)};

\end{axis}
\end{tikzpicture}
        \caption{
            Schematic of cylinder domain and boundary conditions. Boundary
            conditions: supersonic inlet (\ref{line:cyl0:inflow}), supersonic
            outlet (\ref{line:cyl0:outflow}), slip wall
            (\ref{line:cyl0:wall}).
        }
        \label{fig:ns2_geom0}
    \end{figure}
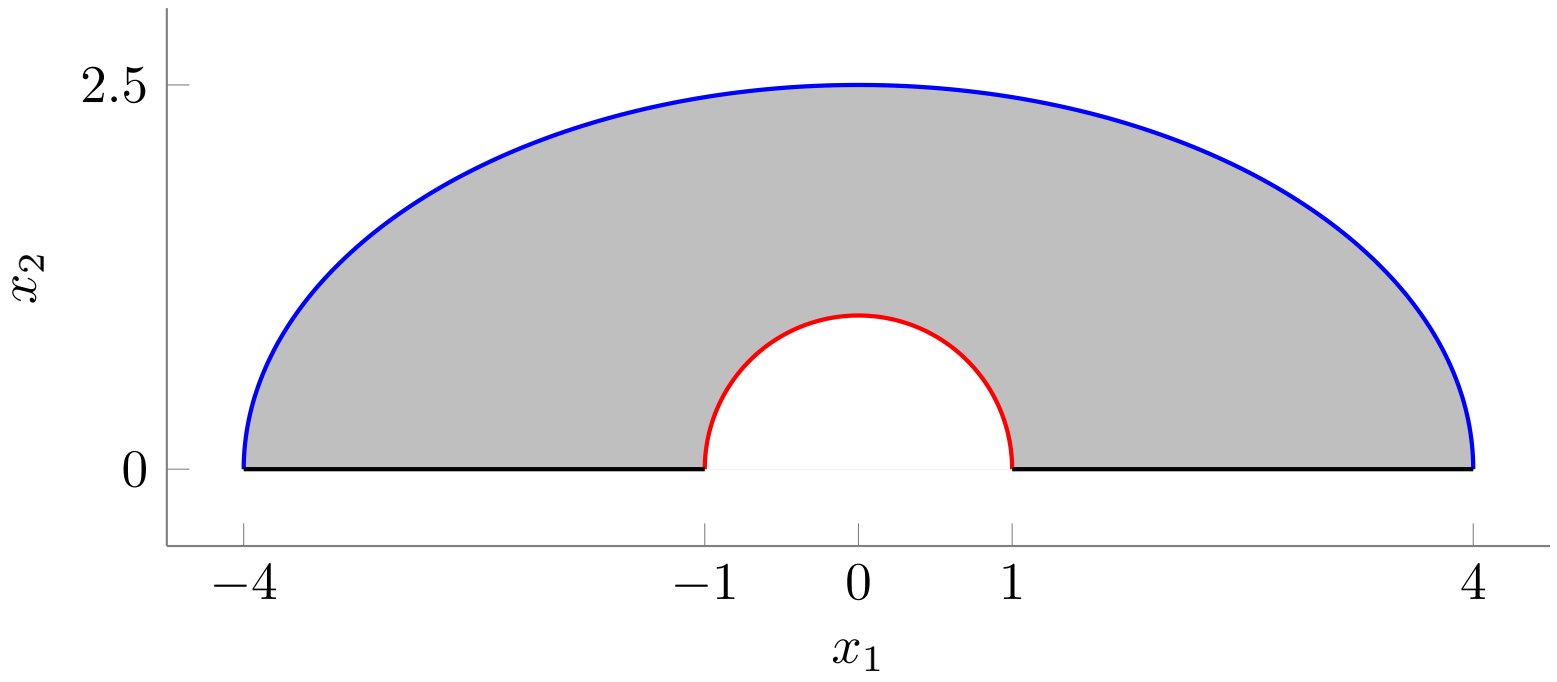
    
    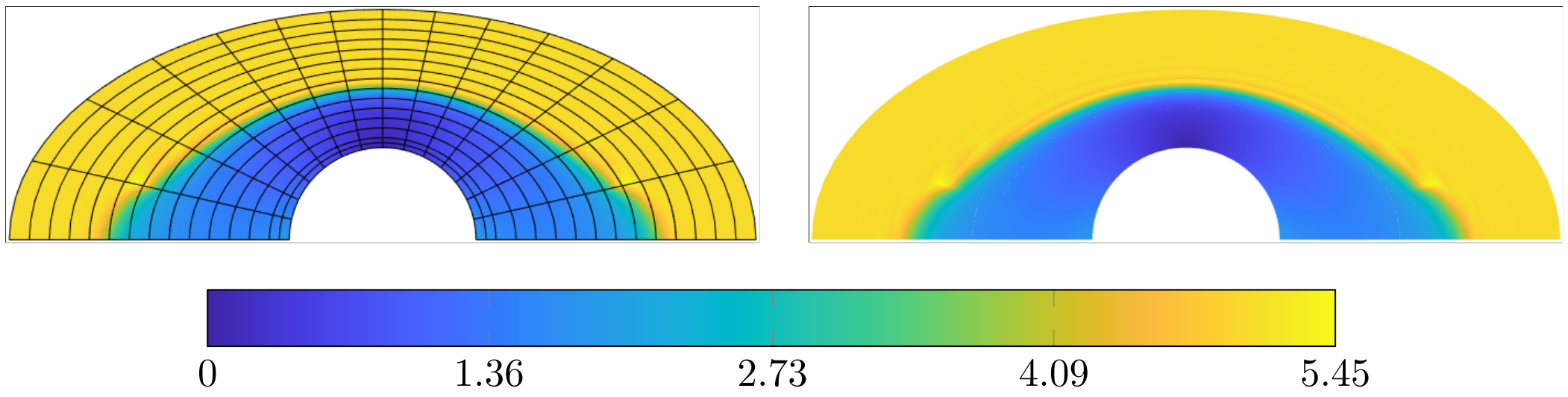
\begin{figure}[!htbp]
    	\centering
    	\begin{tikzpicture}[scale=1.0]
\begin{groupplot}[
  group style={
      group size=2 by 1,
      horizontal sep=0.45cm,
      vertical sep=1.5cm
  },
  width=0.5\textwidth,
  axis equal image,
  xtick = {-4.0, 0.0, 4.0},
  xticklabels={-4.0, 0.0, 4.0},
  ytick = \empty,
  xmin=-4.0, xmax=4.0,
  ymin=0, ymax=2.5
]

\nextgroupplot[xtick=\empty, xlabel={}]
\addplot graphics [xmin=-4.0, xmax=4.0, ymin=0, ymax=2.5] {{_img/cyl_Euler2dM5_Mach_wmsh}.png};

\nextgroupplot[xtick=\empty, xlabel={}]
\addplot graphics [xmin=-4.0, xmax=4.0, ymin=0, ymax=2.5] {{_img/cyl_Euler2dM5_Mach_womsh}.png};

\end{groupplot}

\node[anchor=north] at ($(group c1r1.south)!0.5!(group c2r1.south)$) {\colorbarMatlabParula{0.0}{1.36}{2.73}{4.09}{5.45}};

\end{tikzpicture}
    	\caption{
    		Mesh and Mach number field of the shock-capturing solution at $M_\infty = 5$ used to initialize the HOIST method for the cylinder problem.
    	}
    	\label{fig:ns2_soln0}
    \end{figure}

    \begin{figure}[!htbp]
        \centering
        \begin{tikzpicture}[scale=1.0]
\begin{groupplot}[
  group style={
      group size=2 by 2,
      horizontal sep=0.35cm,
      vertical sep=1.5cm
  },
  width=0.5\textwidth,
  axis equal image,
  xtick = {-4.0, 0.0, 4.0},
  xticklabels={-4.0, 0.0, 4.0},
  ytick = \empty,
  xmin=-4.0, xmax=4.0,
  ymin=0, ymax=2.5
]

\nextgroupplot[xtick=\empty, xlabel={}]
\addplot graphics [xmin=-4.0, xmax=4.0, ymin=0, ymax=2.5] {{_img/cyl_Euler2dM5IST_Mach_wmsh}.png};

\nextgroupplot[xtick=\empty, xlabel={}]
\addplot graphics [xmin=-4.0, xmax=4.0, ymin=0, ymax=2.5] {{_img/cyl_Euler2dM5IST_Mach_womsh}.png};

\nextgroupplot[xtick=\empty, xlabel={}]
\addplot graphics [xmin=-4.0, xmax=4.0, ymin=0, ymax=2.5] {{_img/cyl_Euler2dM17p6IST_Mach_wmsh}.png};

\nextgroupplot[xtick=\empty, xlabel={}]
\addplot graphics [xmin=-4.0, xmax=4.0, ymin=0, ymax=2.5] {{_img/cyl_Euler2dM17p6IST_Mach_womsh}.png};

\end{groupplot}

\node[anchor=north] at ($(group c1r1.south)!0.5!(group c2r1.south)$) {\colorbarMatlabParula{0.0}{1.25}{2.5}{3.75}{5}};
\node[anchor=north] at ($(group c1r2.south)!0.5!(group c2r2.south)$) {\colorbarMatlabParula{0.0}{4.4}{8.81}{13.21}{17.61}};
\end{tikzpicture}
        \caption{
        	HOIST mesh (\textit{left}) and solution (Mach number, \textit{right}) at the initial ($M_\infty = 5.0$, \textit{top}) and final ($M_\infty = 17.605$, \textit{bottom}) continuation stages for the cylinder problem.
        }
        \label{fig:ns2_soln1}
    \end{figure}
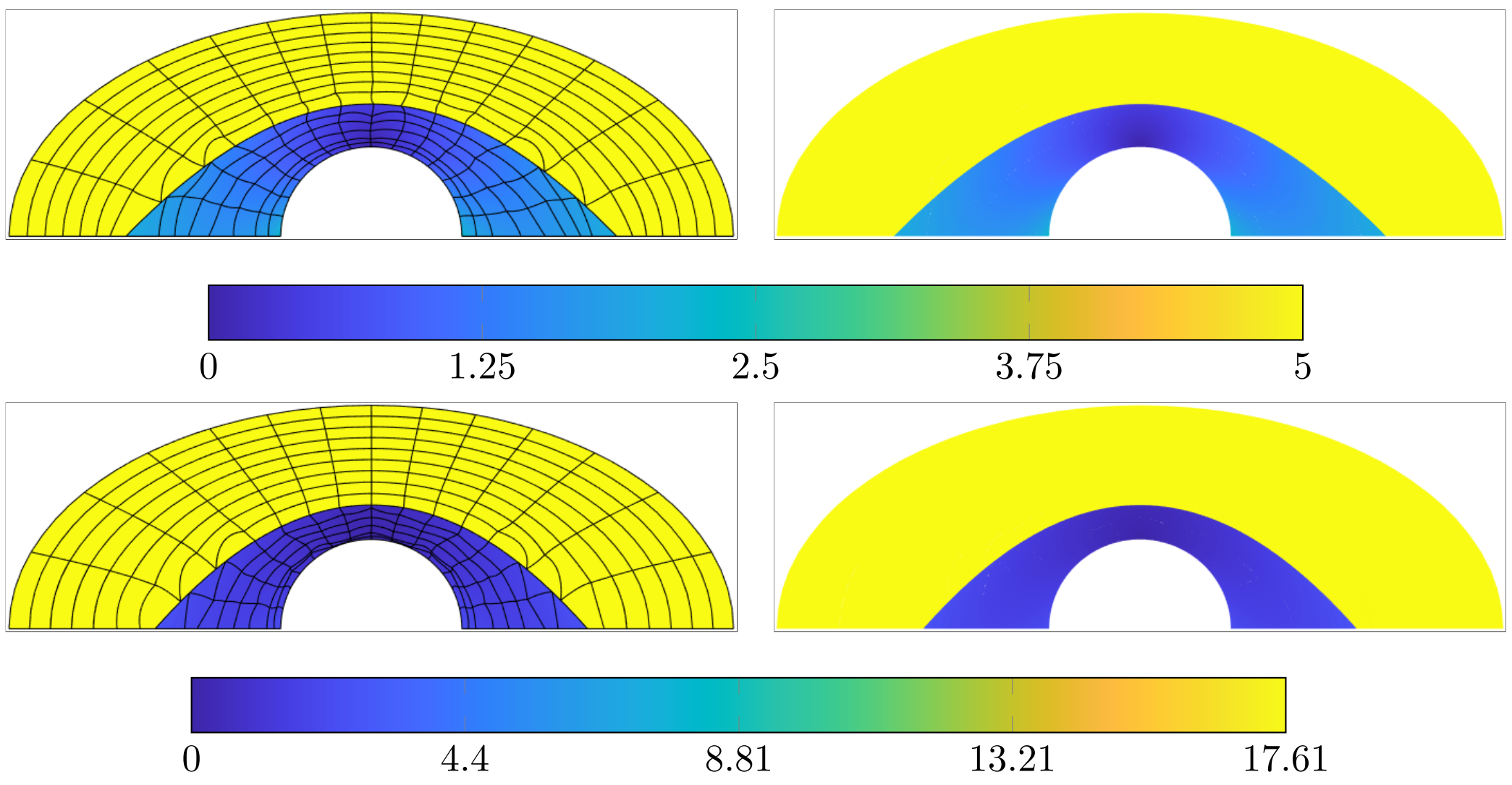

    \begin{figure}[!htbp]
    	\centering
    	\input{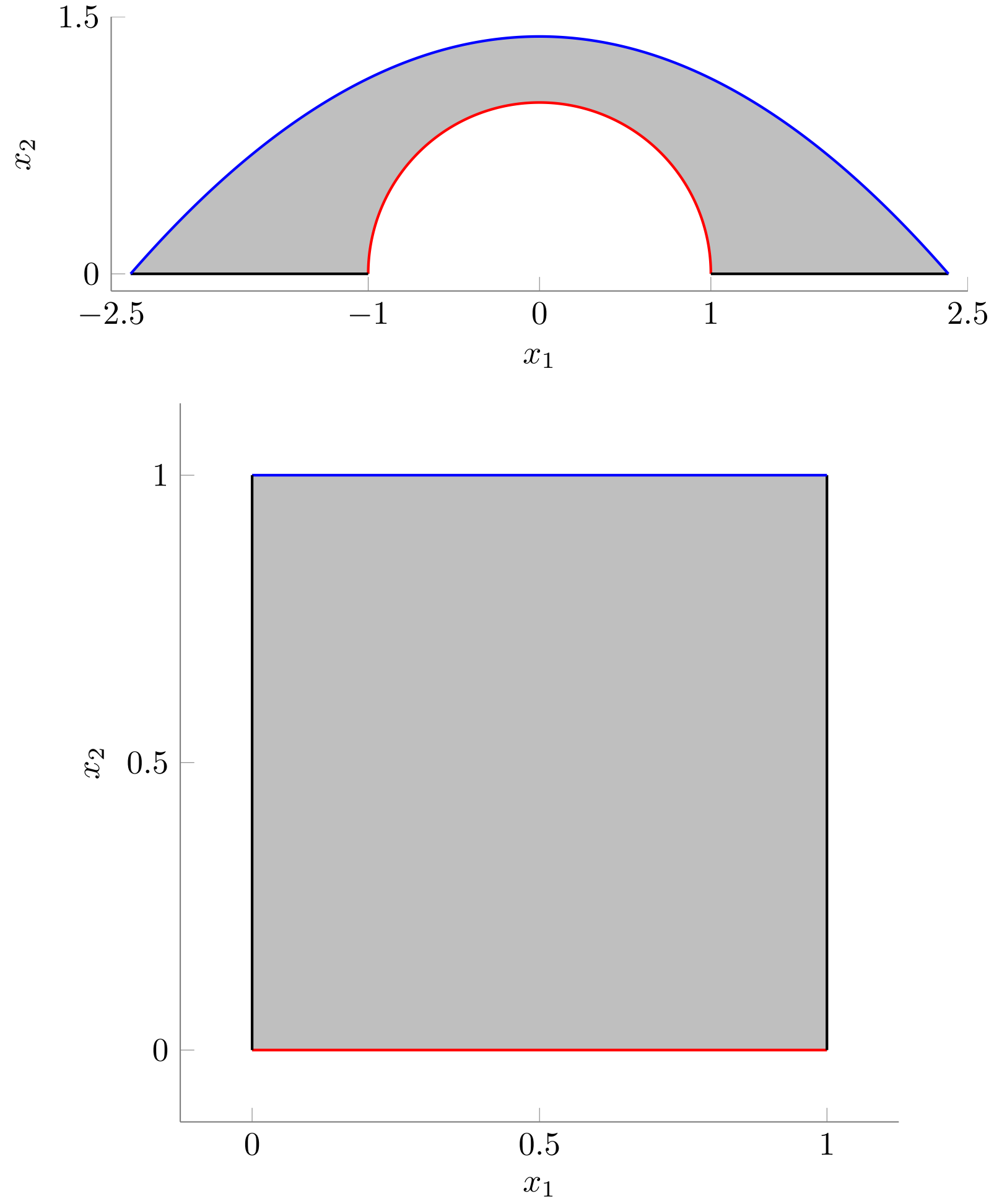}
    	\caption{
    		Schematic of cylinder post-shock domain (\textit{top}), and corresponding transfinite 
    		unit block domain (\textit{bottom}). Boundary conditions: shock boundary condition in Section \ref{subsec:oblshk} (\ref{line:cyl1:shkbc}), 
    		supersonic
    		outlet (\ref{line:cyl1:outflow}), isothermal no-slip wall
    		(\ref{line:cyl1:wall}).
    	}
    	\label{fig:ns2_geom1}
    \end{figure}
    
    \begin{figure}[!htbp]
    	\centering
    	\begin{tikzpicture}[scale=1.0]
\begin{groupplot}[
  group style={
      group size=2 by 2,
      horizontal sep=0.35cm,
      vertical sep=0.35cm
  },
  width=0.5\textwidth,
  axis equal image,
  xtick = {-2.4, 0.0, 2.4},
  xticklabels={-2.4, 0.0, 2.4},
  ytick = \empty,
  xmin=-2.4, xmax=2.4,
  ymin=0, ymax=1.4
]

\nextgroupplot[xtick=\empty, xlabel={}]
\addplot graphics [xmin=-2.4, xmax=2.4, ymin=0, ymax=1.4] {{_img/cyl_M17p6shkdwns1msh1_Mach_wmsh}.png};

\nextgroupplot[xtick=\empty, xlabel={}]
\addplot graphics [xmin=-2.4, xmax=2.4, ymin=0, ymax=1.4] {{_img/cyl_M17p6shkdwns1msh1_Mach_womsh}.png};

\nextgroupplot[xtick=\empty, xlabel={}]
\addplot graphics [xmin=-2.4, xmax=2.4, ymin=0, ymax=1.4] {{_img/cyl_M17p6shkdwns6_Mach_wmsh}.png};

\nextgroupplot[xtick=\empty, xlabel={}]
\addplot graphics [xmin=-2.4, xmax=2.4, ymin=0, ymax=1.4] {{_img/cyl_M17p6shkdwns6_Mach_womsh}.png};

\end{groupplot}

\node[anchor=north] at ($(group c1r2.south)!0.5!(group c2r2.south)$) {\colorbarMatlabParula{0.0}{0.65}{1.30}{1.95}{2.60}};
\end{tikzpicture}
    	\caption{
    		Post-shock mesh (\textit{left}) and solution (Mach number, \textit{right}) at the initial (\textit{top}) and final (\textit{bottom}) mesh refinement levels for the $M_\infty = 17.605$, $\text{Re}_\infty = 376930$ cylinder problem.
    	}
    	\label{fig:ns2_soln2}
    \end{figure}
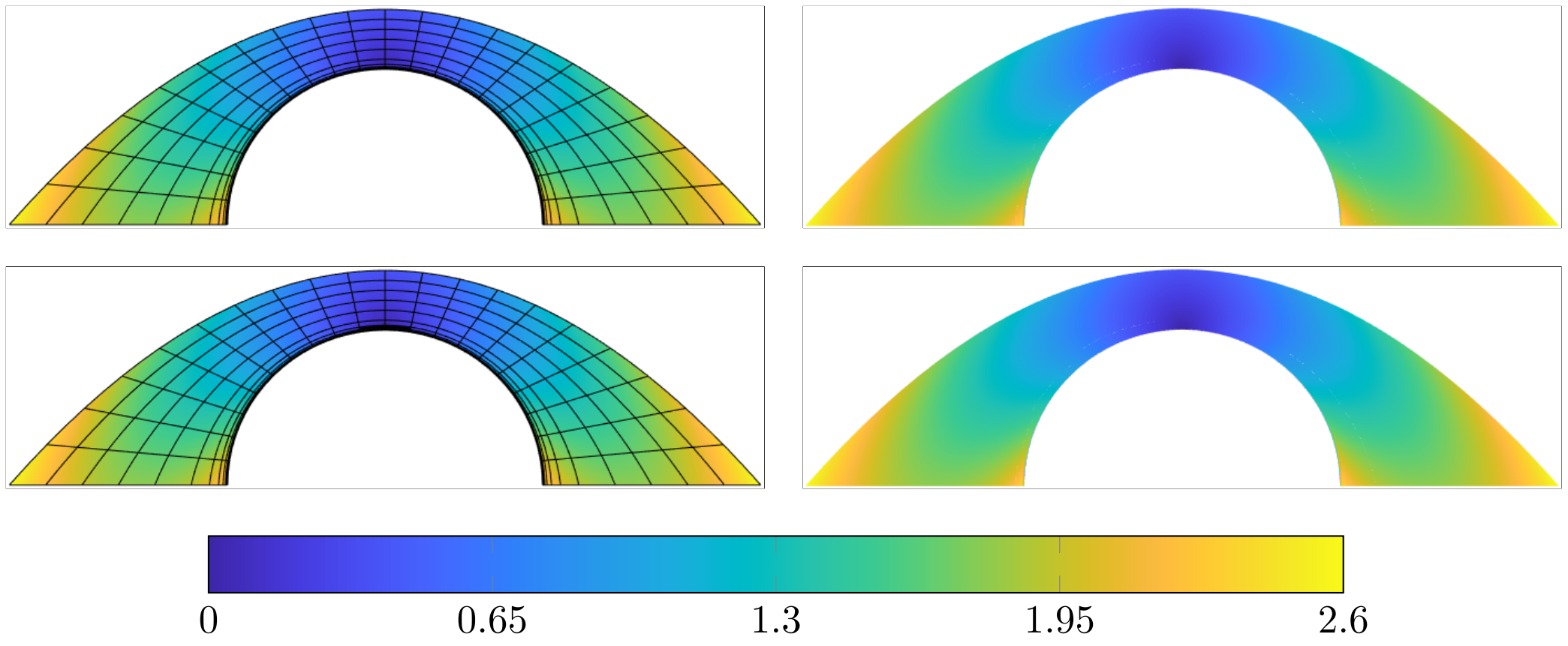

    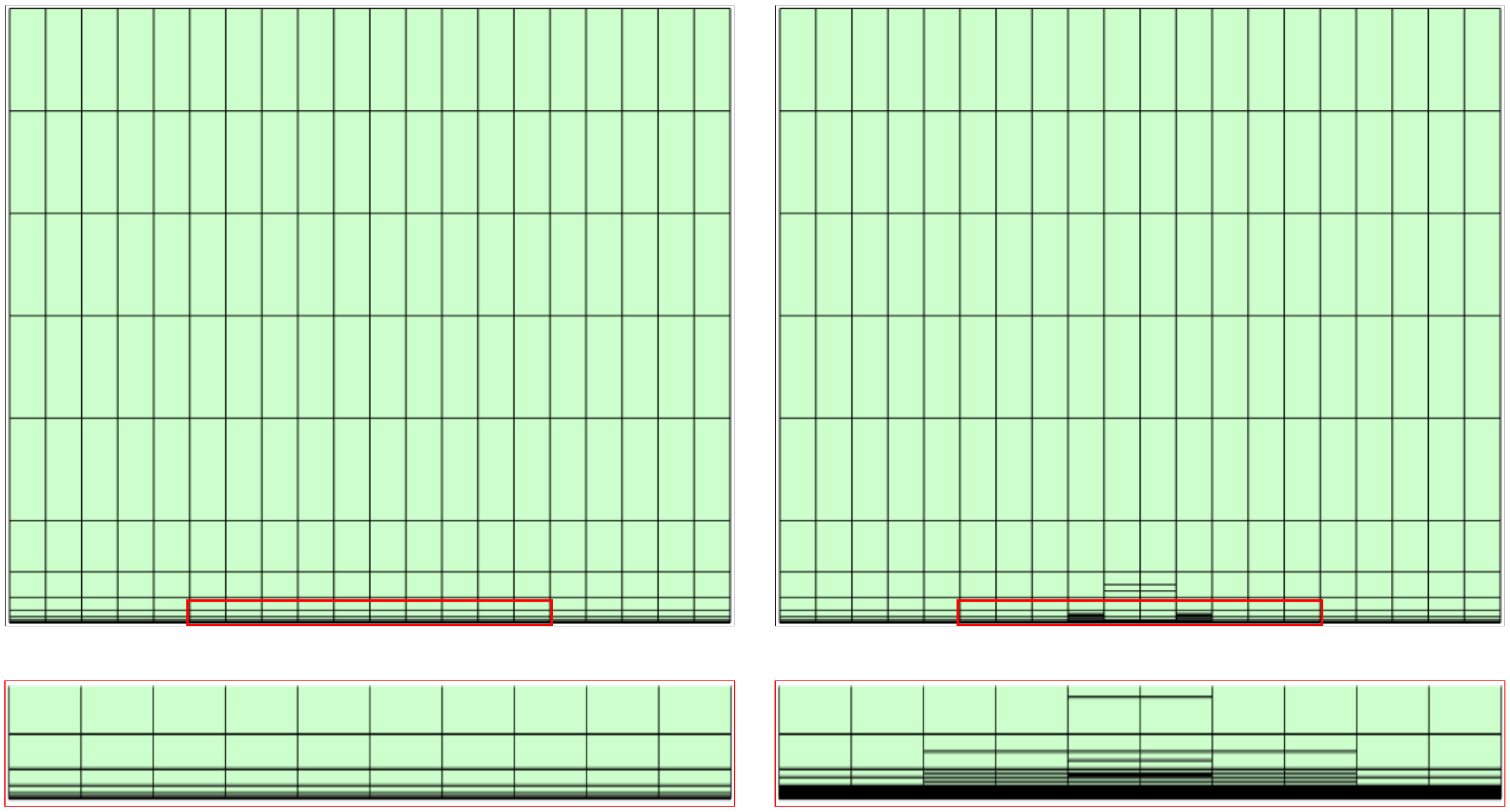
\begin{figure}[!htbp]
    	\centering
    	
    	\begin{tikzpicture}[scale=1.0]
\begin{groupplot}[
  group style={
      group size=2 by 2,
      horizontal sep=0.45cm,
      vertical sep=0.6cm
  },
  width=0.47\textwidth,
  scale only axis,
  enlargelimits=false,
  xlabel={},
  ylabel={},
  xtick=\empty,
  ytick=\empty
]

\nextgroupplot[height=0.4\textwidth, xmin=0.0, xmax=1.0, ymin=0.0, ymax=1.0]
\addplot graphics [xmin=0.0, xmax=1.0, ymin=0, ymax=1.0] {{_img/cyl_NSM17p6_mshgls1}.png};
\draw[red, thick] (axis cs:0.25, 0.0) rectangle (axis cs:0.75, 0.04);

\nextgroupplot[height=0.4\textwidth, xmin=0.0, xmax=1.0, ymin=0.0, ymax=1.0]
\addplot graphics [xmin=0.0, xmax=1.0, ymin=0, ymax=1.0] {{_img/cyl_NSM17p6_mshgls6}.png};
\draw[red, thick] (axis cs:0.25, 0.0) rectangle (axis cs:0.75, 0.04);

\nextgroupplot[height=0.08\textwidth, xmin=0.25, xmax=0.75, ymin=0.0, ymax=0.04, axis line style={red, thick}]
\addplot graphics [xmin=0.25, xmax=0.75, ymin=0.0, ymax=0.04] {{_img/cyl_NSM17p6_refmshs1zoom}.png};

\nextgroupplot[height=0.08\textwidth, xmin=0.25, xmax=0.75, ymin=0.0, ymax=0.04, axis line style={red, thick}]
\addplot graphics [xmin=0.25, xmax=0.75, ymin=0.0, ymax=0.04] {{_img/cyl_NSM17p6_refmshs8zoom}.png};

\end{groupplot}
\end{tikzpicture}
    	\caption{
    		Post-shock mesh in the transfinite block reference domain at the initial (\textit{left}) and final (\textit{right}) refinement levels for the $M_\infty = 17.605$, $\text{Re}_\infty = 376930$ cylinder problem (\textit{top}), and a magnified view of the near-wall region marked by the red box, which covers the middle half of the wall-tangential coordinate and the first four percent of the wall-normal coordinate, at the same refinement levels (\textit{bottom}). The wall is the bottom edge of the block. The magnified view shows the anisotropic $h$-refinement of the boundary layer beneath the stagnation region.
    	}
    	\label{fig:ns2_msh}
    \end{figure}

    \begin{figure}[!htbp]
    	\centering
    	\input{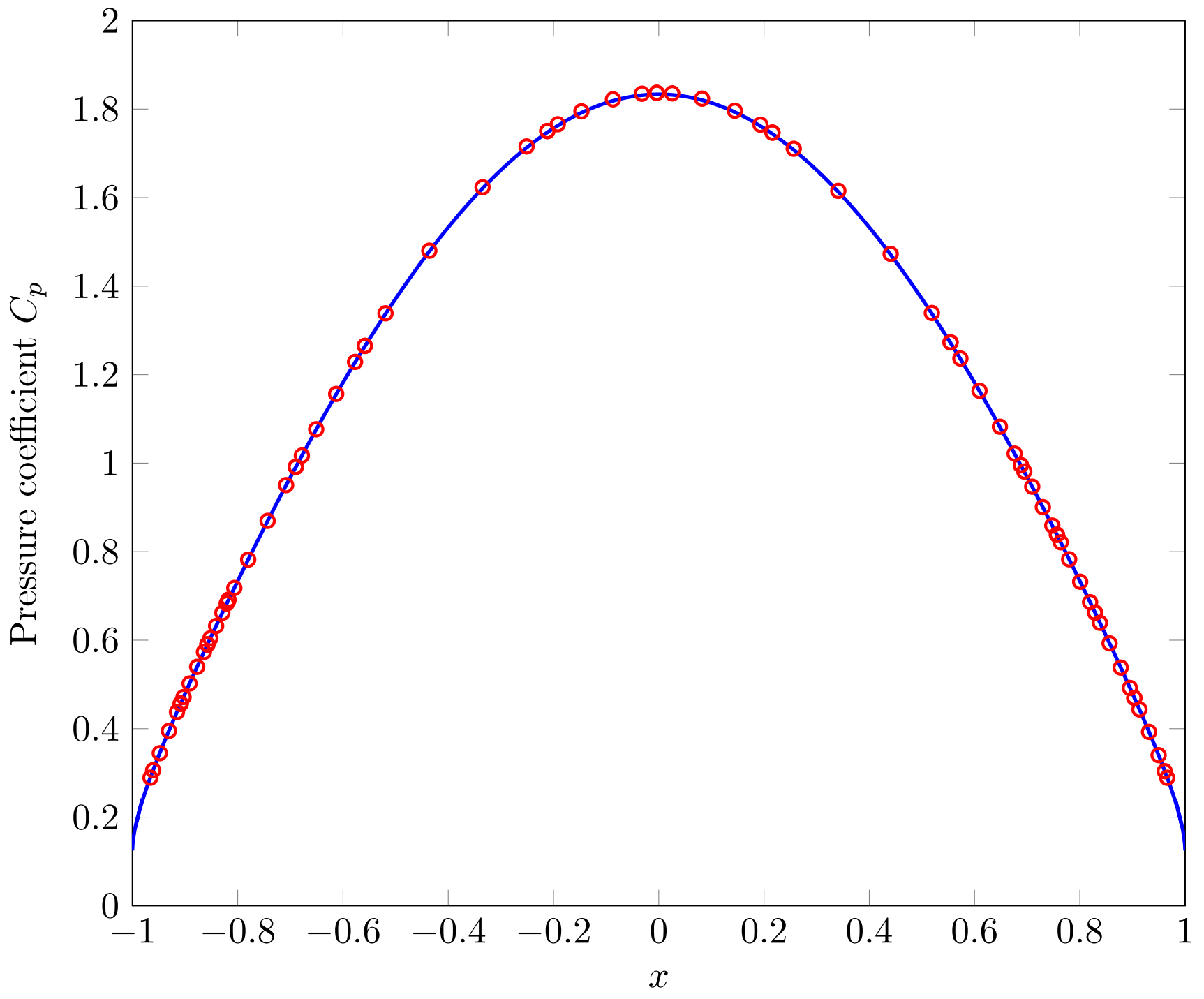}
    		\caption{
    			Pressure coefficient over the cylinder wall surface for the $M_\infty = 17.605$, $\text{Re}_\infty = 376930$ cylinder problem at the final mesh refinement level (\ref{dots:Cp}) and MDG-ICE results~\cite{ching2024moving} (\ref{dots:MDG-ICE_Cp}). Only the final refinement level is shown for clarity; the pressure coefficients at all refinement levels visually coincide.
    			}
    	\label{fig:ns2_precoeff}
    \end{figure}
    
    \begin{figure}[!htbp]
    	\centering
        	\input{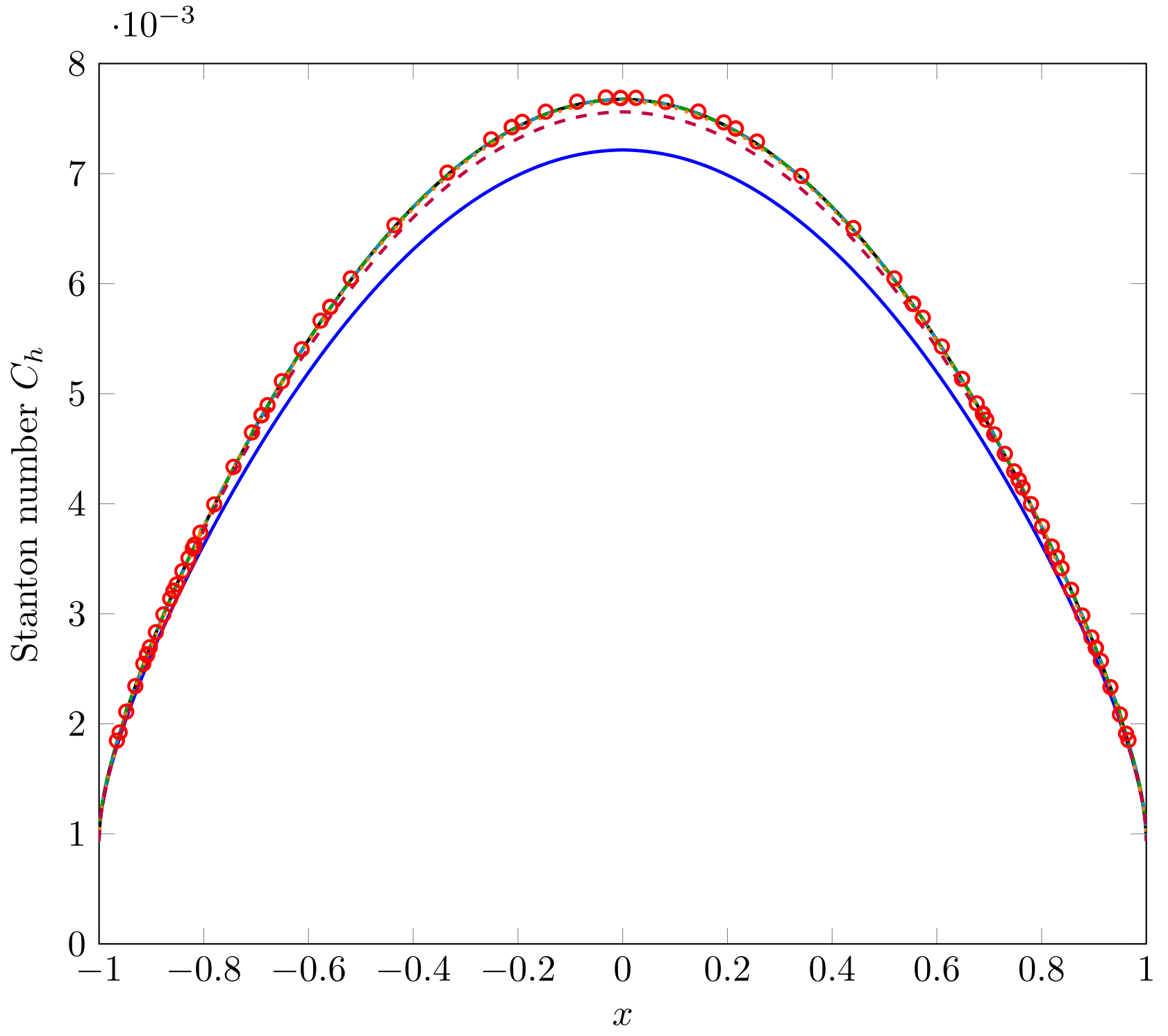}
    	\caption{
             Stanton number over cylinder wall surface prior to $h$-adaptation (\ref{dots:0th}) and after the first (\ref{dots:1st}), 
             second (\ref{dots:2nd}), third (\ref{dots:3rd}), fourth (\ref{dots:4th}), and fifth (\ref{dots:5th}) $h$-adaptation refinement, and MDG-ICE results~\cite{ching2024moving} (\ref{dots:MDG-ICE}) for the cylinder problem. The third through fifth refinement levels visually lie on top of one another and agree well with the reference MDG-ICE data.
    	}
    	\label{fig:ns2_hflx}
    \end{figure}

}

\subsection{Mach \texorpdfstring{$5$}{5} flow over a sphere}
\label{subsec:sph3d}

We consider a steady hypersonic flow over a sphere. The freestream flow conditions are
$M_{\infty}=5$, $\text{Re}_{\infty}=943750$, and a wall temperature ratio $T_{w}/T_{\infty}=1.308$. This problem is taken from the axisymmetric sphere simulation of Blottner~\cite{blottner1989accurate}, with a correction to the Reynolds number based on the experimental data~\cite{widhopf1972transitional} on which Blottner's computations are based. The geometry was nondimensionalized to a sphere of radius $r=1$ and only one octant of the sphere is used to reduce computational cost. The geometry and boundary conditions are illustrated in Figure~\ref{fig:ns3_geom0}. A staged approach is taken for this simulation. In the first stage, the HOIST method is applied with a piecewise constant ($p = 0$) solution approximation on a $q = 2$ mesh to roughly locate the shock; in the second stage, the polynomial degree is increased to $p = 2$ and the HOIST method is applied again, initialized from the first-stage solution. The first stage is inexpensive: a $p = 0$ tetrahedral element carries one solution node, versus ten for a $p = 2$ element, so each $p = 0$ SQP iteration involves roughly ten times fewer flow unknowns than a $p = 2$ iteration. 

We discretize the domain using an unstructured mesh with 3625 curved ($q=2$) tetrahedral elements. The $p = 0$ HOIST solve is initialized from a $p = 0$ DG solution on this mesh, i.e., a first-order finite volume solution that requires no additional stabilization (Figure~\ref{fig:ns3_ist}, \textit{left}). The $p = 0$ HOIST solve takes 200 SQP iterations over 39 outer iterations, during which 45 elements are collapsed, and yields a $p = 0$ HOIST solution whose element faces are approximately aligned with the shock. Farfield elements are then removed from the mesh of this $p = 0$ HOIST solution by filtering with a Mach cutoff of $M_{\mathrm{cut}} = 4.5$; a buffer band of elements is retained wherever an element's immediate neighbor is above the cutoff, so that exactly one layer of upstream elements remains next to the shock (Figure~\ref{fig:ns3_ist}, \textit{middle}), for a total of 800 curved tetrahedral elements. In the second stage, the polynomial degree of the solution representation on this mesh is increased to $p = 2$ and the HOIST method is run for 200 SQP iterations, initialized from the $p = 0$ HOIST solution. This $p = 2$ HOIST solve reduces the DG residual by 
three orders of magnitude from $\Ocal(10^{-1})$ to $\Ocal(10^{-4})$ as the mesh faces align with the
detached bow shock, yielding a stagnation shock standoff of $\delta_{\mathrm{sh}} = 0.1576\,r$
($\delta_{\mathrm{sh}}/D = 0.0788$) (Figure~\ref{fig:ns3_ist}).

Once the shock-aligned mesh is obtained, the pre-shock elements are removed with a Mach cutoff of
$M_{\mathrm{cut}} = 5.0$ to isolate the post-shock domain, reducing the mesh from $696$ elements (after element collapses
during the $p = 2$ HOIST solve) to $447$ curved elements. A boundary-layer mesh is then generated by extruding $60$ geometrically-clustered
prismatic layers (growth ratio $1.2$) from the sphere wall out to the fitted shock boundary, giving
$3600$ curved ($q = 2$) triangular prism elements that resolve the momentum boundary layer (Figure~\ref{fig:ns3_geom1}).

The Navier-Stokes equations are solved on this boundary-layer mesh with a Sutherland viscosity law
at $\text{Re}_\infty = 943750$ and Prandtl number $0.72$. The no-slip isothermal wall uses $T_w/T_\infty = 1.308$, the shock boundary condition imposes the oblique Rankine-Hugoniot post-shock
state through a Roe flux, the symmetry planes are slip walls, and the outflow enforces zero tangential viscous traction. The boundary viscous flux is zero at the shock boundary and on the symmetry planes, and at the wall it is the physical viscous flux evaluated with the wall state (interior density, zero velocity, wall temperature) and the interior gradient, contracted with the normal~\cite{hartmann2008optimal}. The solution is advanced through a
$p = 0 \to 1 \to 2$ polynomial-degree ladder (Figure~\ref{fig:ns3_soln}).

The resulting wall heat flux (Figure~\ref{fig:ns3_hflx}) reaches a value of
$108.07~\mathrm{kW/m^2}$ at the stagnation point, within $0.5\%$ of Blottner's reference value of
$107.572~\mathrm{kW/m^2}$~\cite{blottner1989accurate}, and decays monotonically toward the shoulder.
The heat flux is dimensionalized with $q_{\mathrm{ref}} = \rho_\infty c_\infty^3 = 465.4~\mathrm{kW/m^2}$,
the value corresponding to the freestream conditions used by Blottner~\cite{blottner1989accurate}.
Figure~\ref{fig:ns3_cmp} compares the computed
shock standoff distance, wall pressure, and surface heat flux against Blottner's data along the
sphere, showing close agreement across the windward arc. Evaluated along the three meridians
$\phi = 0^\circ$, $45^\circ$, and $90^\circ$---the two symmetry planes and the plane between them---the
standoff, pressure, and heat-flux profiles are in close agreement with one another (to within $1\%$ at
every station), confirming that the three-dimensional computation preserves the axisymmetry of the
problem. In particular, the agreement in shock standoff distance confirms that the shock position obtained from the inviscid tracking stage is adequate for the viscous computation at this Reynolds number.

Finally, we compare the shock-fitted viscous Navier-Stokes simulation with a shock-capturing
solution on a $43{,}008$-element structured quadratic hexahedral mesh ($q = 2$) with a $p = 2$
solution (Figure~\ref{fig:ns3_capture}), computed with the PDE-based artificial viscosity method
of Marques et al.~\cite{marques2026pdeav} for the same freestream conditions and wall temperature. The two solutions are compared quantitatively in
Figure~\ref{fig:ns3_cmp} and Table~\ref{tab:ns3_stag}. The shock-capturing stagnation heat flux is
$106.90~\mathrm{kW/m^2}$ ($-0.63\%$ from Blottner), bracketing the reference together with the
shock-tracking value of $108.07~\mathrm{kW/m^2}$ ($+0.46\%$); the two surface heat-flux profiles
agree to within a few percent along the entire windward arc, as do the wall-pressure profiles
(stagnation $p_w/p_\infty = 32.685$ for capturing and $32.672$ for tracking, both within $0.1\%$ of
Blottner's $32.655$). A captured shock has no sharp position, so the standoff of the shock-capturing solution is
taken as the ridge of the artificial viscosity field~\cite{marques2026pdeav}; it lies $1.5$--$3\%$
above Blottner's values along the body ($0.0811$ at the stagnation point, $+2.7\%$), whereas the
fitted shock agrees to within $0.2\%$ at the stagnation point.

\ifbool{fastcompile}{}{

	\begin{figure}[!htbp]
		\centering
		\raisebox{-0.5\height}{\begin{tikzpicture}
\begin{axis}[
axis equal image,
axis line style={gray},
axis x line*=bottom,
axis y line*=right,
xtick={-1.75, -1, 0},
ytick={-2, -1, 0},
width=0.42\textwidth,
xmin=-1.88125,
xmax=0.13125,
ymin=-2.15,
ymax=0.15,
xlabel={$x_1$},
ylabel={$x_3$}]
\addplot [fill=black!30!white, opacity=0.6, forget plot]
coordinates {
(-1.75000000e+00, -2.00000000e+00)
( 0.00000000e+00, -2.00000000e+00)
( 0.00000000e+00, -1.00000000e+00)
(-3.20515776e-02, -9.99486216e-01)
(-6.40702200e-02, -9.97945393e-01)
(-9.60230259e-02, -9.95379113e-01)
(-1.27877162e-01, -9.91790014e-01)
(-1.59599895e-01, -9.87181783e-01)
(-1.91158629e-01, -9.81559157e-01)
(-2.22520934e-01, -9.74927912e-01)
(-2.53654584e-01, -9.67294863e-01)
(-2.84527587e-01, -9.58667853e-01)
(-3.15108218e-01, -9.49055747e-01)
(-3.45365054e-01, -9.38468422e-01)
(-3.75267005e-01, -9.26916757e-01)
(-4.04783343e-01, -9.14412623e-01)
(-4.33883739e-01, -9.00968868e-01)
(-4.62538290e-01, -8.86599306e-01)
(-4.90717552e-01, -8.71318704e-01)
(-5.18392568e-01, -8.55142763e-01)
(-5.45534901e-01, -8.38088105e-01)
(-5.72116660e-01, -8.20172255e-01)
(-5.98110530e-01, -8.01413622e-01)
(-6.23489802e-01, -7.81831482e-01)
(-6.48228395e-01, -7.61445958e-01)
(-6.72300890e-01, -7.40277997e-01)
(-6.95682551e-01, -7.18349350e-01)
(-7.18349350e-01, -6.95682551e-01)
(-7.40277997e-01, -6.72300890e-01)
(-7.61445958e-01, -6.48228395e-01)
(-7.81831482e-01, -6.23489802e-01)
(-8.01413622e-01, -5.98110530e-01)
(-8.20172255e-01, -5.72116660e-01)
(-8.38088105e-01, -5.45534901e-01)
(-8.55142763e-01, -5.18392568e-01)
(-8.71318704e-01, -4.90717552e-01)
(-8.86599306e-01, -4.62538290e-01)
(-9.00968868e-01, -4.33883739e-01)
(-9.14412623e-01, -4.04783343e-01)
(-9.26916757e-01, -3.75267005e-01)
(-9.38468422e-01, -3.45365054e-01)
(-9.49055747e-01, -3.15108218e-01)
(-9.58667853e-01, -2.84527587e-01)
(-9.67294863e-01, -2.53654584e-01)
(-9.74927912e-01, -2.22520934e-01)
(-9.81559157e-01, -1.91158629e-01)
(-9.87181783e-01, -1.59599895e-01)
(-9.91790014e-01, -1.27877162e-01)
(-9.95379113e-01, -9.60230259e-02)
(-9.97945393e-01, -6.40702200e-02)
(-9.99486216e-01, -3.20515776e-02)
(-1.00000000e+00, -6.12323400e-17)
(-1.75000000e+00,  0.00000000e+00)
(-1.75000000e+00, -2.00000000e+00)};

\addplot [black, thick]
coordinates {
( 0.00000000e+00, -1.00000000e+00)
(-3.20515776e-02, -9.99486216e-01)
(-6.40702200e-02, -9.97945393e-01)
(-9.60230259e-02, -9.95379113e-01)
(-1.27877162e-01, -9.91790014e-01)
(-1.59599895e-01, -9.87181783e-01)
(-1.91158629e-01, -9.81559157e-01)
(-2.22520934e-01, -9.74927912e-01)
(-2.53654584e-01, -9.67294863e-01)
(-2.84527587e-01, -9.58667853e-01)
(-3.15108218e-01, -9.49055747e-01)
(-3.45365054e-01, -9.38468422e-01)
(-3.75267005e-01, -9.26916757e-01)
(-4.04783343e-01, -9.14412623e-01)
(-4.33883739e-01, -9.00968868e-01)
(-4.62538290e-01, -8.86599306e-01)
(-4.90717552e-01, -8.71318704e-01)
(-5.18392568e-01, -8.55142763e-01)
(-5.45534901e-01, -8.38088105e-01)
(-5.72116660e-01, -8.20172255e-01)
(-5.98110530e-01, -8.01413622e-01)
(-6.23489802e-01, -7.81831482e-01)
(-6.48228395e-01, -7.61445958e-01)
(-6.72300890e-01, -7.40277997e-01)
(-6.95682551e-01, -7.18349350e-01)
(-7.18349350e-01, -6.95682551e-01)
(-7.40277997e-01, -6.72300890e-01)
(-7.61445958e-01, -6.48228395e-01)
(-7.81831482e-01, -6.23489802e-01)
(-8.01413622e-01, -5.98110530e-01)
(-8.20172255e-01, -5.72116660e-01)
(-8.38088105e-01, -5.45534901e-01)
(-8.55142763e-01, -5.18392568e-01)
(-8.71318704e-01, -4.90717552e-01)
(-8.86599306e-01, -4.62538290e-01)
(-9.00968868e-01, -4.33883739e-01)
(-9.14412623e-01, -4.04783343e-01)
(-9.26916757e-01, -3.75267005e-01)
(-9.38468422e-01, -3.45365054e-01)
(-9.49055747e-01, -3.15108218e-01)
(-9.58667853e-01, -2.84527587e-01)
(-9.67294863e-01, -2.53654584e-01)
(-9.74927912e-01, -2.22520934e-01)
(-9.81559157e-01, -1.91158629e-01)
(-9.87181783e-01, -1.59599895e-01)
(-9.91790014e-01, -1.27877162e-01)
(-9.95379113e-01, -9.60230259e-02)
(-9.97945393e-01, -6.40702200e-02)
(-9.99486216e-01, -3.20515776e-02)
(-1.00000000e+00, -6.12323400e-17)
(-1.75000000e+00,  0.00000000e+00)};\label{line:sph0:wall}

\addplot [blue, thick]
coordinates {
(-1.75000000e+00, -2.00000000e+00)
( 0.00000000e+00, -2.00000000e+00)};\label{line:sph0:inlet}

\addplot [blue, thick, forget plot]
coordinates {
(-1.75000000e+00,  0.00000000e+00)
(-1.75000000e+00, -2.00000000e+00)};

\addplot [red, thick]
coordinates {
( 0.00000000e+00, -2.00000000e+00)
( 0.00000000e+00, -1.00000000e+00)};\label{line:sph0:outlet}

\end{axis}
\end{tikzpicture}} \quad
		\raisebox{-0.5\height}{\input{_tikz/sph_geom.tikz}} \qquad\quad
		\raisebox{-0.425\height}{\includegraphics[width=0.19\textwidth]{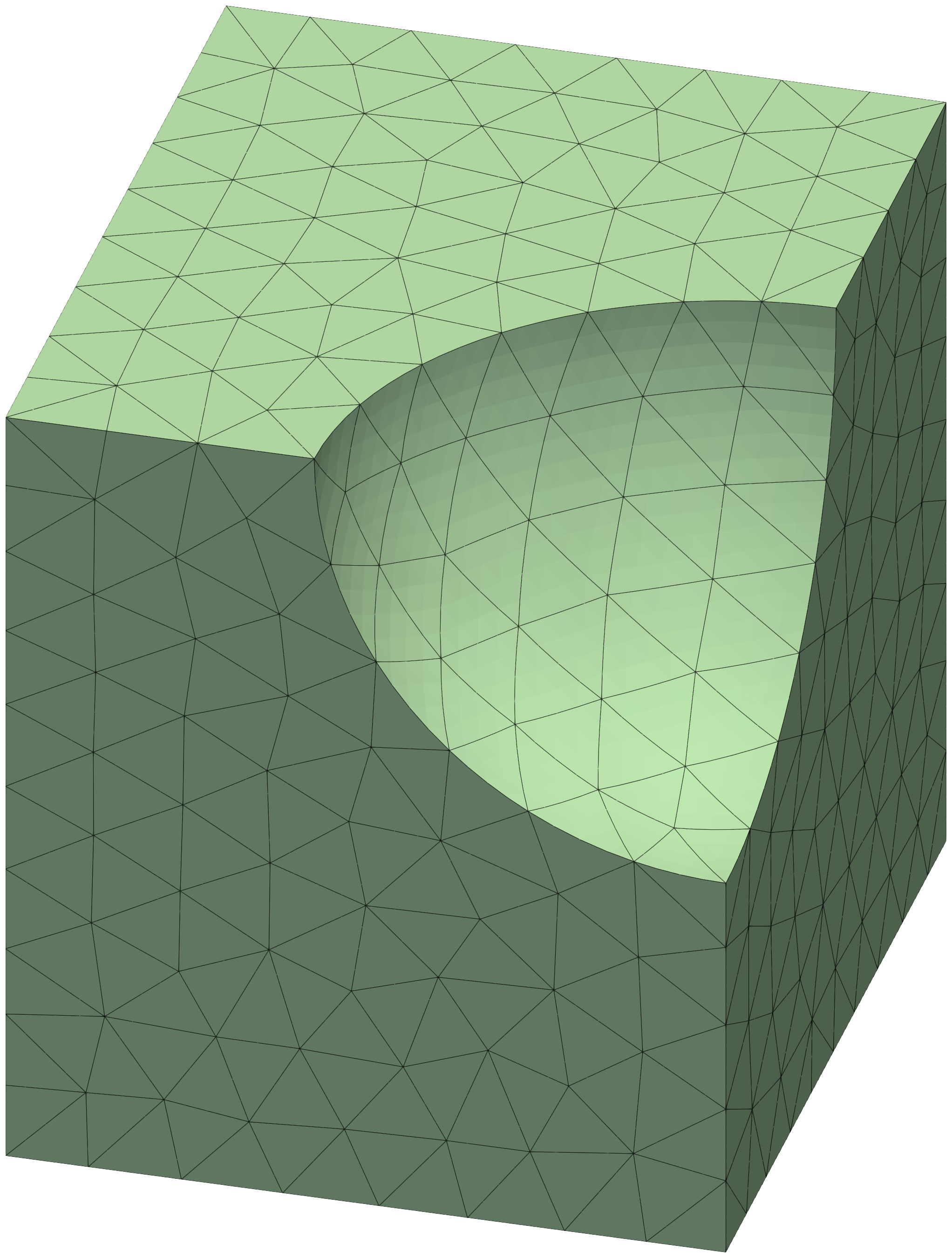}}
		\caption{
			Geometry of the sphere octant-section test case including a slice along the $x_2 = 0$ plane
			(\textit{left}) and a three-dimensional view (\textit{middle}), as well as the unstructured,
			high-order mesh with $3625$ quadratic tetrahedral elements (\textit{right}). The boundary
			conditions are indicated in the leftmost figure: slip wall (\ref{line:sph0:wall}), Mach~$5$
			supersonic inflow (\ref{line:sph0:inlet}), and supersonic outflow (\ref{line:sph0:outlet});
			the $x_2 = 0$ and $x_3 = 0$ planes use symmetry (slip wall) conditions.
		}
		\label{fig:ns3_geom0}
	\end{figure}

	\begin{figure}[!htbp]
		\centering
		\input{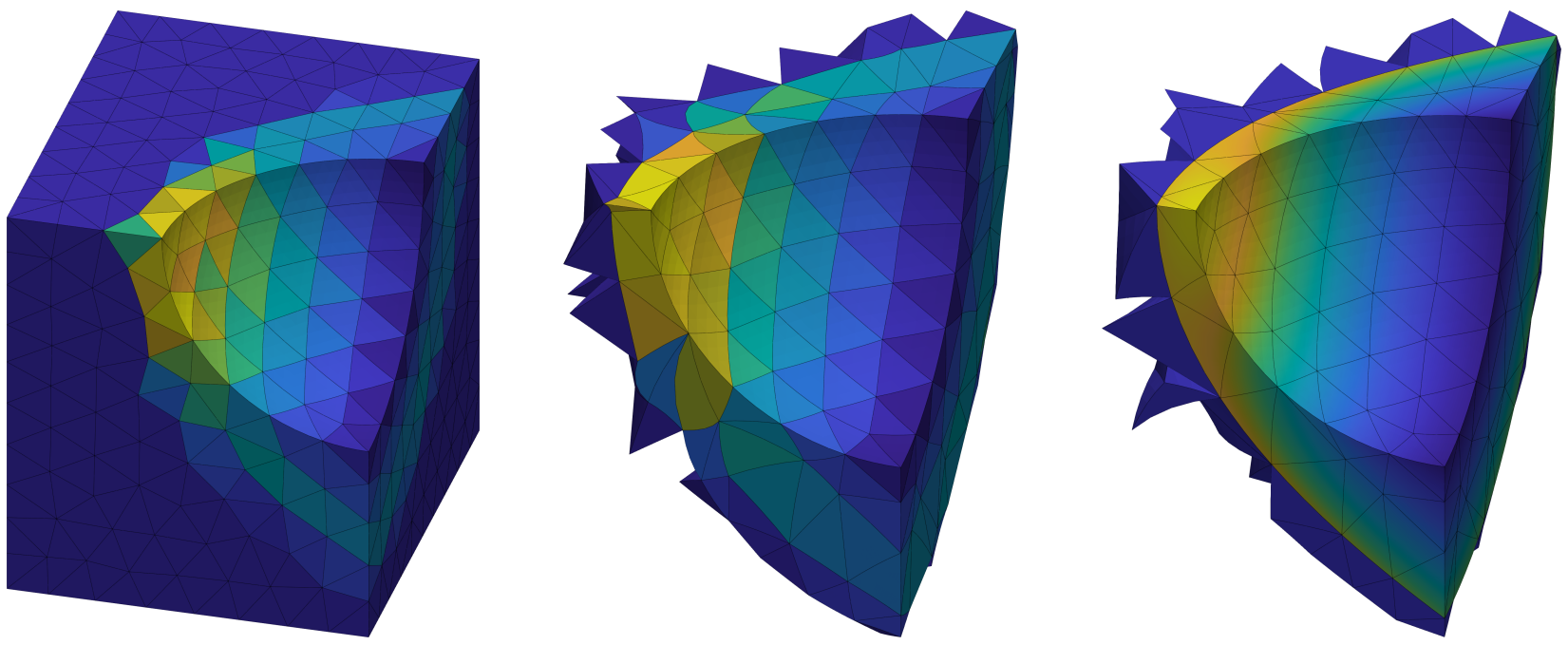}
		\caption{
			The $p = 0$ DG solution used to initialize the $p = 0$ HOIST solve (\textit{left}), the $p = 0$ HOIST solution after farfield element removal (\textit{middle}), and the final $p = 2$ HOIST solution (\textit{right}) for the Mach~$5$ sphere test case, colored by density. The $p = 0$ HOIST solution provides an approximate bow-shock location; elements whose Mach number exceeds the cutoff ($M > 4.5$, the pre-shock freestream) are then removed, while a single layer of pre-shock elements (a buffer band) is retained across the discontinuity. The resulting $p = 0$ state initializes the $p = 2$ HOIST solve, which aligns the mesh faces with the shock.
		}
		\label{fig:ns3_ist}
	\end{figure}

	\begin{figure}[!htbp]
		\centering
		\raisebox{-0.5\height}{\begin{tikzpicture}
\begin{axis}[
axis equal image,
axis line style={gray}, axis x line*=bottom, axis y line*=left,
xtick={-1,0}, ytick={0,1},
width=0.42\textwidth,
xmin=-1.35, xmax=0.12, ymin=-0.12, ymax=1.85,
xlabel={$x_1$}, ylabel={$x_2$}]
\addplot [fill=black!20!white, opacity=0.6, draw=none, forget plot] coordinates {(-1.157643e+00,0.000000e+00) (-1.157588e+00,1.794700e-02) (-1.157299e+00,3.563700e-02) (-1.156775e+00,5.307000e-02) (-1.153574e+00,1.086610e-01) (-1.150072e+00,1.469600e-01) (-1.145509e+00,1.851430e-01) (-1.139885e+00,2.232110e-01) (-1.133200e+00,2.611630e-01) (-1.125455e+00,2.989990e-01) (-1.116649e+00,3.367200e-01) (-1.106782e+00,3.743250e-01) (-1.094169e+00,4.178490e-01) (-1.080067e+00,4.613110e-01) (-1.064476e+00,5.047100e-01) (-1.047396e+00,5.480470e-01) (-1.028827e+00,5.913220e-01) (-1.008768e+00,6.345350e-01) (-9.872210e-01,6.776850e-01) (-9.641840e-01,7.207740e-01) (-9.508100e-01,7.444490e-01) (-9.369490e-01,7.681330e-01) (-9.226010e-01,7.918270e-01) (-9.077670e-01,8.155290e-01) (-8.916970e-01,8.405320e-01) (-8.749470e-01,8.656130e-01) (-8.575180e-01,8.907730e-01) (-8.394100e-01,9.160130e-01) (-8.206230e-01,9.413320e-01) (-8.011560e-01,9.667300e-01) (-7.810100e-01,9.922070e-01) (-7.601850e-01,1.017764e+00) (-7.307540e-01,1.052666e+00) (-7.003160e-01,1.087441e+00) (-6.688730e-01,1.122090e+00) (-6.364250e-01,1.156612e+00) (-6.029710e-01,1.191008e+00) (-5.685110e-01,1.225278e+00) (-5.330460e-01,1.259421e+00) (-4.965760e-01,1.293438e+00) (-4.786180e-01,1.309743e+00) (-4.604970e-01,1.325970e+00) (-4.422130e-01,1.342120e+00) (-4.237660e-01,1.358193e+00) (-4.051550e-01,1.374189e+00) (-3.863820e-01,1.390108e+00) (-3.674460e-01,1.405949e+00) (-3.483470e-01,1.421713e+00) (-3.130040e-01,1.450358e+00) (-2.773820e-01,1.478603e+00) (-2.414810e-01,1.506448e+00) (-2.053030e-01,1.533893e+00) (-1.802740e-01,1.552561e+00) (-1.550640e-01,1.571118e+00) (-1.296730e-01,1.589563e+00) (-1.041010e-01,1.607897e+00) (-7.834700e-02,1.626120e+00) (-5.241300e-02,1.644232e+00) (-2.629700e-02,1.662233e+00) (0.000000e+00,1.680122e+00) (0.000000e+00,1.000000e+00) (-6.123234e-17,1.000000e+00) (-1.764848e-02,9.998443e-01) (-3.529146e-02,9.993771e-01) (-5.292345e-02,9.985986e-01) (-7.053896e-02,9.975090e-01) (-8.813249e-02,9.961088e-01) (-1.056986e-01,9.943982e-01) (-1.232317e-01,9.923779e-01) (-1.407265e-01,9.900485e-01) (-1.581774e-01,9.874107e-01) (-1.755791e-01,9.844653e-01) (-1.929261e-01,9.812133e-01) (-2.102129e-01,9.776556e-01) (-2.274343e-01,9.737934e-01) (-2.445849e-01,9.696279e-01) (-2.616593e-01,9.651603e-01) (-2.786521e-01,9.603921e-01) (-2.955582e-01,9.553247e-01) (-3.123722e-01,9.499598e-01) (-3.290889e-01,9.442990e-01) (-3.457031e-01,9.383440e-01) (-3.622096e-01,9.320967e-01) (-3.786032e-01,9.255591e-01) (-3.948790e-01,9.187331e-01) (-4.110317e-01,9.116210e-01) (-4.270564e-01,9.042250e-01) (-4.429481e-01,8.965472e-01) (-4.587018e-01,8.885902e-01) (-4.743127e-01,8.803565e-01) (-4.897757e-01,8.718485e-01) (-5.050863e-01,8.630689e-01) (-5.202394e-01,8.540204e-01) (-5.352306e-01,8.447060e-01) (-5.500550e-01,8.351284e-01) (-5.647081e-01,8.252907e-01) (-5.791853e-01,8.151959e-01) (-5.934820e-01,8.048472e-01) (-6.075939e-01,7.942478e-01) (-6.215166e-01,7.834010e-01) (-6.352456e-01,7.723102e-01) (-6.487768e-01,7.609788e-01) (-6.621058e-01,7.494103e-01) (-6.752287e-01,7.376085e-01) (-6.881412e-01,7.255768e-01) (-7.008393e-01,7.133192e-01) (-7.133192e-01,7.008393e-01) (-7.255768e-01,6.881412e-01) (-7.376085e-01,6.752287e-01) (-7.494103e-01,6.621058e-01) (-7.609788e-01,6.487768e-01) (-7.723102e-01,6.352456e-01) (-7.834010e-01,6.215166e-01) (-7.942478e-01,6.075939e-01) (-8.048472e-01,5.934820e-01) (-8.151959e-01,5.791853e-01) (-8.252907e-01,5.647081e-01) (-8.351284e-01,5.500550e-01) (-8.447060e-01,5.352306e-01) (-8.540204e-01,5.202394e-01) (-8.630689e-01,5.050863e-01) (-8.718485e-01,4.897757e-01) (-8.803565e-01,4.743127e-01) (-8.885902e-01,4.587018e-01) (-8.965472e-01,4.429481e-01) (-9.042250e-01,4.270564e-01) (-9.116210e-01,4.110317e-01) (-9.187331e-01,3.948790e-01) (-9.255591e-01,3.786032e-01) (-9.320967e-01,3.622096e-01) (-9.383440e-01,3.457031e-01) (-9.442990e-01,3.290889e-01) (-9.499598e-01,3.123722e-01) (-9.553247e-01,2.955582e-01) (-9.603921e-01,2.786521e-01) (-9.651603e-01,2.616593e-01) (-9.696279e-01,2.445849e-01) (-9.737934e-01,2.274343e-01) (-9.776556e-01,2.102129e-01) (-9.812133e-01,1.929261e-01) (-9.844653e-01,1.755791e-01) (-9.874107e-01,1.581774e-01) (-9.900485e-01,1.407265e-01) (-9.923779e-01,1.232317e-01) (-9.943982e-01,1.056986e-01) (-9.961088e-01,8.813249e-02) (-9.975090e-01,7.053896e-02) (-9.985986e-01,5.292345e-02) (-9.993771e-01,3.529146e-02) (-9.998443e-01,1.764848e-02) (-1.000000e+00,0.000000e+00) (-1.157643e+00,0.000000e+00)} --cycle;
\addplot [red, very thick, forget plot] coordinates {(-1.000000e+00,0.000000e+00) (-9.998443e-01,1.764848e-02) (-9.993771e-01,3.529146e-02) (-9.985986e-01,5.292345e-02) (-9.975090e-01,7.053896e-02) (-9.961088e-01,8.813249e-02) (-9.943982e-01,1.056986e-01) (-9.923779e-01,1.232317e-01) (-9.900485e-01,1.407265e-01) (-9.874107e-01,1.581774e-01) (-9.844653e-01,1.755791e-01) (-9.812133e-01,1.929261e-01) (-9.776556e-01,2.102129e-01) (-9.737934e-01,2.274343e-01) (-9.696279e-01,2.445849e-01) (-9.651603e-01,2.616593e-01) (-9.603921e-01,2.786521e-01) (-9.553247e-01,2.955582e-01) (-9.499598e-01,3.123722e-01) (-9.442990e-01,3.290889e-01) (-9.383440e-01,3.457031e-01) (-9.320967e-01,3.622096e-01) (-9.255591e-01,3.786032e-01) (-9.187331e-01,3.948790e-01) (-9.116210e-01,4.110317e-01) (-9.042250e-01,4.270564e-01) (-8.965472e-01,4.429481e-01) (-8.885902e-01,4.587018e-01) (-8.803565e-01,4.743127e-01) (-8.718485e-01,4.897757e-01) (-8.630689e-01,5.050863e-01) (-8.540204e-01,5.202394e-01) (-8.447060e-01,5.352306e-01) (-8.351284e-01,5.500550e-01) (-8.252907e-01,5.647081e-01) (-8.151959e-01,5.791853e-01) (-8.048472e-01,5.934820e-01) (-7.942478e-01,6.075939e-01) (-7.834010e-01,6.215166e-01) (-7.723102e-01,6.352456e-01) (-7.609788e-01,6.487768e-01) (-7.494103e-01,6.621058e-01) (-7.376085e-01,6.752287e-01) (-7.255768e-01,6.881412e-01) (-7.133192e-01,7.008393e-01) (-7.008393e-01,7.133192e-01) (-6.881412e-01,7.255768e-01) (-6.752287e-01,7.376085e-01) (-6.621058e-01,7.494103e-01) (-6.487768e-01,7.609788e-01) (-6.352456e-01,7.723102e-01) (-6.215166e-01,7.834010e-01) (-6.075939e-01,7.942478e-01) (-5.934820e-01,8.048472e-01) (-5.791853e-01,8.151959e-01) (-5.647081e-01,8.252907e-01) (-5.500550e-01,8.351284e-01) (-5.352306e-01,8.447060e-01) (-5.202394e-01,8.540204e-01) (-5.050863e-01,8.630689e-01) (-4.897757e-01,8.718485e-01) (-4.743127e-01,8.803565e-01) (-4.587018e-01,8.885902e-01) (-4.429481e-01,8.965472e-01) (-4.270564e-01,9.042250e-01) (-4.110317e-01,9.116210e-01) (-3.948790e-01,9.187331e-01) (-3.786032e-01,9.255591e-01) (-3.622096e-01,9.320967e-01) (-3.457031e-01,9.383440e-01) (-3.290889e-01,9.442990e-01) (-3.123722e-01,9.499598e-01) (-2.955582e-01,9.553247e-01) (-2.786521e-01,9.603921e-01) (-2.616593e-01,9.651603e-01) (-2.445849e-01,9.696279e-01) (-2.274343e-01,9.737934e-01) (-2.102129e-01,9.776556e-01) (-1.929261e-01,9.812133e-01) (-1.755791e-01,9.844653e-01) (-1.581774e-01,9.874107e-01) (-1.407265e-01,9.900485e-01) (-1.232317e-01,9.923779e-01) (-1.056986e-01,9.943982e-01) (-8.813249e-02,9.961088e-01) (-7.053896e-02,9.975090e-01) (-5.292345e-02,9.985986e-01) (-3.529146e-02,9.993771e-01) (-1.764848e-02,9.998443e-01) (-6.123234e-17,1.000000e+00)};            % wall (sphere arc)
\addplot [blue, very thick, forget plot] coordinates {(-1.157643e+00,0.000000e+00) (-1.157588e+00,1.794700e-02) (-1.157299e+00,3.563700e-02) (-1.156775e+00,5.307000e-02) (-1.153574e+00,1.086610e-01) (-1.150072e+00,1.469600e-01) (-1.145509e+00,1.851430e-01) (-1.139885e+00,2.232110e-01) (-1.133200e+00,2.611630e-01) (-1.125455e+00,2.989990e-01) (-1.116649e+00,3.367200e-01) (-1.106782e+00,3.743250e-01) (-1.094169e+00,4.178490e-01) (-1.080067e+00,4.613110e-01) (-1.064476e+00,5.047100e-01) (-1.047396e+00,5.480470e-01) (-1.028827e+00,5.913220e-01) (-1.008768e+00,6.345350e-01) (-9.872210e-01,6.776850e-01) (-9.641840e-01,7.207740e-01) (-9.508100e-01,7.444490e-01) (-9.369490e-01,7.681330e-01) (-9.226010e-01,7.918270e-01) (-9.077670e-01,8.155290e-01) (-8.916970e-01,8.405320e-01) (-8.749470e-01,8.656130e-01) (-8.575180e-01,8.907730e-01) (-8.394100e-01,9.160130e-01) (-8.206230e-01,9.413320e-01) (-8.011560e-01,9.667300e-01) (-7.810100e-01,9.922070e-01) (-7.601850e-01,1.017764e+00) (-7.307540e-01,1.052666e+00) (-7.003160e-01,1.087441e+00) (-6.688730e-01,1.122090e+00) (-6.364250e-01,1.156612e+00) (-6.029710e-01,1.191008e+00) (-5.685110e-01,1.225278e+00) (-5.330460e-01,1.259421e+00) (-4.965760e-01,1.293438e+00) (-4.786180e-01,1.309743e+00) (-4.604970e-01,1.325970e+00) (-4.422130e-01,1.342120e+00) (-4.237660e-01,1.358193e+00) (-4.051550e-01,1.374189e+00) (-3.863820e-01,1.390108e+00) (-3.674460e-01,1.405949e+00) (-3.483470e-01,1.421713e+00) (-3.130040e-01,1.450358e+00) (-2.773820e-01,1.478603e+00) (-2.414810e-01,1.506448e+00) (-2.053030e-01,1.533893e+00) (-1.802740e-01,1.552561e+00) (-1.550640e-01,1.571118e+00) (-1.296730e-01,1.589563e+00) (-1.041010e-01,1.607897e+00) (-7.834700e-02,1.626120e+00) (-5.241300e-02,1.644232e+00) (-2.629700e-02,1.662233e+00) (0.000000e+00,1.680122e+00)};          % shock/inflow (fitted arc)
\addplot [orange, very thick] coordinates {(0,1) (0,1.680122e+00)};\label{line:sph2:outlet}
\addplot [black!55!white, very thick] coordinates {(-1.000000e+00,0) (-1.157643e+00,0)};\label{line:sph2:sym}
\end{axis}
\end{tikzpicture}} \quad
		\raisebox{-0.5\height}{\input{_tikz/sph_carve3d.tikz}} \qquad\quad
		\raisebox{-0.425\height}{\includegraphics[width=0.19\textwidth]{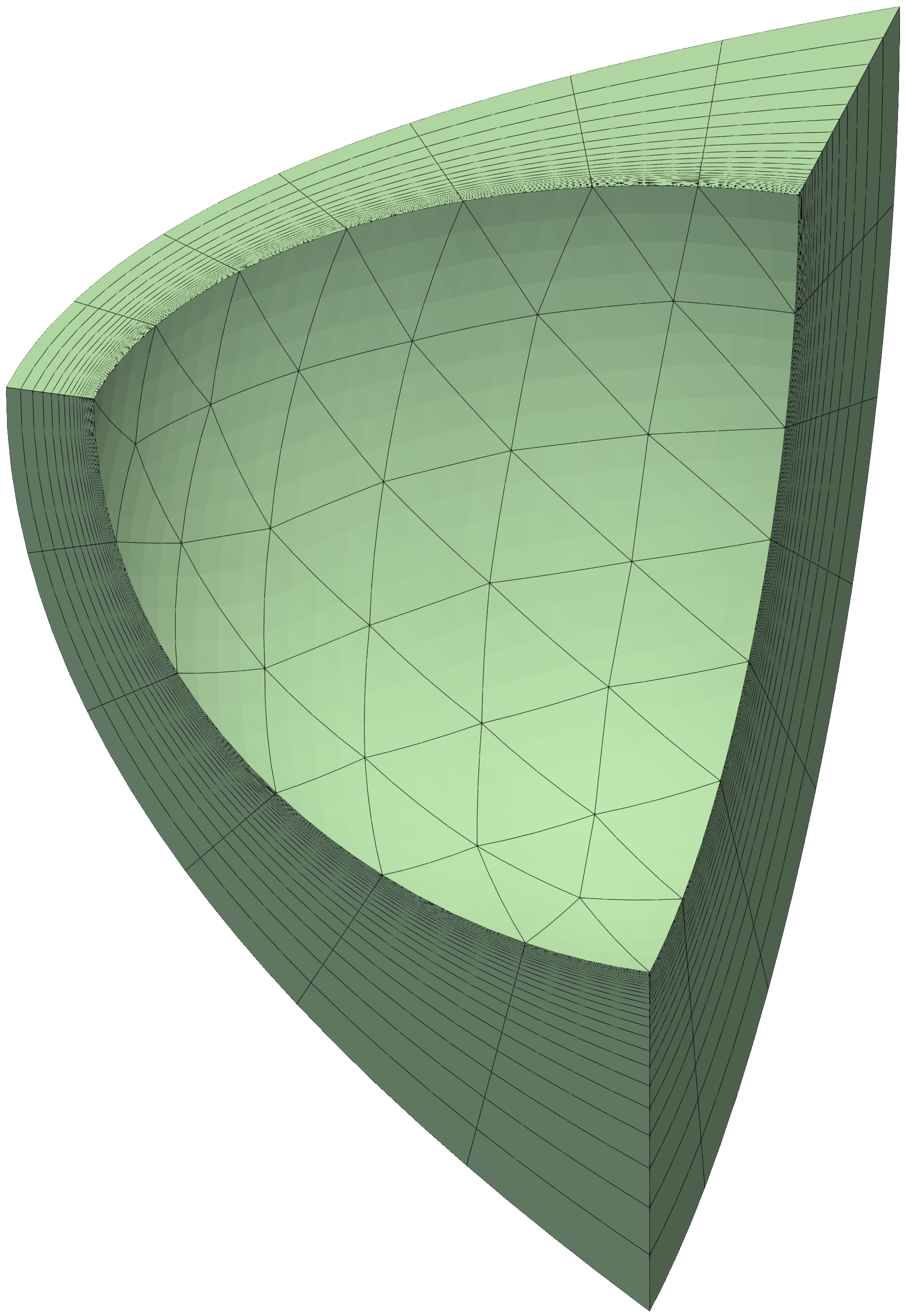}}
		\caption{
			Shock-aligned post-shock domain for the Mach~$5$ sphere test case including a slice along the
			$x_3 = 0$ plane (\textit{left}) and a three-dimensional view (\textit{middle}), as well as the
			boundary-layer mesh with $3600$ quadratic prismatic elements (\textit{right}). The boundary
			conditions are indicated in the left and middle figures: a no-slip, isothermal wall
			(\ref{line:sph2:wall}); the shock boundary condition, across which the oblique
			Rankine-Hugoniot post-shock state is imposed weakly through a Roe flux
			(\ref{line:sph2:shock}); and a subsonic outlet that enforces zero tangential viscous traction (\ref{line:sph2:outlet}). The $x_2 = 0$ and $x_3 = 0$ symmetry planes are slip walls (\ref{line:sph2:sym}).
		}
		\label{fig:ns3_geom1}
	\end{figure}

	\begin{figure}[!htbp]
		\centering
		\input{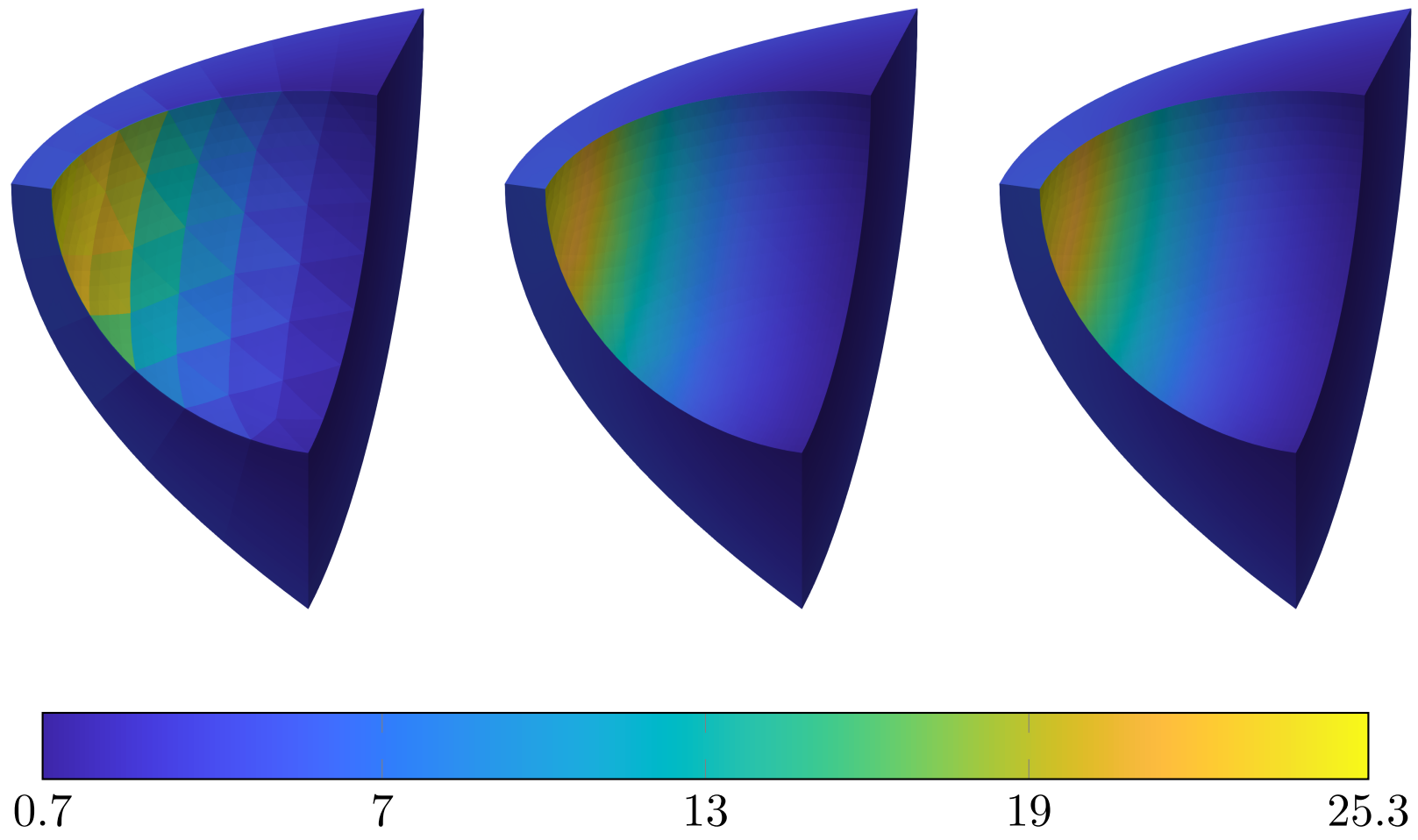}
		\caption{
			Navier-Stokes solution for the Mach~$5$ sphere ($\text{Re}_\infty = 943750$) on
			the boundary-layer mesh, colored by density over a shared range, through the
			$p = 0 \to 1 \to 2$ polynomial-degree ladder: the coarse $p = 0$ stage (\textit{left}), the
			converged $p = 1$ solution (\textit{middle}), and the converged $p = 2$ solution
			(\textit{right}). The post-shock density field sharpens as the degree increases.
		}
		\label{fig:ns3_soln}
	\end{figure}

	\begin{figure}[!htbp]
		\centering
		\input{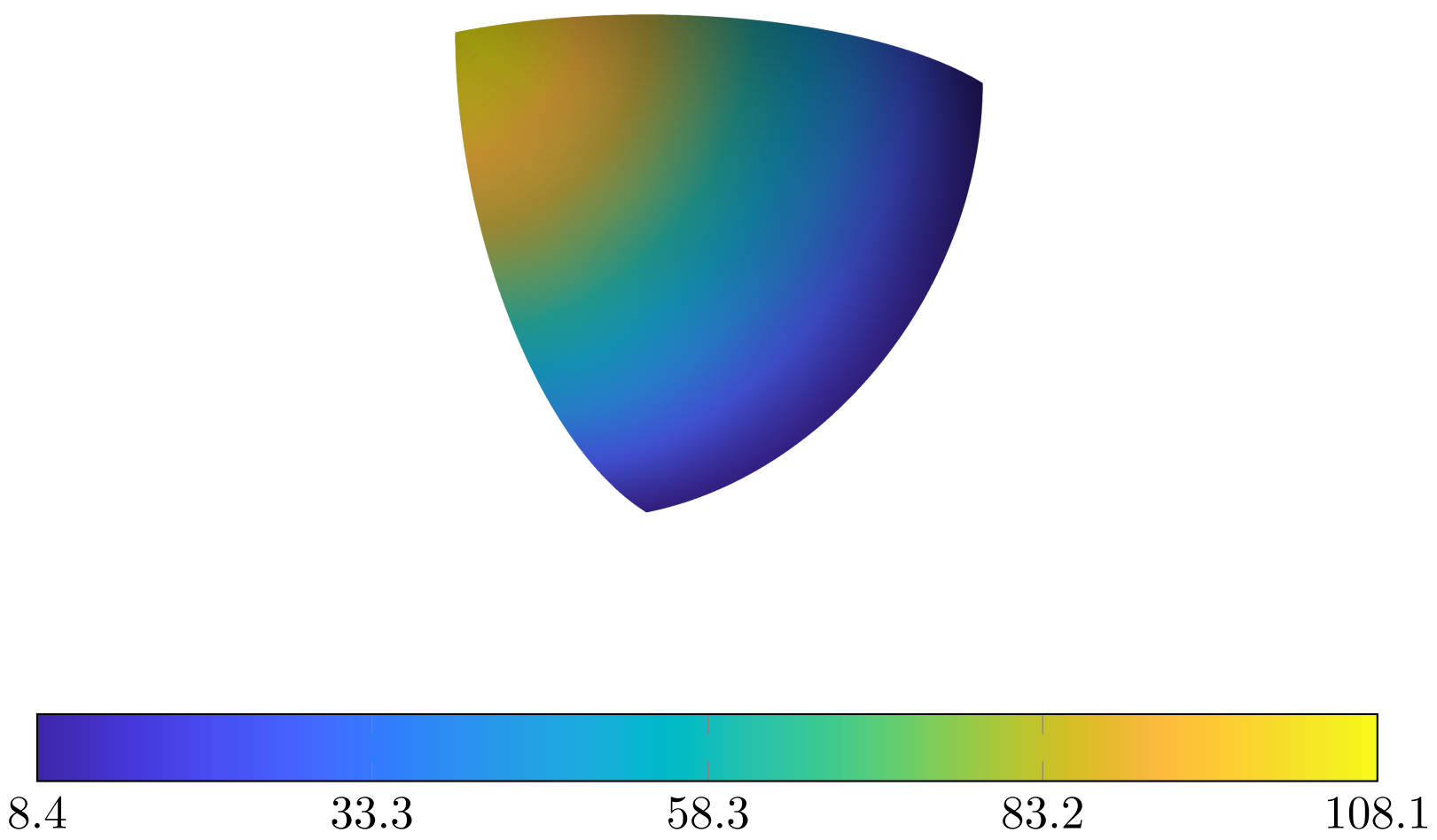}
		\caption{
			Wall heat flux over the sphere surface from the converged $p = 2$ Navier-Stokes solution
			in physical units ($\mathrm{kW/m^2}$, $q_{\mathrm{ref}} = \rho_\infty c_\infty^3 =
			465.4~\mathrm{kW/m^2}$). The stagnation-point heat flux is $108.07~\mathrm{kW/m^2}$.
		}
		\label{fig:ns3_hflx}
	\end{figure}

	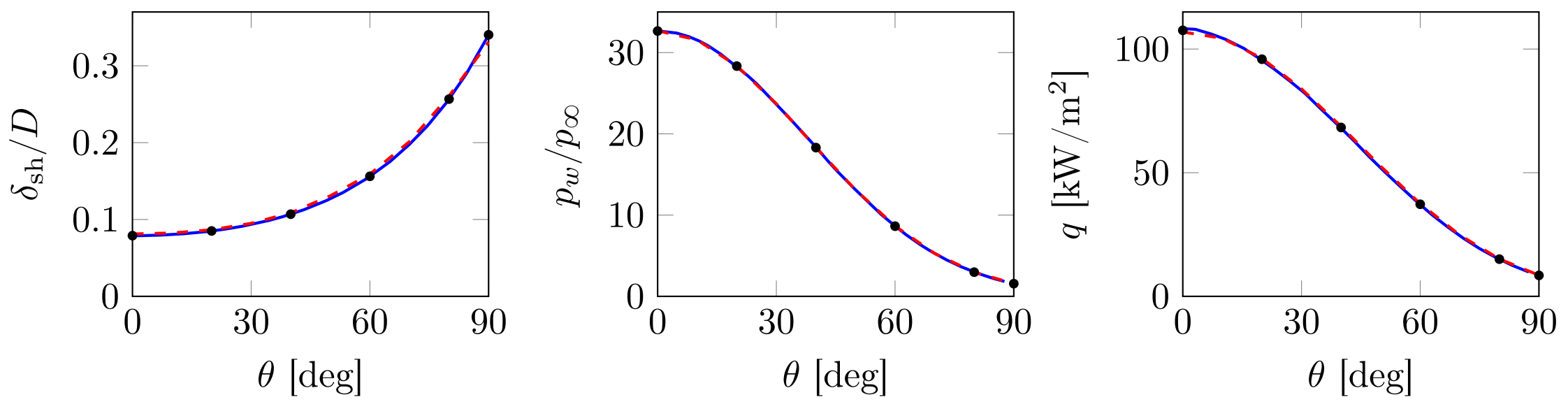
\begin{figure}[!htbp]
		\centering
		\begin{tikzpicture}
\begin{groupplot}[
group style={group size=3 by 1, horizontal sep=1.6cm},
width=0.3\textwidth, xmin=0, xmax=90, xtick={0,30,60,90}, xlabel={$\theta$ [deg]},]
\nextgroupplot[ylabel={$\delta_{\mathrm{sh}}/D$}, ymin=0, ymax=0.37]
\addplot[blue, thick] coordinates {(0.00,7.882e-2) (2.627,7.900e-2) (7.282,7.971e-2) (12.98,8.145e-2) (18.69,8.418e-2) (23.13,8.723e-2) (27.62,9.106e-2) (34.47,9.872e-2) (38.06,1.0379e-1) (43.31,1.1270e-1) (48.92,1.2440e-1) (53.24,1.3516e-1) (59.20,1.5316e-1) (65.11,1.7537e-1) (69.93,1.9743e-1) (74.47,2.2140e-1) (79.38,2.5220e-1) (84.36,2.8938e-1) (90.00,3.4006e-1)};\label{line:sph:ours}
\addplot[red, thick, dashed] coordinates {(0,8.113e-2) (10,8.240e-2) (20,8.662e-2) (30,9.496e-2) (40,1.0864e-1) (50,1.3079e-1) (60,1.5896e-1) (70,2.0138e-1) (80,2.6059e-1) (90,3.3082e-1)};
\addplot[black, only marks, mark=*, mark size=1.1] coordinates {(0,7.898e-2) (20,8.507e-2) (40,1.0691e-1) (60,1.5616e-1) (80,2.5668e-1) (90,3.4029e-1)};\label{mark:sph:blot}
\nextgroupplot[ylabel={$p_w/p_\infty$}, ymin=0, ymax=35]
\addplot[blue, thick] coordinates {(0.00,32.672) (4.78,32.406) (7.95,31.950) (10.56,31.415) (12.71,30.855) (15.90,29.829) (22.72,27.178) (25.07,26.139) (29.54,23.865) (33.67,21.747) (36.88,20.034) (39.81,18.477) (43.42,16.444) (45.33,15.432) (49.81,13.209) (53.65,11.410) (56.46,10.095) (59.19,8.983) (62.69,7.620) (66.41,6.348) (69.01,5.558) (73.01,4.488) (76.81,3.619) (81.19,2.785) (84.87,2.199) (87.65,1.836)};
\addplot[red, thick, dashed] coordinates {(0.0,32.685) (10.0,31.557) (20.0,28.375) (30.0,23.705) (40.0,18.334) (50.0,13.108) (60.0,8.658) (70.0,5.277) (80.0,2.976) (90.0,1.571)};\label{line:sph:capt}
\addplot[black, only marks, mark=*, mark size=1.1] coordinates {(0,32.6546) (20,28.3396) (40,18.3054) (60,8.6363) (80,2.9754) (90,1.5609)};
\nextgroupplot[ylabel={$q$ [kW/m$^2$]}, ymin=0, ymax=115]
\addplot[blue, thick] coordinates {(0.00,108.07) (0.53,108.227) (3.19,107.973) (7.16,106.101) (10.84,103.918) (15.44,100.274) (19.04,96.501) (22.86,92.198) (26.95,87.184) (30.81,82.176) (35.07,75.512) (39.17,69.295) (43.25,62.883) (47.17,56.360) (51.06,50.281) (54.93,44.457) (58.88,38.663) (63.01,32.928) (67.06,27.912) (70.97,23.427) (75.02,19.289) (79.11,15.704) (83.28,12.506) (87.44,9.837)};
\addplot[red, thick, dashed] coordinates {(0.0,106.900) (10.0,104.361) (20.0,96.286) (30.0,83.896) (40.0,68.713) (50.0,52.599) (60.0,37.492) (70.0,24.753) (80.0,15.134) (90.0,8.536)};
\addplot[black, only marks, mark=*, mark size=1.1] coordinates {(0,107.572) (20,95.894) (40,68.293) (60,37.232) (80,15.044) (90,8.472)};
\end{groupplot}
\end{tikzpicture}
		\caption{
			Surface quantities for the Mach~$5$ sphere as functions of the polar angle $\theta$ measured
			from the stagnation point: shock standoff distance normalized by diameter
			$\delta_{\mathrm{sh}}/D$ (\textit{left}), wall pressure ratio $p_w/p_\infty$ (\textit{middle}),
			and wall heat flux $q$ in $\mathrm{kW/m^2}$ (\textit{right}; $q_{\mathrm{ref}} = \rho_\infty
			c_\infty^3 = 465.4~\mathrm{kW/m^2}$). The shock-tracking result (\ref{line:sph:ours}) and the shock-capturing
			result (\ref{line:sph:capt}) are compared with the reference data of
			Blottner~\cite{blottner1989accurate} (\ref{mark:sph:blot}); stagnation-point values are
			tabulated in Table~\ref{tab:ns3_stag}. The solution is axisymmetric: the three meridians
			$\phi = 0^\circ$, $45^\circ$, and $90^\circ$ (the two symmetry planes and the plane between
			them) overlay to within $1\%$ at every station, so a single representative meridian
			($\phi = 0^\circ$) is shown. The standoff of the capturing solution is the ridge of the artificial
			viscosity field~\cite{marques2026pdeav}.
		}
		\label{fig:ns3_cmp}
	\end{figure}

	\begin{table}[!htbp]
		\centering
		\caption{
			Stagnation-point ($\theta = 0$) comparison for the Mach~$5$ sphere. Heat flux is
			dimensionalized with $q_{\mathrm{ref}} = \rho_\infty c_\infty^3 = 465.4~\mathrm{kW/m^2}$.
			The wall temperature is imposed in both computations. The standoff of the capturing
			solution is the ridge of the artificial viscosity field~\cite{marques2026pdeav}.
		}
		\label{tab:ns3_stag}
		{\small\setlength{\tabcolsep}{5pt}\begin{tabular}{l c c c c c}
\toprule
 & & \multicolumn{2}{c}{Shock tracking} & \multicolumn{2}{c}{Shock capturing} \\
\cmidrule(lr){3-4}\cmidrule(lr){5-6}
Quantity & Blottner~\cite{blottner1989accurate} & value & rel.\ err. & value & rel.\ err. \\
\midrule
Heat flux, $q_0$ [kW/m$^2$]                              & $107.572$ & $108.07$  & $+0.46\%$ & $106.90$  & $-0.63\%$ \\
Wall pressure, $p_0/p_\infty$                            & $32.655$  & $32.672$  & $+0.05\%$ & $32.685$  & $+0.09\%$ \\
Shock standoff, $\delta_{\mathrm{sh}}/D$                 & $0.0790$  & $0.0788$  & $-0.20\%$ & $0.0811$  & $+2.7\%$ \\
Wall temperature, $T_w/T_\infty$                         & $1.308$   & $1.308$   & ---       & $1.308$   & --- \\
\bottomrule
\end{tabular}
}
	\end{table}

	\begin{figure}[!htbp]
		\centering
		\input{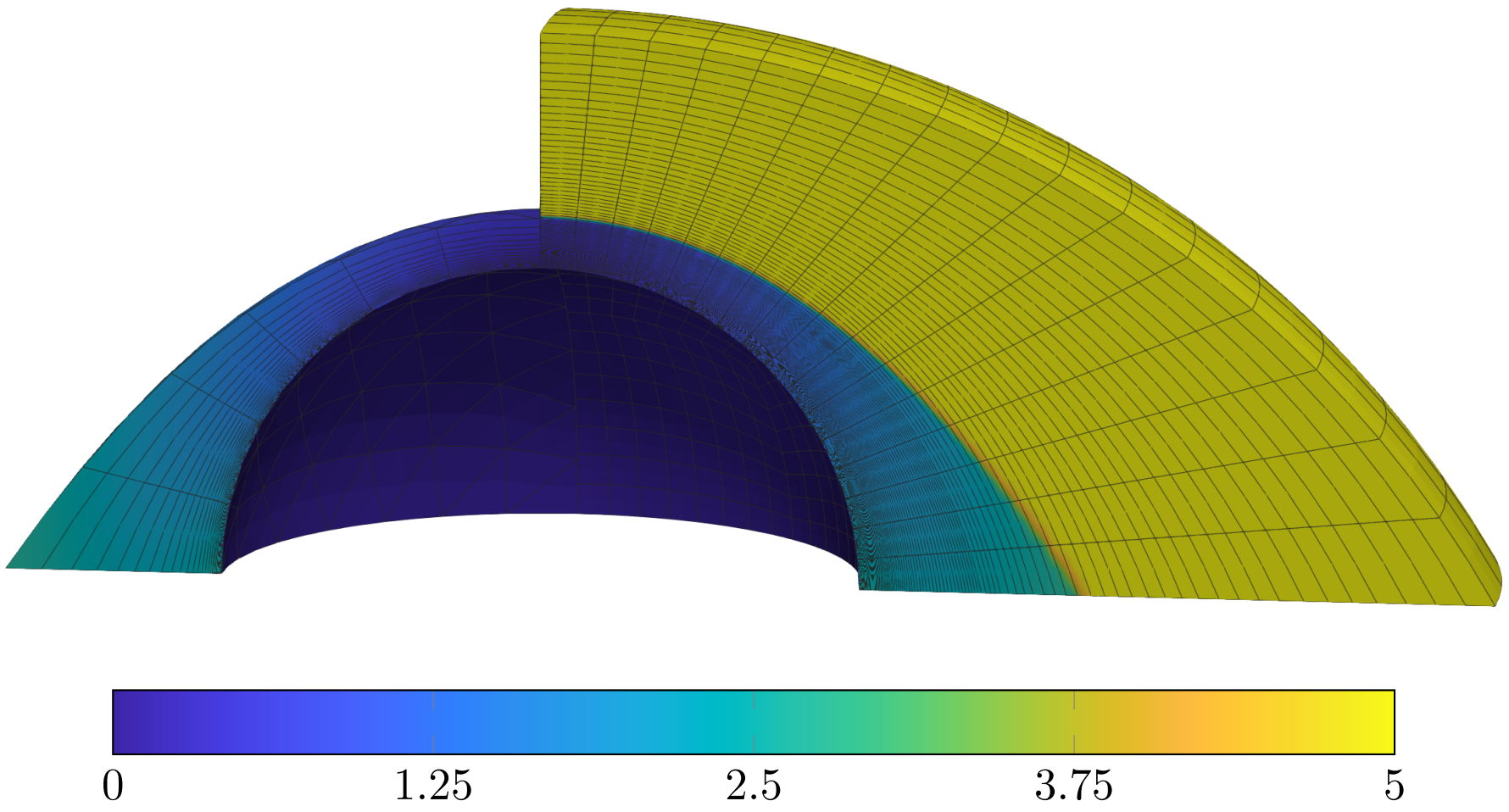}
		\caption{
			Mach number and mesh for the Mach~$5$ sphere ($\text{Re}_\infty = 943750$): the shock-fitted
			solution of Figure~\ref{fig:ns3_soln} on the post-shock domain bounded by the fitted shock (\textit{left})
			and a shock-capturing solution ($p = 2$ with the PDE-based artificial viscosity
			of~\cite{marques2026pdeav}) on an octant mesh extending to the far field (\textit{right}).
		}
		\label{fig:ns3_capture}
	\end{figure}
}

\section{Conclusion}
\label{sec:conclude}
We introduce a numerical framework based on the high-order implicit shock tracking method (HOIST)~\cite{zahr2020implicit,huang2022robust,huang2023high}
for solving hypersonic problems with strong shocks and viscous effects. The approach applies implicit shock 
tracking on a shock-agnostic mesh to generate a shock-aligned mesh and the corresponding flow field. The final 
hypersonic solution is obtained through Mach number continuation, in which the freestream Mach number is 
gradually increased to the target value to ensure robust convergence. 
The computational procedure begins with inviscid HOIST simulations, which align the mesh with the shock on a coarse
Euler mesh without evaluating viscous terms. Starting from the inviscid formulation yields an accurate shock geometry
without resolving the viscous shock structure. This avoids the severe element compression and specialized formulations
that fully coupled viscous tracking requires at these conditions. The shock tracking optimization is confined to the
coarse inviscid mesh, and the viscous computation on the boundary-layer mesh involves no mesh unknowns. Once a shock-aligned mesh is obtained through the HOIST method, all elements upstream of the bow shock 
are removed to form a reduced downstream computational domain, based on the assumption that shock thickness 
can be neglected under high Mach and Reynolds number flow conditions. Oblique shock relations are then used to 
prescribe boundary conditions along the shock-aligned boundary of the post-shock domain. 
Viscous computations are subsequently performed on the boundary-layer-refined post-shock mesh to recover downstream viscous effects, including boundary layers and surface heat-transfer predictions. Finally, 
goal-oriented mesh adaptation is employed to further enhance resolution in regions that most strongly 
influence viscous and thermal quantities of interest.

In this work, we use an axisymmetric blunt cone, a two-dimensional cylinder, and a three-dimensional sphere to demonstrate the robustness and flexibility of the proposed framework for viscous hypersonic problems. These canonical examples feature strong bow shocks, thin boundary layers, and severe thermal gradients, providing a stringent assessment of the method across different geometries, flow conditions, and spatial dimensions. For all three configurations, the framework successfully obtains the shock structure, boundary layers, and post-shock viscous effects on shock-aligned meshes. In particular, the surface heat flux predicted for the blunt cone agrees well with experimental measurements, and the surface pressure and heat flux predicted for the cylinder and sphere agree well with reference numerical results. These results demonstrate that the proposed approach can reliably recover viscous and 
thermal quantities of interest in hypersonic regimes while maintaining robustness and computational efficiency.
\section*{Acknowledgments}
The authors thank Aaron Marques for providing the shock-capturing solution of the Mach~5 sphere
used for comparison in Section~\ref{subsec:sph3d}.
This work is supported by AFOSR award numbers FA9550-20-1-0236,
FA9550-22-1-0002, FA9550-22-1-0004, ONR award number
N00014-22-1-2299, and NSF award number CBET-2338843.
The content of this publication does not necessarily reflect the position
or policy of any of these supporters, and no official endorsement should
be inferred.

\bibliographystyle{plain}
\bibliography{biblio}

\begin{thebibliography}{10}

\bibitem{anderson2006hypersonic}
John~D. Anderson.
\newblock {\em Hypersonic and High-Temperature Gas Dynamics}.
\newblock American Institute of Aeronautics and Astronautics, Reston, VA, 2nd
  edition, 2006.

\bibitem{anderson2011ebook}
John~D. Anderson.
\newblock {\em Fundamentals of Aerodynamics}.
\newblock McGraw-Hill, New York, 5th edition, 2011.

\bibitem{arnold1982interior}
Douglas~N. Arnold.
\newblock An interior penalty finite element method with discontinuous
  elements.
\newblock {\em SIAM Journal on Numerical Analysis}, 19(4):742--760, 1982.

\bibitem{arnold2002unified}
Douglas~N. Arnold, Franco Brezzi, Bernardo Cockburn, and L.~Donatella Marini.
\newblock Unified analysis of discontinuous {G}alerkin methods for elliptic
  problems.
\newblock {\em SIAM Journal on Numerical Analysis}, 39(5):1749--1779, 2002.

\bibitem{barter2010shock}
Garrett~E. Barter and David~L. Darmofal.
\newblock Shock capturing with {PDE}-based artificial viscosity for {DGFEM}:
  {P}art {I}. {F}ormulation.
\newblock {\em Journal of Computational Physics}, 229(5):1810--1827, 2010.

\bibitem{becker1996feed}
Roland Becker and Rolf Rannacher.
\newblock A feed-back approach to error control in finite element methods:
  Basic analysis and examples.
\newblock {\em East-West Journal of Numerical Mathematics}, 4:237--264, 1996.

\bibitem{Becker_2001_DWR}
Roland Becker and Rolf Rannacher.
\newblock An optimal control approach to a posteriori error estimation in
  finite element methods.
\newblock {\em Acta Numerica}, 10:1--102, 2001.

\bibitem{blottner1989accurate}
F.~G. Blottner.
\newblock Accurate {Navier--Stokes} results for the hypersonic flow over a
  spherical nosetip.
\newblock In {\em 24th Thermophysics Conference}, Buffalo, NY, 1989.
\newblock AIAA Paper 89-1671.

\bibitem{bonfiglioli2016unsteady}
Aldo Bonfiglioli, Renato Paciorri, and Lorenzo Campoli.
\newblock Unsteady shock-fitting for unstructured grids.
\newblock {\em International Journal for Numerical Methods in Fluids},
  81(4):245--261, 2016.

\bibitem{candler2007unstructured}
Graham Candler, Michael Barnhardt, Travis Drayna, Ioannis Nompelis, David
  Peterson, and Pramod Subbareddy.
\newblock Unstructured grid approaches for accurate aeroheating simulations.
\newblock In {\em 18th AIAA Computational Fluid Dynamics Conference}, 2007.
\newblock AIAA Paper 2007-3959.

\bibitem{candler2015next}
Graham~V. Candler.
\newblock Next-generation {CFD} for hypersonic and aerothermal flows.
\newblock In {\em 22nd AIAA Computational Fluid Dynamics Conference}, Dallas,
  Texas, 2015.
\newblock AIAA Paper 2015-3048.

\bibitem{candler2009current}
Graham~V. Candler, Dimitri~J. Mavriplis, and Loretta Trevino.
\newblock Current status and future prospects for the numerical simulation of
  hypersonic flows.
\newblock In {\em 47th AIAA Aerospace Sciences Meeting including The New
  Horizons Forum and Aerospace Exposition}, Orlando, Florida, 2009.
\newblock AIAA Paper 2009-153.

\bibitem{candler2015advances}
Graham~V. Candler, Pramod~K. Subbareddy, and Joseph~M. Brock.
\newblock Advances in computational fluid dynamics methods for hypersonic
  flows.
\newblock {\em Journal of Spacecraft and Rockets}, 52(1):17--28, 2015.

\bibitem{ching2024moving}
Eric~J. Ching, Andrew~D. Kercher, and Andrew Corrigan.
\newblock The moving discontinuous {G}alerkin method with interface condition
  enforcement for the simulation of hypersonic, viscous flows.
\newblock {\em Computer Methods in Applied Mechanics and Engineering},
  427:117045, 2024.

\bibitem{ching2019shock}
Eric~J. Ching, Yu~Lv, Peter Gnoffo, Michael Barnhardt, and Matthias Ihme.
\newblock Shock capturing for discontinuous {G}alerkin methods with application
  to predicting heat transfer in hypersonic flows.
\newblock {\em Journal of Computational Physics}, 376:54--75, 2019.

\bibitem{ciallella2020extrapolated}
Mirco Ciallella, Mario Ricchiuto, Renato Paciorri, and Aldo Bonfiglioli.
\newblock Extrapolated shock tracking: bridging shock-fitting and embedded
  boundary methods.
\newblock {\em Journal of Computational Physics}, 412:109440, 2020.

\bibitem{cleary1965experimental}
Joseph~W. Cleary.
\newblock An experimental and theoretical investigation of the pressure
  distribution and flow fields of blunted cones at hypersonic {M}ach numbers.
\newblock Technical Note NASA TN D-2969, National Aeronautics and Space
  Administration, 1965.

\bibitem{cleary1969effects}
Joseph~W. Cleary.
\newblock Effects of angle of attack and bluntness on laminar heating-rate
  distributions of a $15^\circ$ cone at a {M}ach number of 10.6.
\newblock Technical Note NASA TN D-5450, National Aeronautics and Space
  Administration, 1969.

\bibitem{corrigan2019moving}
Andrew Corrigan, Andrew~D. Kercher, and David~A. Kessler.
\newblock A moving discontinuous {G}alerkin finite element method for flows
  with interfaces.
\newblock {\em International Journal for Numerical Methods in Fluids},
  89(9):362--406, 2019.

\bibitem{dong2025rpadapt}
Huijing Dong, Masayuki Yano, Tianci Huang, and Matthew~J. Zahr.
\newblock An $rp$-adaptive method for accurate resolution of shock-dominated
  viscous flow based on implicit shock tracking.
\newblock {\em Journal of Computational Physics}, 543:114337, 2025.

\bibitem{eriksson1982generation}
Lars-Erik Eriksson.
\newblock Generation of boundary-conforming grids around wing-body
  configurations using transfinite interpolation.
\newblock {\em AIAA Journal}, 20(10):1313--1320, 1982.

\bibitem{fernandez2018physics}
Pablo Fernandez, Cuong Nguyen, and Jaime Peraire.
\newblock A physics-based shock capturing method for unsteady laminar and
  turbulent flows.
\newblock In {\em 2018 AIAA Aerospace Sciences Meeting}, Kissimmee, Florida,
  2018.
\newblock AIAA Paper 2018-0062.

\bibitem{fidkowski2011review}
Krzysztof~J. Fidkowski and David~L. Darmofal.
\newblock Review of output-based error estimation and mesh adaptation in
  computational fluid dynamics.
\newblock {\em AIAA Journal}, 49(4):673--694, 2011.

\bibitem{geisenhofer2020extended}
Markus Geisenhofer, Florian Kummer, and Martin Oberlack.
\newblock An extended discontinuous {G}alerkin method for high-order
  shock-fitting.
\newblock {\em arXiv preprint arXiv:2012.08860}, 2020.

\bibitem{gnoffo2003computational}
Peter Gnoffo.
\newblock Computational fluid dynamic technology for hypersonic applications.
\newblock In {\em AIAA International Air and Space Symposium and Exposition:
  The Next 100 Years}, 2003.
\newblock AIAA Paper 2003-2829.

\bibitem{gnoffo2004computational}
Peter Gnoffo and Jeffery White.
\newblock Computational aerothermodynamic simulation issues on unstructured
  grids.
\newblock In {\em 37th AIAA Thermophysics Conference}, 2004.
\newblock AIAA Paper 2004-2371.

\bibitem{gordon1973transfinite}
William~J. Gordon and Charles~A. Hall.
\newblock Transfinite element methods: blending-function interpolation over
  arbitrary curved element domains.
\newblock {\em Numerische Mathematik}, 21(2):109--129, 1973.

\bibitem{harten1997uniformly}
Ami Harten, Bjorn Engquist, Stanley Osher, and Sukumar~R. Chakravarthy.
\newblock Uniformly high order accurate essentially non-oscillatory schemes,
  {III}.
\newblock {\em Journal of Computational Physics}, 131(1):3--47, 1997.

\bibitem{hartmann2008optimal}
Ralf Hartmann and Paul Houston.
\newblock An optimal order interior penalty discontinuous {G}alerkin
  discretization of the compressible {N}avier--{S}tokes equations.
\newblock {\em Journal of Computational Physics}, 227(22):9670--9685, 2008.

\bibitem{hartmann2014higher}
Ralf Hartmann and Tobias Leicht.
\newblock Higher order and adaptive {DG} methods for compressible flows.
\newblock {VKI} Lecture Series 2014-03, von Karman Institute for Fluid
  Dynamics, Rhode-Saint-Gen\`ese, Belgium, 2014.
\newblock 37th Advanced {CFD} Lecture Series: Recent developments in higher
  order methods and industrial application in aeronautics.

\bibitem{huang2023high}
Tianci Huang, Charles~J. Naudet, and Matthew~J. Zahr.
\newblock High-order implicit shock tracking boundary conditions for flows with
  parametrized shocks.
\newblock {\em Journal of Computational Physics}, 495:112517, 2023.

\bibitem{huang2022robust}
Tianci Huang and Matthew~J. Zahr.
\newblock A robust, high-order implicit shock tracking method for simulation of
  complex, high-speed flows.
\newblock {\em Journal of Computational Physics}, 454:110981, 2022.

\bibitem{jiang1996efficient}
Guang-Shan Jiang and Chi-Wang Shu.
\newblock Efficient implementation of weighted {ENO} schemes.
\newblock {\em Journal of Computational Physics}, 126(1):202--228, 1996.

\bibitem{johnsen2010assessment}
Eric Johnsen, Johan Larsson, Ankit~V. Bhagatwala, William~H. Cabot, Parviz
  Moin, Britton~J. Olson, Pradeep~S. Rawat, Santhosh~K. Shankar, Bj{\"o}rn
  Sj{\"o}green, Helen~C. Yee, Xiaolin Zhong, and Sanjiva~K. Lele.
\newblock Assessment of high-resolution methods for numerical simulations of
  compressible turbulence with shock waves.
\newblock {\em Journal of Computational Physics}, 229(4):1213--1237, 2010.

\bibitem{kercher2021least}
Andrew~D. Kercher and Andrew Corrigan.
\newblock A least-squares formulation of the moving discontinuous {G}alerkin
  finite element method with interface condition enforcement.
\newblock {\em Computers \& Mathematics with Applications}, 95:143--171, 2021.

\bibitem{kercher2021moving}
Andrew~D. Kercher, Andrew Corrigan, and David~A. Kessler.
\newblock The moving discontinuous {G}alerkin finite element method with
  interface condition enforcement for compressible viscous flows.
\newblock {\em International Journal for Numerical Methods in Fluids},
  93(5):1490--1519, 2021.

\bibitem{kopriva1993spectral}
David~A. Kopriva.
\newblock Spectral solution of the viscous blunt-body problem.
\newblock {\em AIAA Journal}, 31(7):1235--1242, 1993.

\bibitem{kopriva1996spectral}
David~A. Kopriva.
\newblock Spectral solution of the viscous blunt-body problem. {II}:
  {M}ultidomain approximation.
\newblock {\em AIAA Journal}, 34(3):560--564, 1996.

\bibitem{lee1999spurious}
Theodore~Kai Lee and Xiaolin Zhong.
\newblock Spurious numerical oscillations in simulation of supersonic flows
  using shock-capturing schemes.
\newblock {\em AIAA Journal}, 37(3):313--319, 1999.

\bibitem{liu1994weighted}
Xu-Dong Liu, Stanley Osher, and Tony Chan.
\newblock Weighted essentially non-oscillatory schemes.
\newblock {\em Journal of Computational Physics}, 115(1):200--212, 1994.

\bibitem{marques2026pdeav}
Aaron Marques, Alireza~H. Razavi, Masayuki Yano, and Matthew~J. Zahr.
\newblock A principled framework for {PDE}-based shock capturing in high-order
  discontinuous {G}alerkin methods.
\newblock Submitted to Journal of Computational Physics, 2026.

\bibitem{masatsuka2009like}
Katate Masatsuka.
\newblock {\em I Do Like {CFD}}, volume~1.
\newblock Lulu, 2009.

\bibitem{modisette2011automated}
James~M. Modisette.
\newblock {\em An automated reliable method for two-dimensional
  {R}eynolds-averaged {N}avier-{S}tokes simulations}.
\newblock PhD thesis, Massachusetts Institute of Technology, 2011.

\bibitem{moretti1969blunt}
Gino Moretti.
\newblock The blunt body problem for a viscous rarefied gas flow.
\newblock In {\em 7th Aerospace Sciences Meeting}, 1969.
\newblock AIAA Paper 69-139.

\bibitem{moretti2002thirty}
Gino Moretti.
\newblock Thirty-six years of shock fitting.
\newblock {\em Computers \& Fluids}, 31(4-7):719--723, 2002.

\bibitem{moretti1970numerical}
Gino Moretti and Manuel~D. Salas.
\newblock Numerical analysis of viscous one-dimensional flows.
\newblock {\em Journal of Computational Physics}, 5(3):487--506, 1970.

\bibitem{nocedal2006numerical}
Jorge Nocedal and Stephen~J. Wright.
\newblock {\em Numerical Optimization}.
\newblock Springer, New York, 2nd edition, 2006.

\bibitem{nompelis2004development}
Ioannis Nompelis, Travis Drayna, and Graham Candler.
\newblock Development of a hybrid unstructured implicit solver for the
  simulation of reacting flows over complex geometries.
\newblock In {\em 34th AIAA Fluid Dynamics Conference and Exhibit}, 2004.
\newblock AIAA Paper 2004-2227.

\bibitem{paciorri2009shock}
Renato Paciorri and Aldo Bonfiglioli.
\newblock A shock-fitting technique for {2D} unstructured grids.
\newblock {\em Computers \& Fluids}, 38(3):715--726, 2009.

\bibitem{peery1988blunt}
K.~M. Peery and S.~T. Imlay.
\newblock Blunt-body flow simulations.
\newblock In {\em 24th Joint Propulsion Conference}, Boston, Massachusetts,
  1988.
\newblock AIAA Paper 88-2904.

\bibitem{persson2006sub}
Per-Olof Persson and Jaime Peraire.
\newblock Sub-cell shock capturing for discontinuous {G}alerkin methods.
\newblock In {\em 44th AIAA Aerospace Sciences Meeting and Exhibit}, Reno,
  Nevada, 2006.
\newblock AIAA Paper 2006-112.

\bibitem{salas2009shock}
Manuel~D. Salas.
\newblock {\em A Shock-Fitting Primer}.
\newblock CRC Press, 2009.

\bibitem{van1979towards}
Bram van Leer.
\newblock Towards the ultimate conservative difference scheme. {V}. {A}
  second-order sequel to {G}odunov's method.
\newblock {\em Journal of Computational Physics}, 32(1):101--136, 1979.

\bibitem{widhopf1972transitional}
George~F. Widhopf and Robert Hall.
\newblock Transitional and turbulent heat-transfer measurements on a yawed
  blunt conical nosetip.
\newblock {\em AIAA Journal}, 10(10):1318--1325, 1972.

\bibitem{zahr2018shktrk}
Matthew~J. Zahr and Per-Olof Persson.
\newblock An optimization-based approach for high-order accurate discretization
  of conservation laws with discontinuous solutions.
\newblock {\em Journal of Computational Physics}, 365:105--134, 2018.

\bibitem{zahr2020implicit}
Matthew~J. Zahr, Andrew Shi, and Per-Olof Persson.
\newblock Implicit shock tracking using an optimization-based high-order
  discontinuous {G}alerkin method.
\newblock {\em Journal of Computational Physics}, 410:109385, 2020.

\bibitem{zou2021embedded}
Dongyang Zou, Aldo Bonfiglioli, Renato Paciorri, and Jun Liu.
\newblock An embedded shock-fitting technique on unstructured dynamic grids.
\newblock {\em Computers \& Fluids}, 218:104847, 2021.

\end{thebibliography}

\end{document}